%% file: main.tex
\documentclass[journal]{IEEEtran}

\usepackage{cite}
\usepackage{array}
\usepackage{color}
\usepackage{float}
\usepackage[cmex10]{amsmath}
\usepackage{nomencl}						
\usepackage[normalem]{ulem}			        
\usepackage{diagbox}						
\usepackage{colortbl}						
\usepackage{multirow}						
\usepackage{tabularx}
\usepackage{placeins}
\usepackage{algorithm}
\usepackage{mathrsfs}
\usepackage{mathdots}
\usepackage{amssymb}
\usepackage{dblfloatfix}
\usepackage{amsthm}
\usepackage{arydshln}
\usepackage{color}
\usepackage[noend]{algpseudocode}
\usepackage{bm}
\usepackage{url}
\usepackage[hidelinks]{hyperref}
\usepackage[T1]{fontenc}
\newcommand{\subparagraph}{}
\usepackage{fancyhdr} 
\usepackage{extarrows}
\usepackage{enumitem}
\usepackage{booktabs}
\usepackage{threeparttable}

\usepackage{booktabs,siunitx,makecell}
\usepackage{pifont}

\ifCLASSINFOpdf
  \usepackage[pdftex]{graphicx}
	\graphicspath{{./figure/}}
  \DeclareGraphicsExtensions{.pdf,.jpeg,.png}
\else
  \usepackage[dvips]{graphicx}
  \graphicspath{{./figure/}}
  \DeclareGraphicsExtensions{.eps}
\fi

\ifCLASSOPTIONcompsoc
  \usepackage[caption=false,font=normalsize,labelfont=sf,textfont=sf]{subfig}
\else
  \usepackage[caption=false,font=footnotesize]{subfig}
\fi

\newcommand{\figref}[1]{\figurename~\ref{#1}}
\newcommand{\tabref}[1]{Table~\ref{#1}}

\ifCLASSOPTIONonecolumn

\else

\renewcommand{\arraystretch}{1.5}
\fi

\begin{document}
\setcounter{page}{1}
\vspace{-2\baselineskip}
\title{Large-Signal Stability of Power Systems with \\Mixtures of GFL, GFM and GSP Inverters}
\author{Yifan~Zhang, \IEEEmembership{Student Member, IEEE}, Yaoxin~Wang, \IEEEmembership{Student Member, IEEE}, Yunjie~Gu, \IEEEmembership{Senior Member, IEEE}, Yitong Li, \IEEEmembership{Member, IEEE}, Sijia Geng, \IEEEmembership{Member, IEEE}, Yue Zhu, \IEEEmembership{Member, IEEE},\\ Hsiao-Dong~Chiang, \IEEEmembership{Fellow, IEEE}, Timothy~C.~Green, \IEEEmembership{Fellow, IEEE} 
\thanks{This work was supported by the Royal Society under the award ICA{\textbackslash}R2{\textbackslash}242103 and the Ralph O'Connor Sustainable Energy Institute at Johns Hopkins University.
Yifan Zhang, Yaoxin Wang, Yunjie Gu, and Timothy~C.~Green are with Imperial College, London, UK, Yitong Li is with Xi'an Jiaotong University, Xi'an, China, Sijia Geng is with Johns Hopkins University, Baltimore, MD, USA, Yue Zhu is with City University of Hong Kong, Hong Kong, China, and Hsiao-Dong~Chiang is with Cornell University, Ithaca, NY, USA.
(e-mails: 
\url{yifan.zhang21@imperial.ac.uk};
\url{y.wang24@imperial.ac.uk};
\url{yunjie.gu@imperial.ac.uk};
\url{yitongli@xjtu.edu.cn};
\url{sgeng@jhu.edu};
\url{yue.zhu@cityu.edu.hk};
\url{hc63@cornell.edu};
\url{t.green@imperial.ac.uk}.
Corresponding author: Yunjie Gu.)
}
}
\IEEEaftertitletext{\vspace{-2.5\baselineskip}}
\ifCLASSOPTIONpeerreview
	\maketitle 
\else
	\maketitle
\fi
\thispagestyle{fancy}
\lhead{}
\rhead{\thepage}
\cfoot{}
\renewcommand{\headrulewidth}{0pt}
\pagestyle{fancy}

\begin{abstract}
Grid-following (GFL) inverters have very different large-signal stability characteristics to synchronous generators, and convenient concepts such as the equal-area criterion and global energy function do not apply in the same way. Existing studies mainly focus on the synchronization stability of an individual GFL inverter, and interactions between multiple inverters are less often addressed. This paper elucidates the interaction mechanisms of inverters with heterogeneous control, covering  GFL, grid-forming (GFM), and grid-supporting (GSP) inverters, to determine the stability boundaries of a system with mixtures of inverters. The generalized model for large-signal dynamics of two-inverter systems with various inverter combinations are derived. This establishes that no global energy function exists for systems containing heterogeneous inverters, implying that traditional direct methods of stability assessment cannot be applied to such systems. To overcome this barrier, a manifold method is employed to accurately determine the region of attraction (ROA) of such systems. To address the computational complexity of the manifold method, reduced-order models of inverter are used based on multiscale analysis. The large-signal stability margin is assessed by the shortest distance from a stable equilibrium point (SEP) to the boundary of the ROA, which is called the stability radius (SR). The manifold analysis, verified by EMT simulation, demonstrates that GFM and GSP inverters significantly enhance the large-signal stability of a two-inverter system where the other inverter is GFL, with GFM being slightly better than GSP. Such stability enhancement is attributed to the voltage support effects of GSP and GFM inverters. The improvement is maximized if the GFM or GSP inverter is placed at the mid-point of the transmission line, where the voltage is lowest. All findings in this paper are also validated through power hardware-in-the-loop (PHIL).
\end{abstract}


\input{paper}
\input{appendix}


\ifCLASSOPTIONcaptionsoff
  \newpage
\fi
\bibliographystyle{IEEEtran}
\bibliography{References}

\end{document}

%% file: paper.tex
\section{Introduction} \label{section_intro}
Grid-following (GFL) inverters, which rely on a phase-locked loop (PLL) for synchronization with the grid, are widely used to interface renewable energy generation with the power grid. The incidents of PLL losing synchronization when encountering a grid fault have been frequently reported in recent years \cite{Gu_proceeding, Xiongfei_overview,9309174, 8632731}. While the synchronization challenges of an individual GFL inverter are well-documented, the dynamics become more complex when multiple inverters are interconnected via transmission lines. This complexity is further compounded by the presence of other types of inverters, such as grid-forming (GFM) \cite{Xiongfei_overview} and grid-supporting (GSP) inverters \cite{GSP}, whose interactions can significantly influence synchronization neighboring GFL inverters. The discussion  here will be of threats to synchronization induced by large perturbations such as line faults, so the issue is termed large-signal stability and the analysis must capture the  nonlinear behaviors present.

The large-signal stability of traditional power systems dominated by synchronous generators (SGs) has been extensively studied. The direct method, which avoids computationally expensive time-domain integration of post-fault system dynamics, is well-established and has become one of the most widely used approaches for traditional systems\cite{chiang2011direct}. The method estimates the region of attraction (ROA) of the operating point to a stable equilibrium point (SEP) through the level set of the global energy function through an unstable equilibrium point (UEP). The construction of global energy functions is a vital part of the direct method. However, \cite{chiang2011direct, chiang1989study} prove that global energy functions do not exist for lossy power systems which have a transfer conductance introduced by the losses. Because the transfer conductance is typically small, there are various approaches that can mitigate this issue, such as numerical approximations \cite{chiang2011direct}, extended invariant principle method \cite{pota1989new}, and dissipative systems theory \cite{bretas2003lyapunov}.

However, in power systems dominated by inverter-based resources (IBRs), the construction of global energy functions presents a great challenge because of the differences in dynamics between IBR and SG. Although a global energy function can be constructed for a single-inverter-infinite-bus system \cite{9309174, hu2019large}, for multi-inverter systems,  GFL inverters will introduce cosine interaction terms in differential equations, as will be demonstrated in this paper. This effect is similar to the characteristics of transfer conductance in conventional power systems \cite{chiang1989study}. Unfortunately, the cosine interaction terms induced by GFL inverters are much larger than those introduced by transfer conductance in traditional SG systems, exacerbating the challenge. As a result, global energy functions generally do not exist in power systems containing GFL inverters. There are no effective means to date to mitigate this issue, which prevents the application of the direct method to IBR-dominated systems.

\begin{figure*}[th]
\centering
\subfloat[modelling of GFM inverters.]{\includegraphics[width=0.3\textwidth]{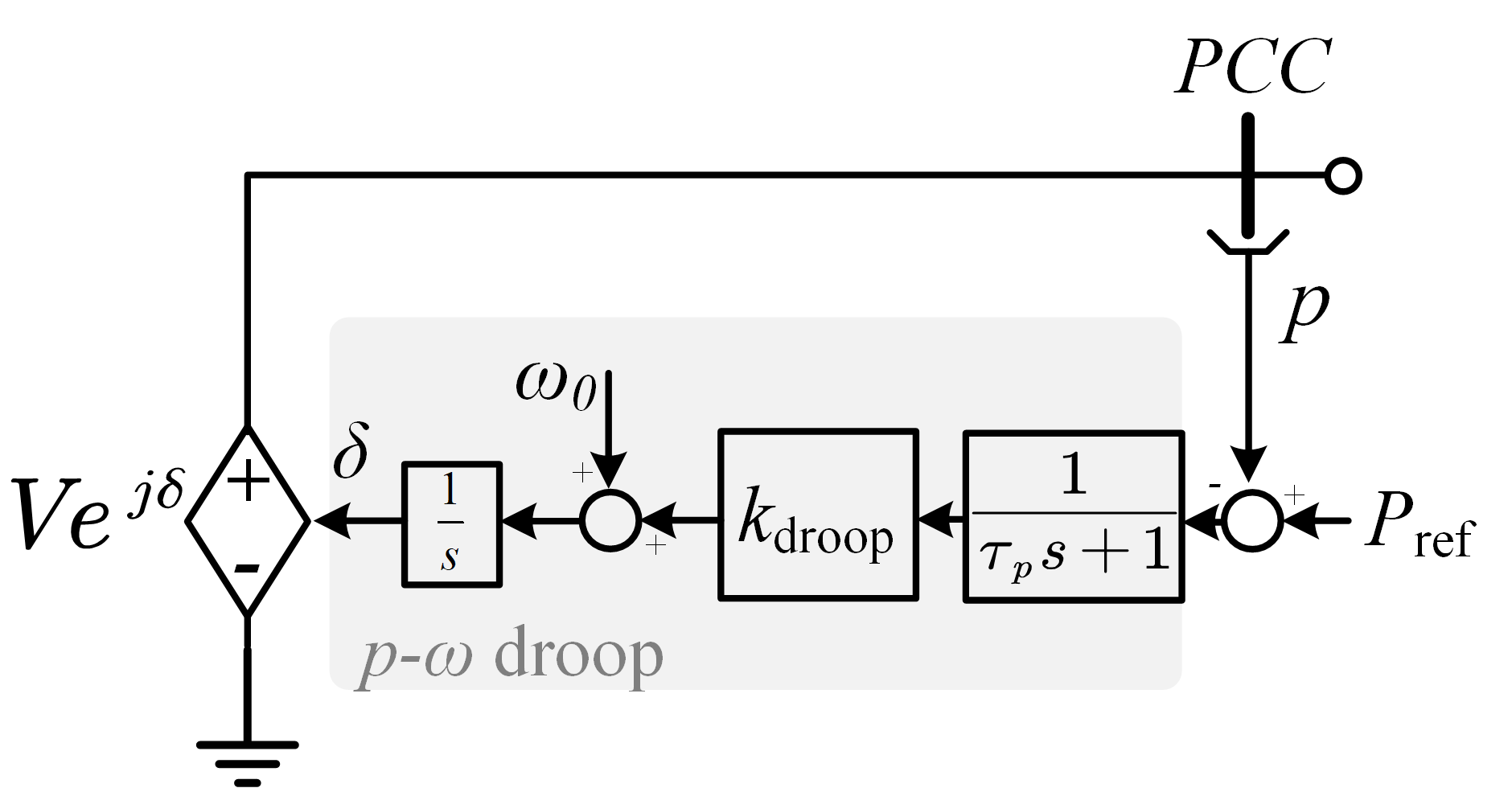}}
\hfil
\subfloat[modelling of GFL inverters.]{\includegraphics[width=0.26\textwidth]{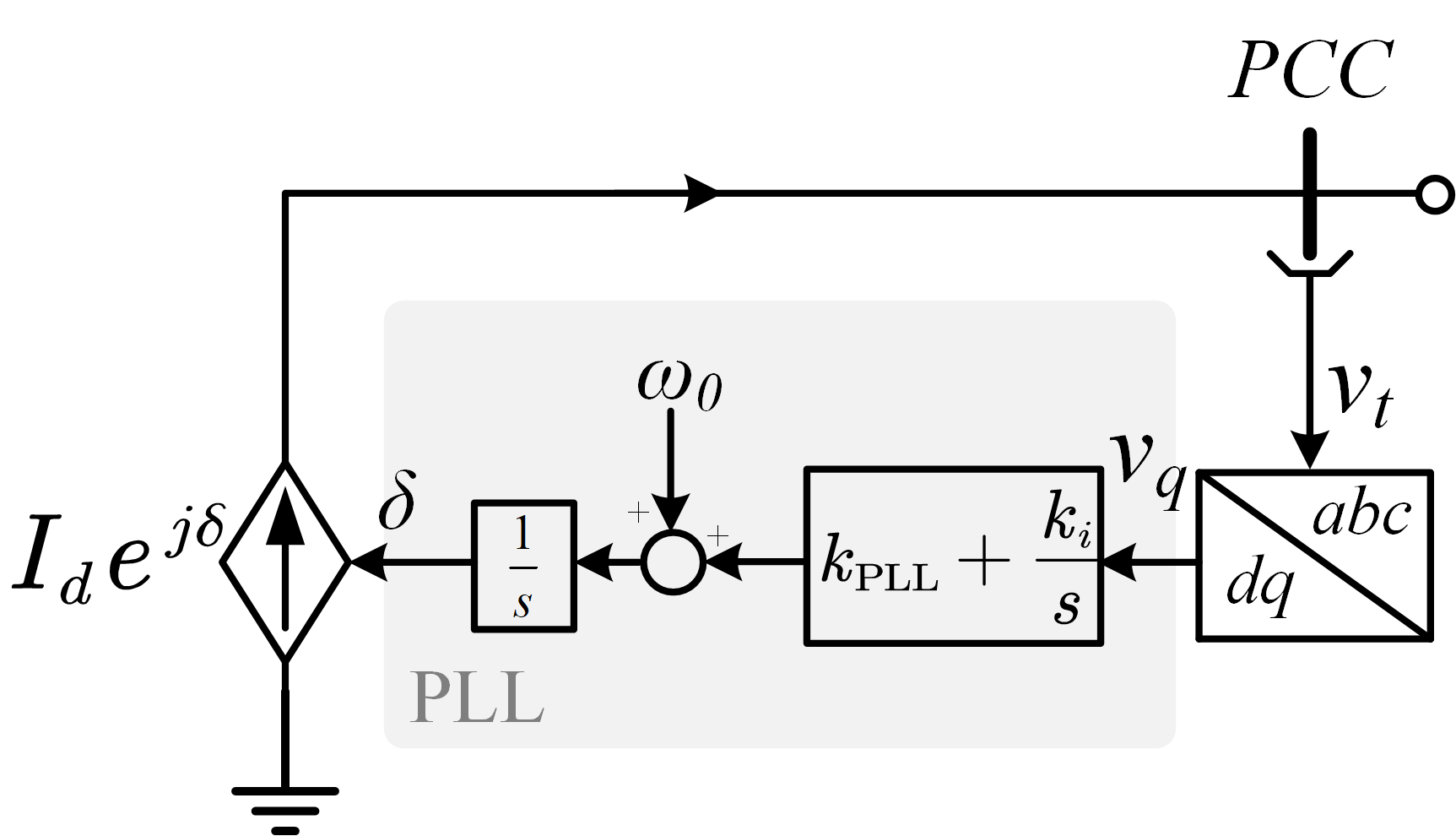}}
\hfil
\subfloat[modelling of GSP inverters.]{\includegraphics[width=0.28\textwidth]{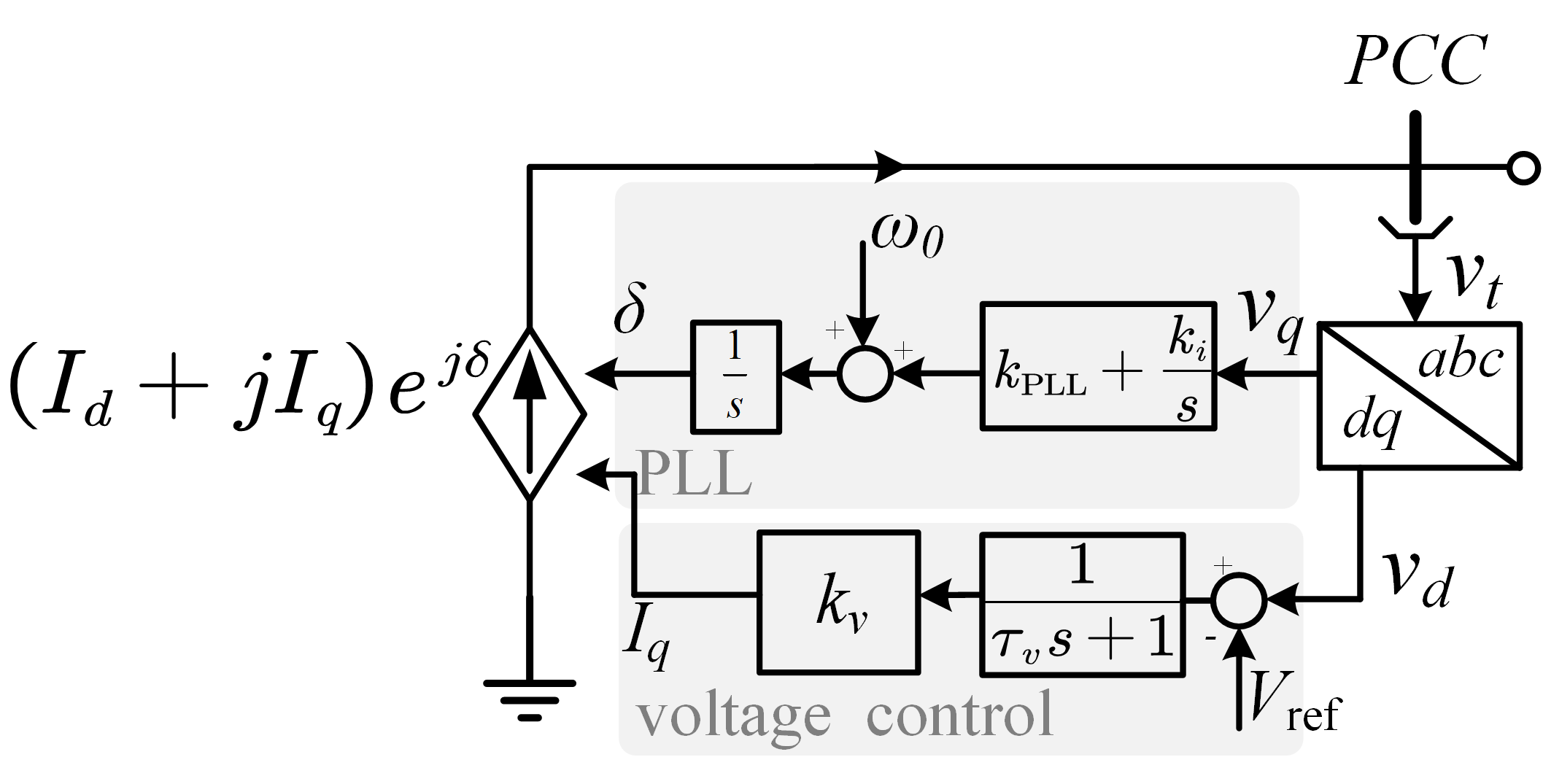}}
\caption{Large-signal modeling of different types of inverters.}
\vspace{-1mm} 
\label{fig1}
\end{figure*}

The famous results in \cite{xue1989extended} show that multi-machine systems can often be characterized by the interaction between two clusters. This implies that an investigation into two-inverter systems is a reasonable first step towards studying multi-inverter systems. The two-inverter large-signal stability has been studied in \cite{Genghua_pll,fu2022cascading, yi2022transient, shen2020transient, me2023transient, xue2023transient, wang2023fixed, zhao2018transient}. Among these studies, \cite{Genghua_pll, zhao2018transient, wang2023fixed} perform quasi-static analysis, focusing on the existence of a SEP, but the ROA around the SEP is not assessed. Other studies transform two-inverter systems into single-inverter systems, by neglecting the dynamics of the second inverter \cite{shen2020transient,xue2023transient,me2023transient}, or treating the second inverter a source disturbances \cite{yi2022transient, fu2022cascading}. These methods are based on very significant assumptions, that is, the two inverters are not tightly coupled, or the second inverter is less significant than the first inverter. 

To address the shortcomings above, this paper presents the manifold method to identify the ROA of two-inverter systems, based on the theory proposed by Chiang et al. \cite{chiang2011direct, chiang2015stability}. Chiang's theory indicates that the ROA of a wide class of nonlinear systems (including power systems) is the region encircled by the stable manifolds emanating from all UEPs of the system. The manifold method is not dependent on a global energy function and therefore can be applied to IBR-dominated systems. The manifold method has been applied to single-inverter systems in \cite{hu2019large}, but obstacles are encountered in multi-inverter systems due to the difficulties in computing and visualizing the stable manifolds of high-order systems. In this paper, we perform model order reduction on inverters based on multiscale analysis and through that extend the manifold method to two-inverter systems. This extension enables us to investigate the interaction of heterogeneous inverters (GFL, GFM, and GSP) and therefore provides deeper insights into the large-signal stability of systems with combinations of different inverters. Additionally, a metric, called stability radius (SR), is used to assess the large-signal stability margin. SR provides a conservative but generic estimation of the stability margin, as it is independent of the location and type of fault. Using SR, we quantify the contributions of different inverters in improving the large-signal stability amid inverter interactions.

\input{figures/Table1}

\section{Large-Signal Stability Assessment of Two-Inverter Systems Using Manifolds}
This section first establishes a generic reduced-order model for two-inverter systems with various combinations of inverter type, and then applies the manifold method to identify the ROA of these systems to assess their large-signal stability.
\subsection{Order reduction of two-inverter models}
The control blocks and large-signal modeles of a GFM, GFL, and GSP inverter are depicted in \figref{fig1}~(a), (b), and (c) respectively. The order reduction method used is based on the multi-timescale characteristics of inverter control \cite{Gu_reduction}. The inner loops of inverters are usually very fast and therefore can be represented as controlled current or voltage sourcesc for the sake of this analysis. Furthermore, we assume that the PLL of the GFL inverters and the droop control of GFM inverters are overdamped. Specifically, in the GFM inverter, the time constant $\tau_p$ of the low-pass filter in the $p\text{-}\omega$ droop control is set to be small to make the angle dynamics overdamped \cite{Xiongfei_overview}. As such, the droop control is treated as an algebraic gain with no state, so the order of the model is reduced to a single state $\delta$ as below
\begin{equation}
\dot\delta = k_\text{droop}(P_\text{ref}-p) -\omega_0
\end{equation}
where $\delta$ is the internal voltage angle of the GFM inverter, and $\omega_0$ is the nominal angular frequency of the grid. In the GFL inverter, the PLL is designed as proportion dominant, i.e. $k_{i} \ll k_\text{PLL}$, where $k_i$ and $k_\text{PLL}$ are the integral and proportional gain of the PLL. This makes PLL overdamped helpful for the angle stability of GFL inverters \cite{wu2019design, li2022whole}. Under this assumption, the output of the integral controller in the PLL, $\omega_i$, can be assumed fixed to the prefault value during the transient, that is, $\omega_i \approx \omega_0$. As a result, the model of an GFL inverter is also reduced to a single state
\begin{equation}
\dot\delta=k_\text{PLL} v_q + \omega_i - \omega_0 \approx k_\text{PLL} v_q.
\label{PLL_ori}
\end{equation}
where $\delta$ is the internal PLL angle. On top of the GFL inverter, the GSP inverter incorporates a $v_{d}\text{-}i_{q}$ voltage control with gain $k_v$ to enhance voltage support through reactive current injection. This is an approximation of $V\text{-}i_q$ droop as per grid codes \cite{zhao2020reactive, netz2006grid}, provided that PLL angle is aligned with grid voltage implying $v_d \approx v_t$ where $v_t$ is the magnitude of grid voltage measured at the inverter terminal. The time constant $\tau_v$ of the low-pass filter in the $v_d\text{-}i_q$ voltage control is also assumed to be small \cite{Xinshuo2022}, which means that the voltage control does not introduce extra states in the model.
\begin{figure}[t]
\centering
\includegraphics[width=3.2in]{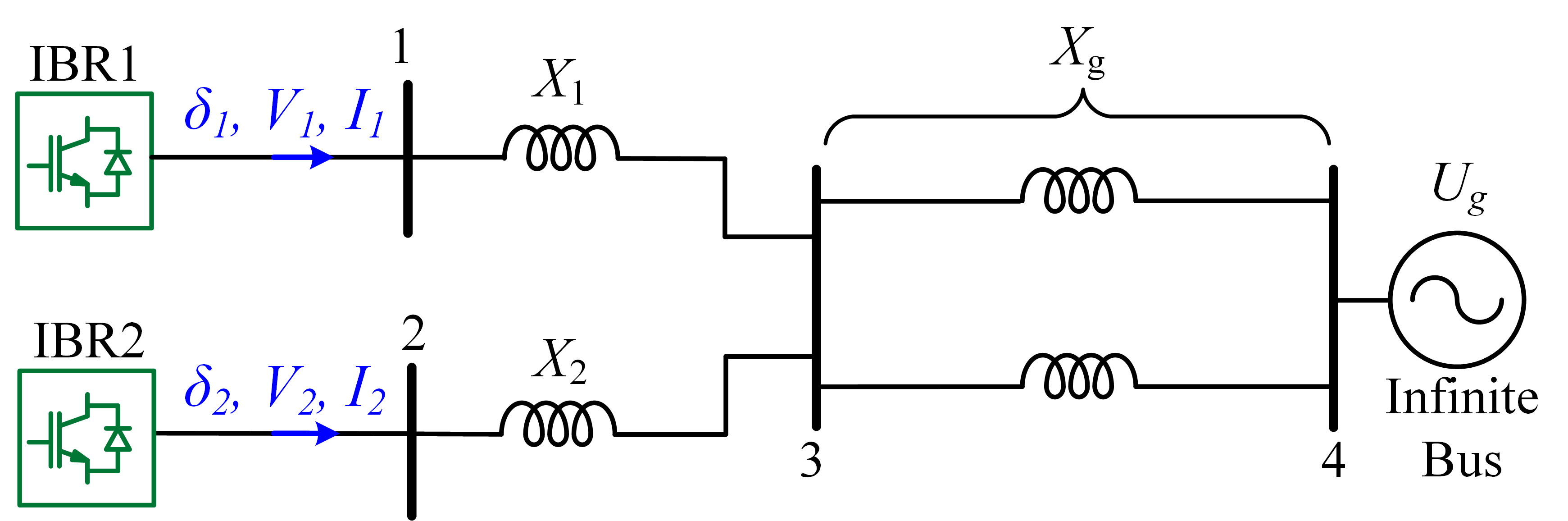}
\caption{Two-inverter-infinite-bus system.}
\vspace{-3mm} 
\label{fig2}
\end{figure} 

In summary, the GFM, GFL, and GSP inverters can all be reduced to first-order models with $\delta$ being the sole state for the purpose of estimating the ROA. The impact of this approximation will be assessed through full-order system simulation later. This simplification leads to the reduce-order model for the two-inverter-infinite-bus system displayed in \figref{fig2}: 
\begin{equation}
        \left\{ \begin{aligned}
	\dot{\delta}_1=&k_1\left[ C_1-A_1\sin \left( \delta _1-\delta _2 \right) -B_1\sin \delta _1 \right. \\
    & \left. +D_1\cos(\delta_1-\delta_2) + \varepsilon \right]\\
      \dot{\delta}_2=&k_2\left[ C_2-A_2\sin \left( \delta _2-\delta _1 \right) -B_2\sin \delta _2  \right. \\
      & \left. +D_2\cos(\delta_1-\delta_2)\right ]\\
     \end{aligned}\right.
     \label{general}
\end{equation}
This system contains two IBRs, i.e., IBR1 and IBR2, each configured to operate in either GFL, GFM, or GSP mode. For all combinations of GFL, GFM, and GSP, the model has the same format but the coefficients are different, which are summarized in \tabref{Table1}. The detailed derivation can be found in Appendix \ref{Expression}. In the GFL inverter, $I_d$ denotes the $d$-axis current reference. The terms $V_1$ or $V_2$ represent the terminal voltages of the GFM inverters, while $V_{\text{ref}}$ denotes the voltage reference used in the voltage control of the GSP inverter. Some auxiliary symbols are used in (\ref{general}) to simplify notation, as defined below 
\begin{equation}
\begin{array}{c}
        X_{\Delta 12}=\frac{X_1X_2+X_1X_g+X_2X_g}{X_g}\\
        X_{\Delta g1}=\frac{X_1X_2+X_1X_g+X_2X_g}{X_2}\\
        X_{\Delta g2}=\frac{X_1X_2+X_1X_g+X_2X_g}{X_1}\\
        X_{\Sigma1}=X_1 + X_g, X_{\Sigma2}=X_2 + X_g\\
        X_{1+2//g}=X_1+\frac{X_gX_2}{X_{\Sigma2}}, X_{2+1//g}=X_1+\frac{X_gX_1}{X_{\Sigma1}}
\end{array}.
\end{equation}
If a GSP inverter is present in the system, a small residual term $\varepsilon$ appears in Eq. (\ref{general}). These terms are typically small and are associated with the voltage control gain $k_v$. Specifically, $\varepsilon_{lp}$ and $\varepsilon_{mp}$ in \tabref{Table1} represent the residual term $\varepsilon$ for GFL-GSP and GFM-GSP combinations, respectively, whose explicit forms are given in \eqref{GFL+GFLVS_spA} and \eqref{GFM+GFLVS_spA} in Appendix \ref{Expression}. These expressions also indicate that the GSP inverter behaves equivalently to a GFM inverter under ideal voltage regulation.

The global energy function has been viewed as an extension of, or a special case of, the Lyapunov function, which requires that its derivative to be non-positive globally, while its value does not need to be strictly positive. For classical synchronous-generator (SG) systems, the existence of a global energy function is well established. Likewise, for systems comprising only GFM inverters—whose dynamics involve only sine terms (e.i, $A_i\sin(\delta_1-\delta_2)$), a global energy function exists in the form of
    \begin{equation}
    \begin{aligned}
        W\left( \delta _1,\delta _2 \right) =&-\frac{C_1}{A_1}\delta _1-\frac{C_2}{A_2}\delta _2-\cos \left( \delta _1-\delta _2 \right) \\
        &-\frac{B_1}{A_1}\cos \delta _1-\frac{B_2}{A_2}\cos \delta _2\\
    \end{aligned}
    \end{equation}
On the other hand, if the system contains at least one GFL inverter, cosine term $D_i\cos(\delta_1-\delta_2)$ appears in the nonlinear differential equations. Chiang \cite{chiang1989study} demonstrated that if the dynamics of a system, as described by \eqref{general}, include cosine terms, no global energy function exists. This necessitates the development of a new approach for assessing large-signal stability with a high penetration of GFL inverters.
\begin{figure}[b!]
	\centering
	{\includegraphics[width=3.5in]{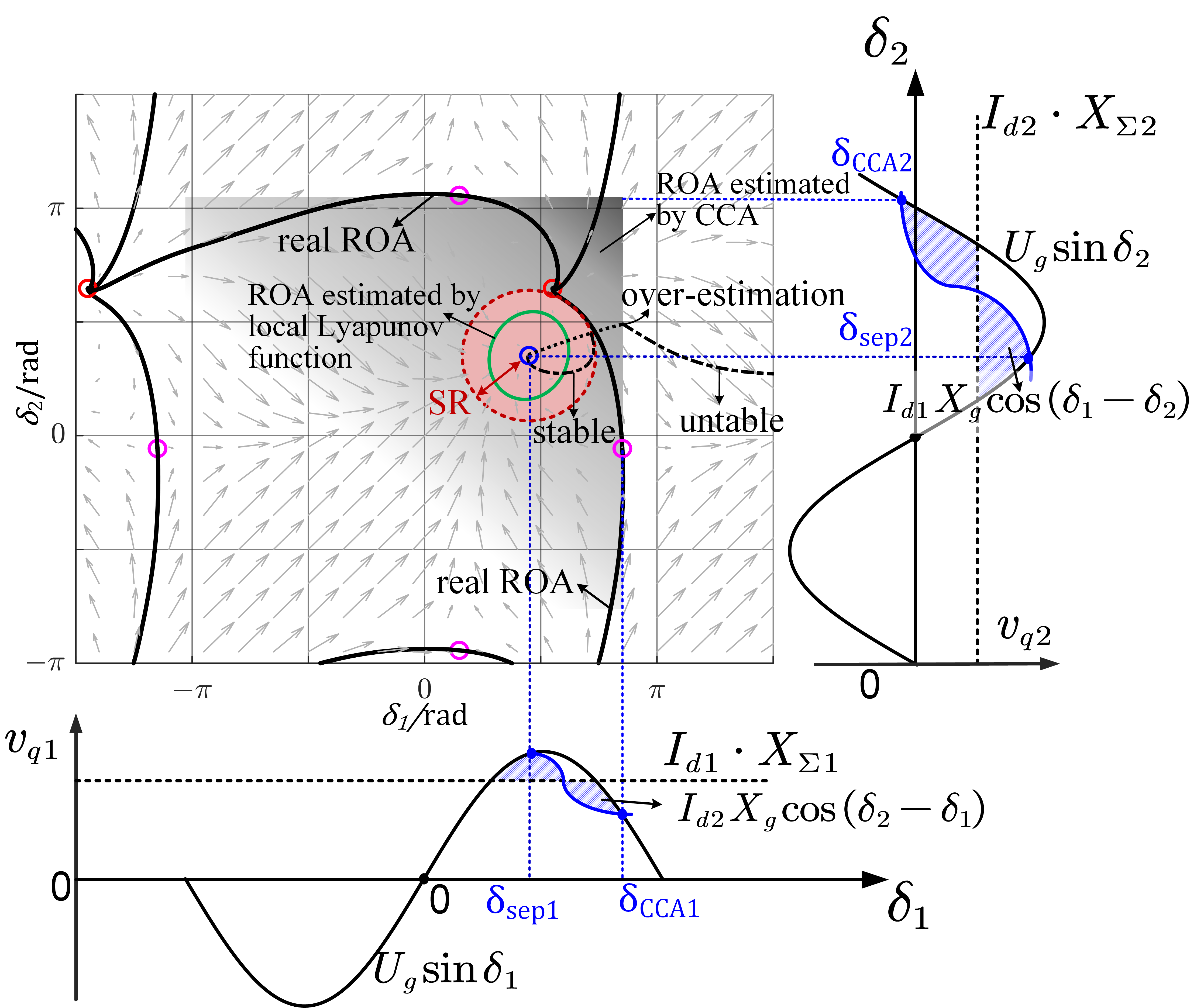}}
	\caption{Comparison between SR and other large-signal stability assessment methods. On the phase-plane plot, the ROA is shown with a solid black line (manifold method), gray rectangular shading (CCA method), and green line (local Lyapunov method). The SEP is an open blue circle and the SR is a red dashed circle centered on the SEP.}
    \vspace{-1mm} 
	\label{fig_ccr}	
\end{figure}
The model order reduction approach used here seems radical, but it turns out to have good accuracy, as is to be validated later in Section III-B. This is due to the fact that the angle dynamics of inverters (PLL of GFL and angle swing of GFM) can be designed to be overdamped thanks to the high controllability of inverters. 
\begin{figure*}[b!]
    \centering
    \addtocounter{figure}{1}
    \subfloat[Schematic diagram of the two-inverter hardware-in-the-loop test system.]{\includegraphics[width=0.8\textwidth]{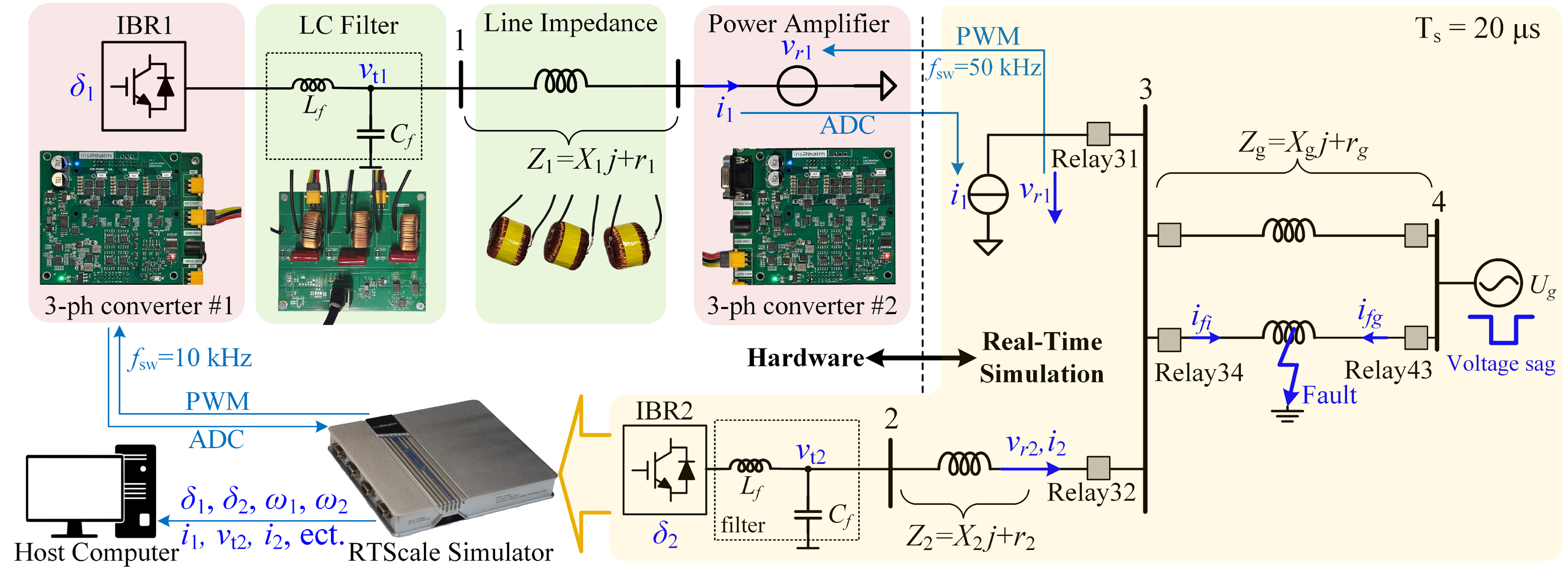}}
    \caption{PHIL experimental platform.}
    \vspace{-3mm}
    \label{philplatform}	
\end{figure*}
\begin{figure}[t!]
	\centering
    \addtocounter{figure}{-2}
	{\includegraphics[width=2in]{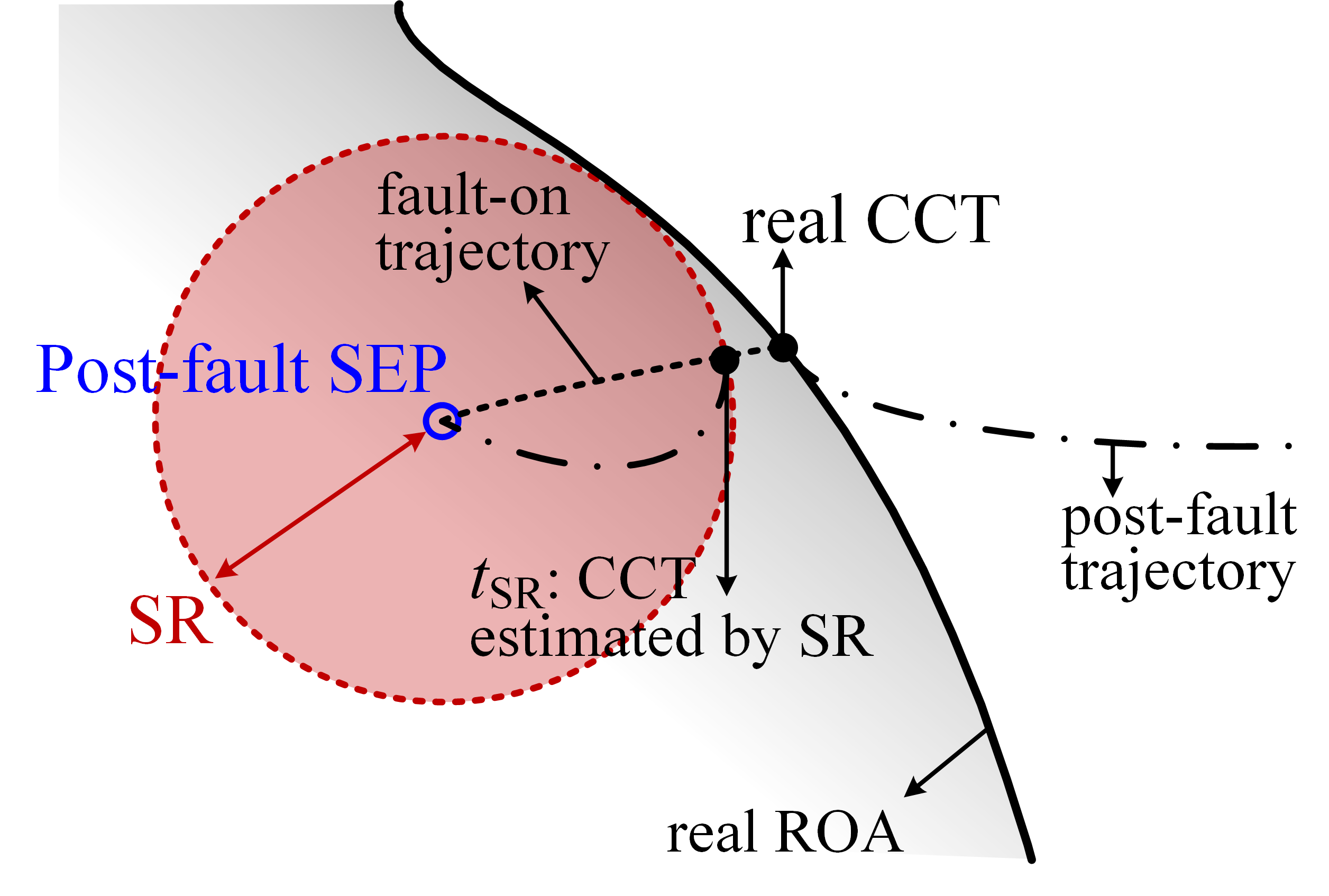}}
	\caption{Relationships among proposed metrics.}
    \vspace{-3mm}
	\label{fig_ccr2}	
\end{figure}

\subsection{Stability radius (SR) based on manifold method}
For the two-dimensional system analyzed in this paper, the UEPs on the ROA are either type-I or type-II\footnote{A type-I UEP has exactly one unstable eigenvalue, whereas a type-II UEP has two unstable eigenvalues.}. The real ROA can be depicted by reverse integration along the stable eigenvector of the type-I UEP in the phase plane, as shown by the black curve in the top left panel of \figref{fig_ccr}. The stability radius (SR) is then used to measure the large-signal stability margin, defined as the shortest distance from the post-fault SEP to the ROA boundary, highlighted by the red circle in \figref{fig_ccr}. 

\figref{fig_ccr} contrasts the SR with other stability assessment methods for the two-inverter system. Notably, the critical clearing angle (CCA) is indicated by the projection of the real ROA onto \(\delta_1, \delta_2\) coordinates, depicted as \(\delta_{\text{CCA}1,2}\) and can also by the intersection between the sine function $U_g\sin\delta$ and the blue solid curve. Due to the existence of interaction terms [e.g. $\cos(\delta_2-\delta_1)$ in \eqref{general}], there is the difference between the blue solid curve and the ideal dotted lines [e.g. constant value $C_{1,2}$ in \eqref{general}], highlighted by the blue shaded area. Previous methods, cited in \cite{shen2020transient, yi2022transient}, use the CCA of one critical GFL as a large-signal stability index, but this often overestimates the ROA and overlooks the other inverter, as shown in the grey rectangle area in \figref{fig_ccr}. In particular, if two inverters are connected to an infinite bus individually, then the ROA is square, decoupling $\delta_1$ and $\delta_2$, and the SR being the minimum of $\delta_{\text{CCA}1}$ and $\delta_{\text{CCA}2}$. The local Lyapunov function to estimate the ROA is also plotted as the green curve in \figref{fig_ccr}, which is an approach to address the situation where the global energy function does not exist. It can be seen that the local Lyapunov method usually produces an under-estimate of the ROA, and is more conservative than the real ROA and SR.

The calculation of SR is straightforward: it requires recording the shortest distance to the post-fault SEP during the process of reverse integration, and no need for storing shape information of stable boundaries which is convenient to implement. Furthermore, critical clearing time (CCT) can be estimated based on SR. The specific steps to estimate CCT are: 1) Integrate the fault-on trajectory $\phi _f\left( x,t \right) $; 2) At the time $t_\text{SR}$, $\phi _f\left( x,t_\text{SR} \right) $ is outside of the circle of SR, but $\phi _f( x,t_\text{SR}-\varDelta t ) $ of the previous time step is inside of the circle of SR; 3) The $t_\text{SR}$ is the CCT estimated by SR. \figref{fig_ccr2} illustrates the relationships between SR, $t_\text{SR}$, real ROA, and the estimated and real CCT. Since the SR is entirely inside the real ROA, the CCT estimated by SR, $t_\text{SR}$, is conservative and guaranteed. 

If fault clearing occurs at a time that gives a position outside the ROA of the SEP, the trajectory will not converge to this SEP. It should be noted that, even though sometimes the trajectory may converge to a different SEP that is one period away from the original SEP, this is considered unstable in engineering terms \cite{Zaborszky1988, XiuqiangMultiGFM}. This case is analogous to the pole slipping of a synchronous generator and can compromise distance protection\cite{fang2019distanceprotection, Quisepe_distance}, cause reverse power flow \cite{li2022whole}, over-voltage, and voltage oscillations \cite{wang2023transient} in inverters. This pole slipping phenomenon of inverters will be explained in detail in the following large-signal interaction between different types of inverters.

In summary, the manifold method completely describes the dynamic behavior in the phase plane, and the SR is utilized on the basis of the manifold method, which can access the large-signal stability, as an extension of CCA in a high-dimensional space.

\section{PHIL Platform and EMT Simulation Setup}
\begin{figure}[t!]
\centering
\setcounter{subfigure}{1}
\subfloat[Platform setup.]{\includegraphics[width=0.45\textwidth]{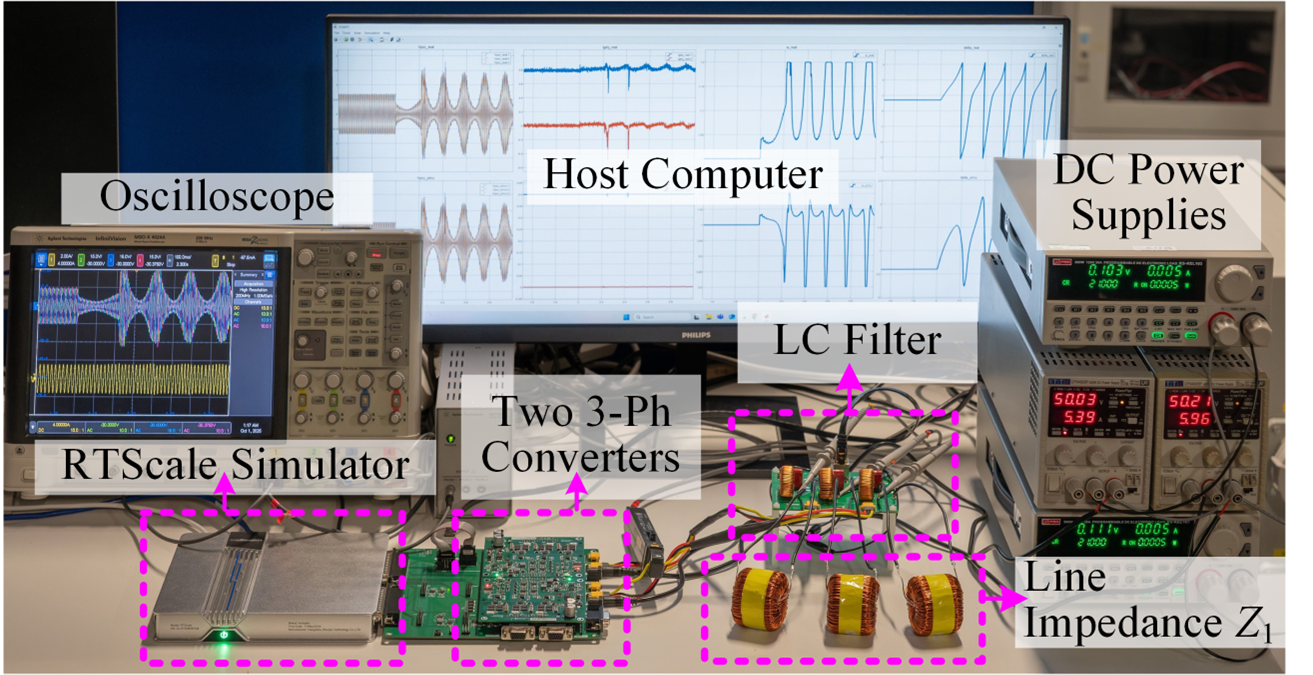}}
\caption{PHIL experimental platform. (continued)}
\vspace{-3mm}
\end{figure} 

To validate the findings in the following sections, a power-hardware-in-the-loop (PHIL) platform was built, as shown in \figref{philplatform}. The two-inverter configuration in \figref{fig2} is implemented in \figref{philplatform}~(a). In the PHIL setup, the system is partitioned into two parts: the portion to the left of Bus~3 in \figref{fig2}—namely IBR1 and its line impedance \(X_{1}\) (i.e., \(Z_{1}\))—is implemented as physical hardware, while the remaining network is executed on a real-time simulator. The platform employs two three-phase power converters built with BSC052N08NS5 MOSFETs. One serves as the physical IBR1, operating at a switching frequency of \(10~\mathrm{kHz}\). The other operates as a power amplifier, interfacing the physical plant with the real-time simulation. Both the power amplifier and the real-time simulator run at \(50~\mathrm{kHz}\), ensuring high-fidelity emulation. The real-time simulator is RTScale~\cite{InsRealm}, which also executes the IBR1 controller and communicates with a host computer for controller programming and system monitoring. All experimental data are logged on the host computer, and electrical quantities (such as, output current of phase A -- $i_{\text{1,}a}$ and terminal voltage of IBR1 -- $v_{\text{t1}}$) in the physical plant are measured with an oscilloscope. The system base power and voltage are 60~V$\cdot$A and 22.8~V; additional parameters are listed in \tabref{PHILPF} in Appendix \ref{parameters}.

In addition, EMT simulations are implemented in \textsc{Matlab}/Simulink using the same per-unit (pu) values as the PHIL experiment. The EMT simulation covers additional cases beyond the experiment, and is cross-validated against the PHIL experiment results and theory. The detailed simulation model also follows the configuration shown in \figref{philplatform}~(a), which also annotates the relay indices and the symbols for voltages and currents. The time axis in the following sections is referenced to the fault event, with \(t=0\) denoting the fault inception.
\begin{figure}[t]
\centering
\subfloat[$X_g=0.35$~pu, $t_{\text{fault}} = 0.212$~s.]{\includegraphics[width=0.23\textwidth]{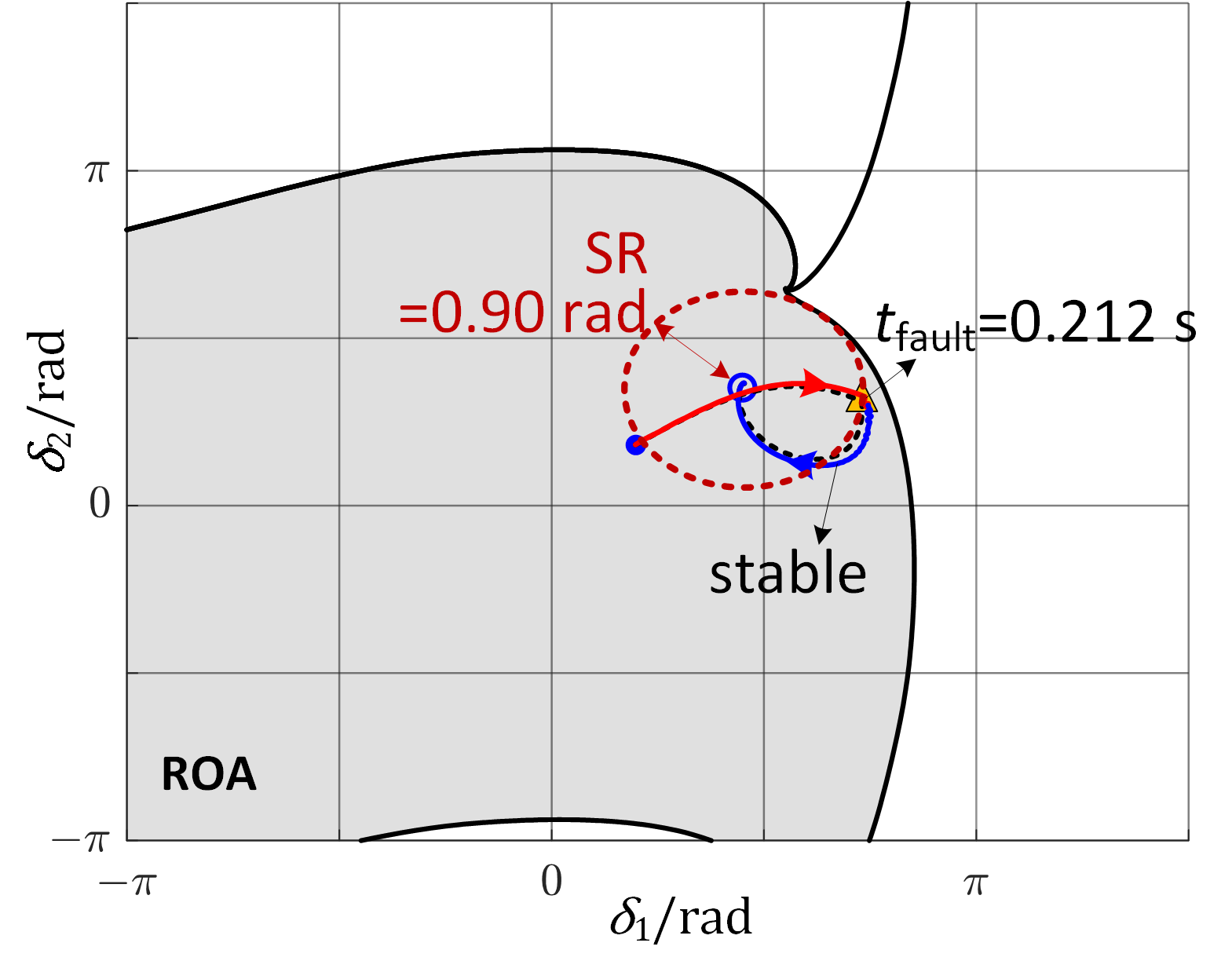}} \quad
\subfloat[$X_g=0.35$~pu, $t_{\text{fault}} = 0.25$~s.]{\includegraphics[width=0.23\textwidth]{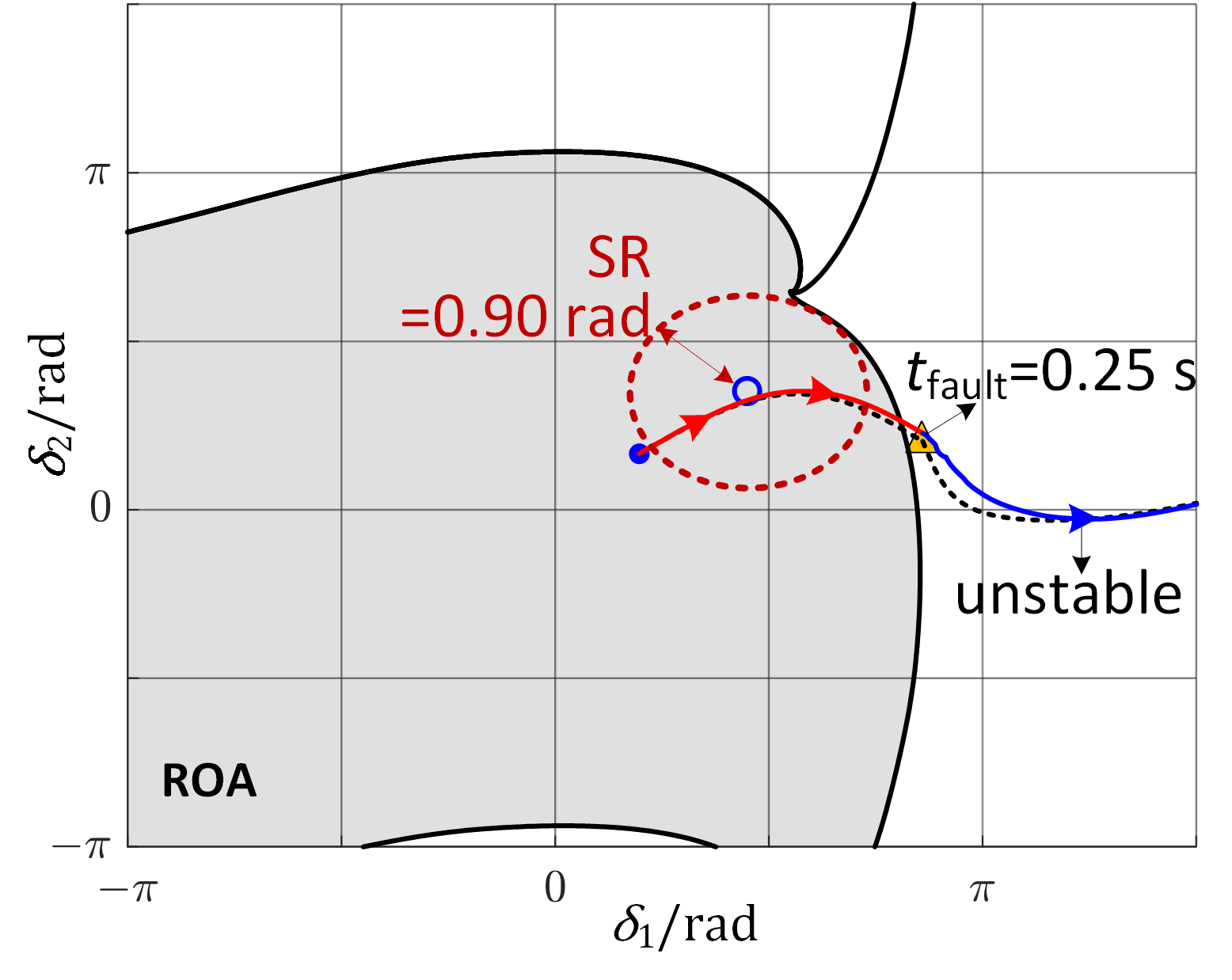}}\\
\subfloat[$X_g=0.4$~pu, $t_{\text{fault}} = 0.131$~s.]{\includegraphics[width=0.22\textwidth]{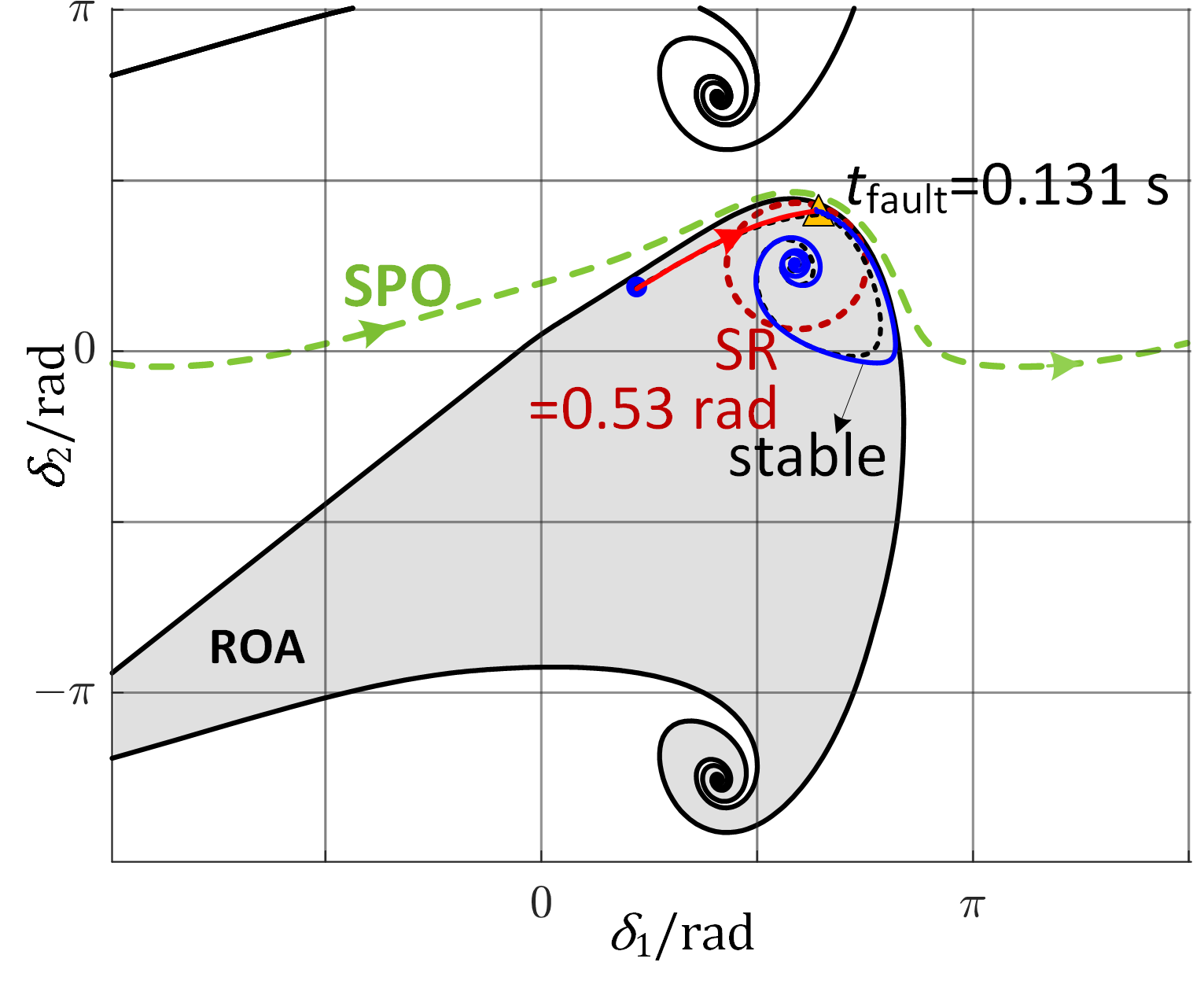}}\hspace{1.3em}
\subfloat[$X_g=0.4$~pu, $t_{\text{fault}} = 0.212$~s.]{\includegraphics[width=0.235\textwidth]{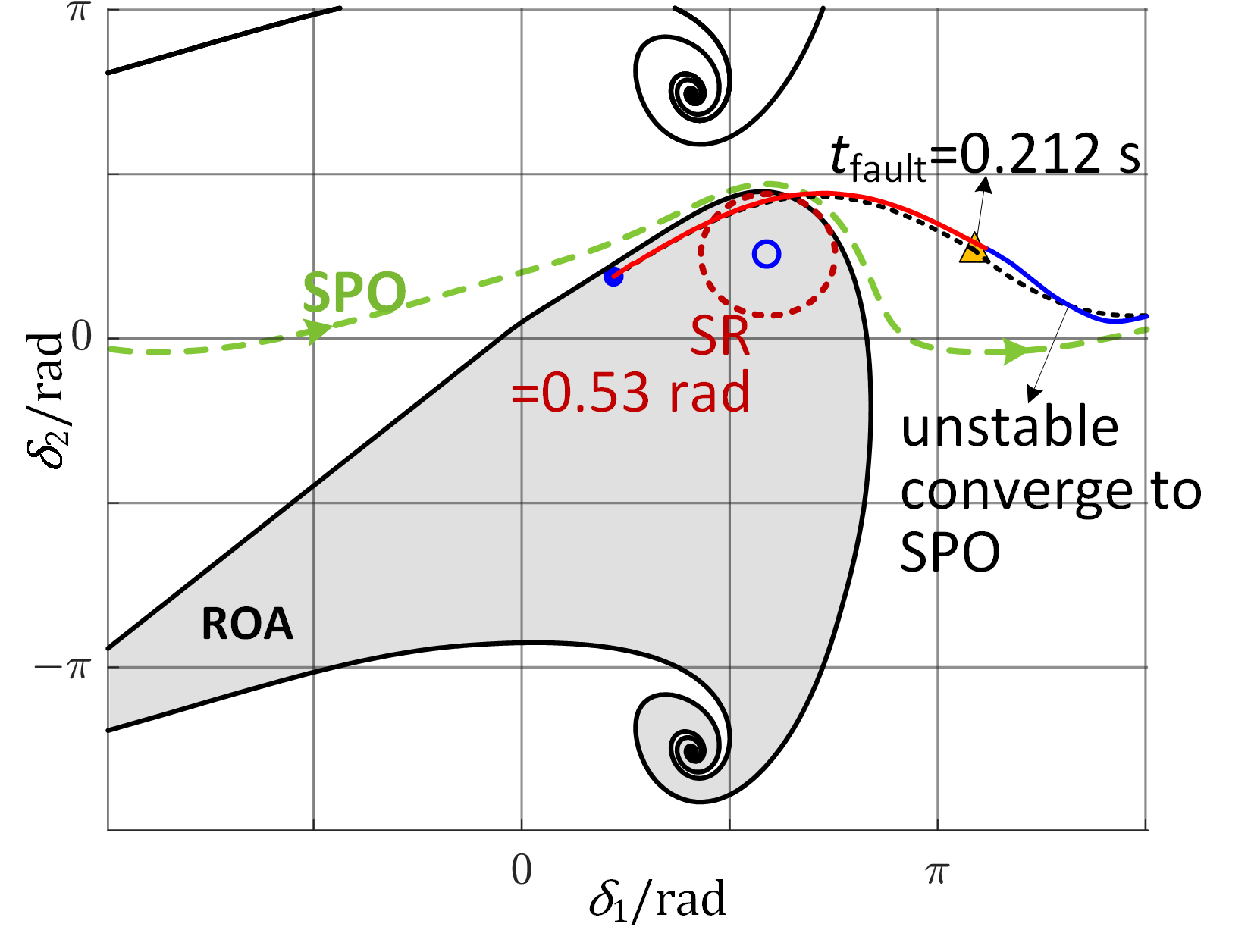}}
\caption{Phase portraits of two GFL inverters for four test conditions as assessed by the manifold method. Pre-fault SEP is marked with a solid blue dot, the post-fault SEP with a hollow blue dot, and the fault clearance point with a yellow triangle. The dashed black line represents the analytical trajectories. Trajectories generated by EMT simulations are represented as red lines during the fault and blue lines post-fault. The green dashed line, where it exits, is a stable periodic orbit. This color convention is applied in subsequent similar figures.}
\label{Sim_1_1}
\end{figure} 

\section{GFL-GFL Interaction} \label{2PLL}
In this section, both IBR1 and IBR2 of \figref{fig2} \(\bigl(\text{or \figref{philplatform}~(a)}\bigr)\) are initially configured as GFL inverters.
 
The PLLs of two inverters are designed to be over-damped, with the integral gain $k_i$ set to the one-tenth of the proportional gain $k_\text{PLL}$. Under this configuration, the contribution of the integral term is negligible, and the proportional term primarily governs the dynamic behavior. The line impedance $X_g$ is set to either $0.35$~pu or $0.4$~pu and is composed of two parallel lines of $2X_g$. A short-circuit fault is applied in the middle of one of the two lines and that line is removed at fault clearance giving a doubling of the effective impedance between busses 3 and 4 and a new operating point for the system. The fault initiates at time $0$~s and is of variable duration. Detailed system parameters for the PHIL experiment and the EMT simulation are given in \tabref{PHILPF} (mentioned before and applicable to all subsequent PHIL cases) and \tabref{TableA1} in Appendix \ref{parameters}, respectively. Both the EMT simulation and the PHIL experiment are conducted to validate the manifold method for finding SR, and the estimate of CCT from SR ($t_\text{SR}$).

\begin{figure}[t]
\centering
\subfloat[$t_\text{fault}=0.212$~s.]{\includegraphics[width=0.4\textwidth]{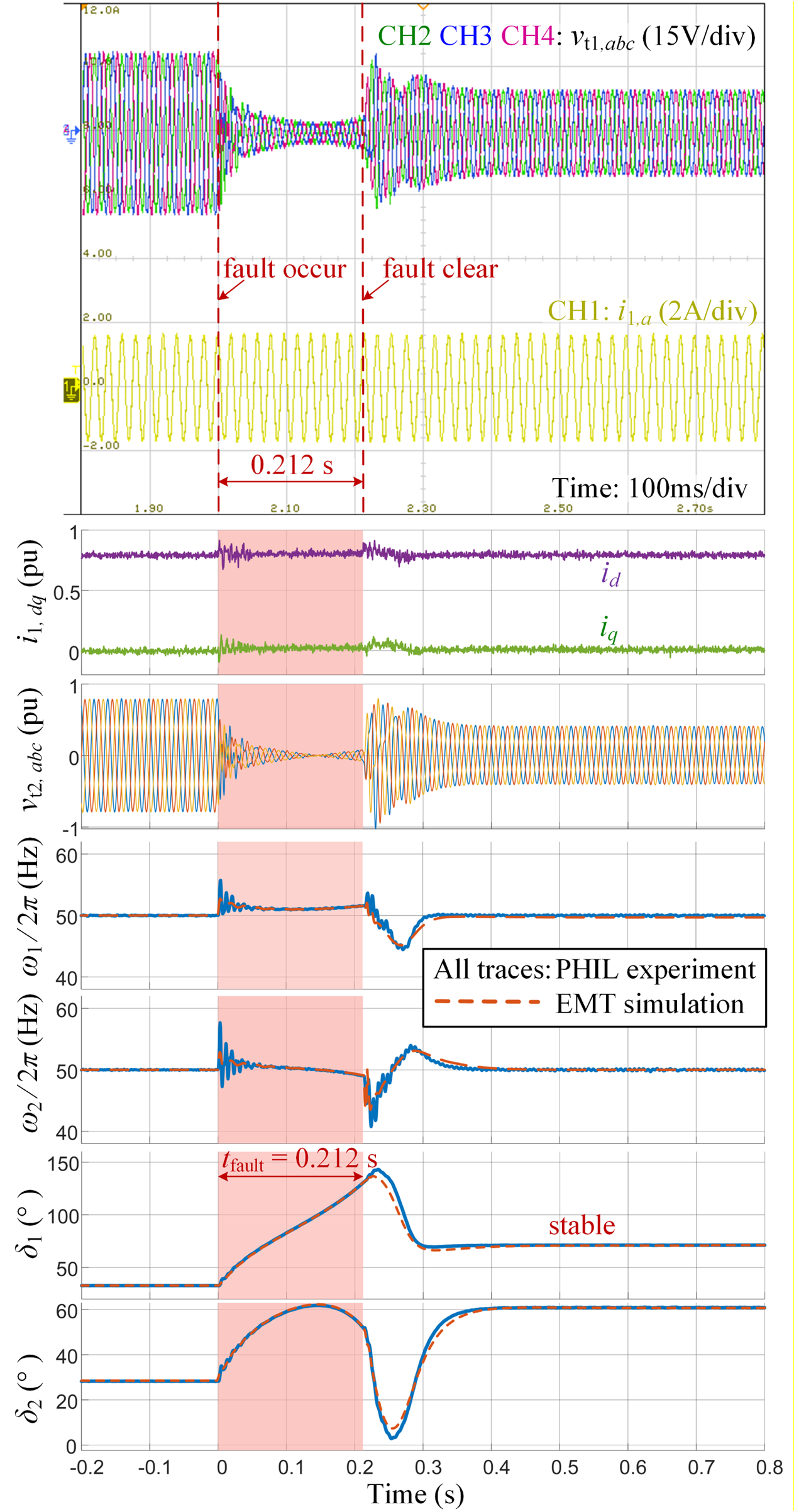}}
\caption{PHIL experiment results of two paralleled GFL inverters under $X_g=0.35$~pu.}
\vspace{-3mm}
\label{Sim_1_2}
\end{figure} 

\begin{figure}[t]
\centering
\addtocounter{figure}{-1}
\setcounter{subfigure}{1}
\subfloat[$t_\text{fault}=0.25$~s.]{\includegraphics[width=0.4\textwidth]{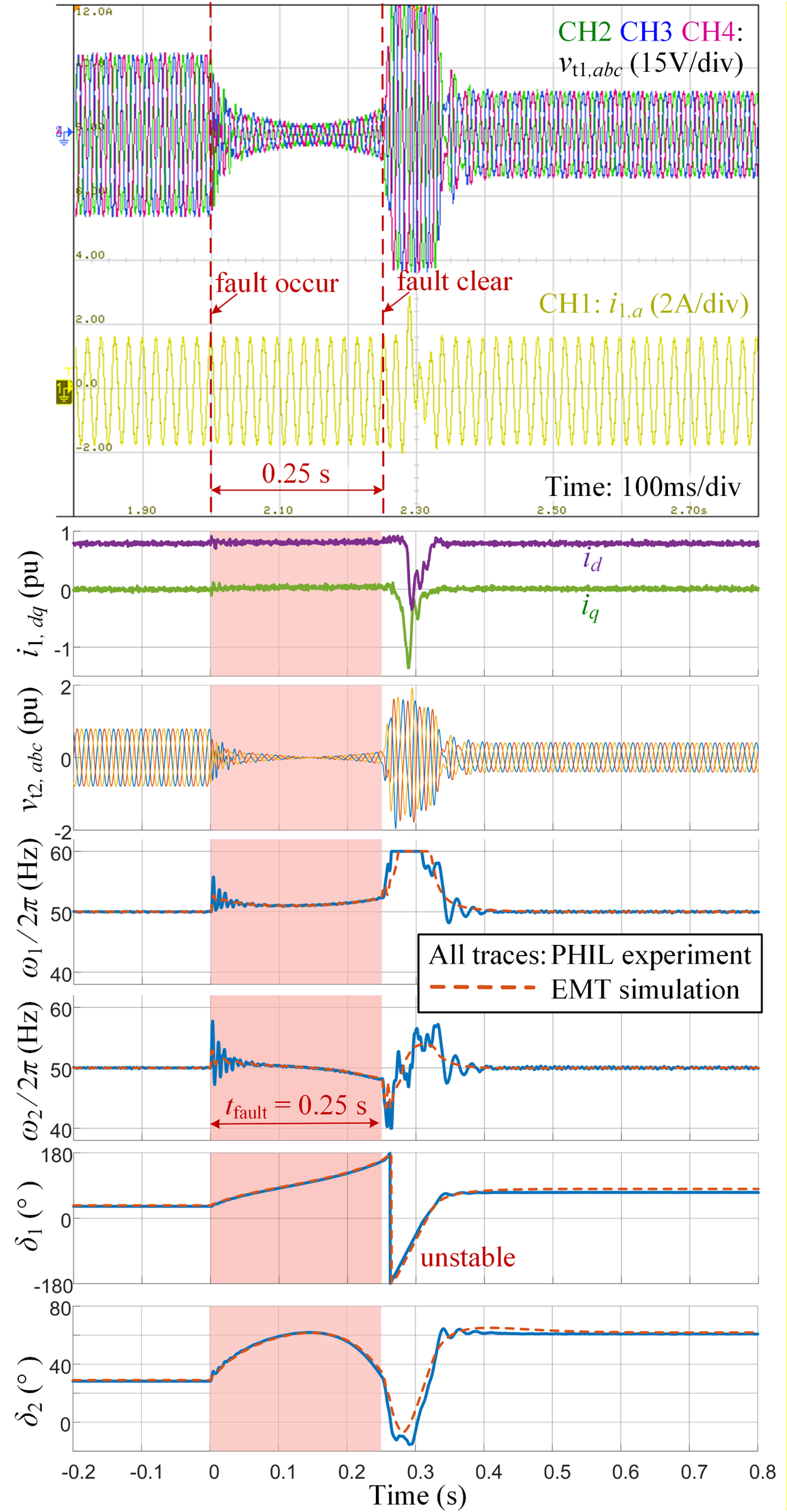}}\\
\subfloat[Measurements of relays under $t_\text{fault}=0.25$~s.]{\includegraphics[width=0.38\textwidth]{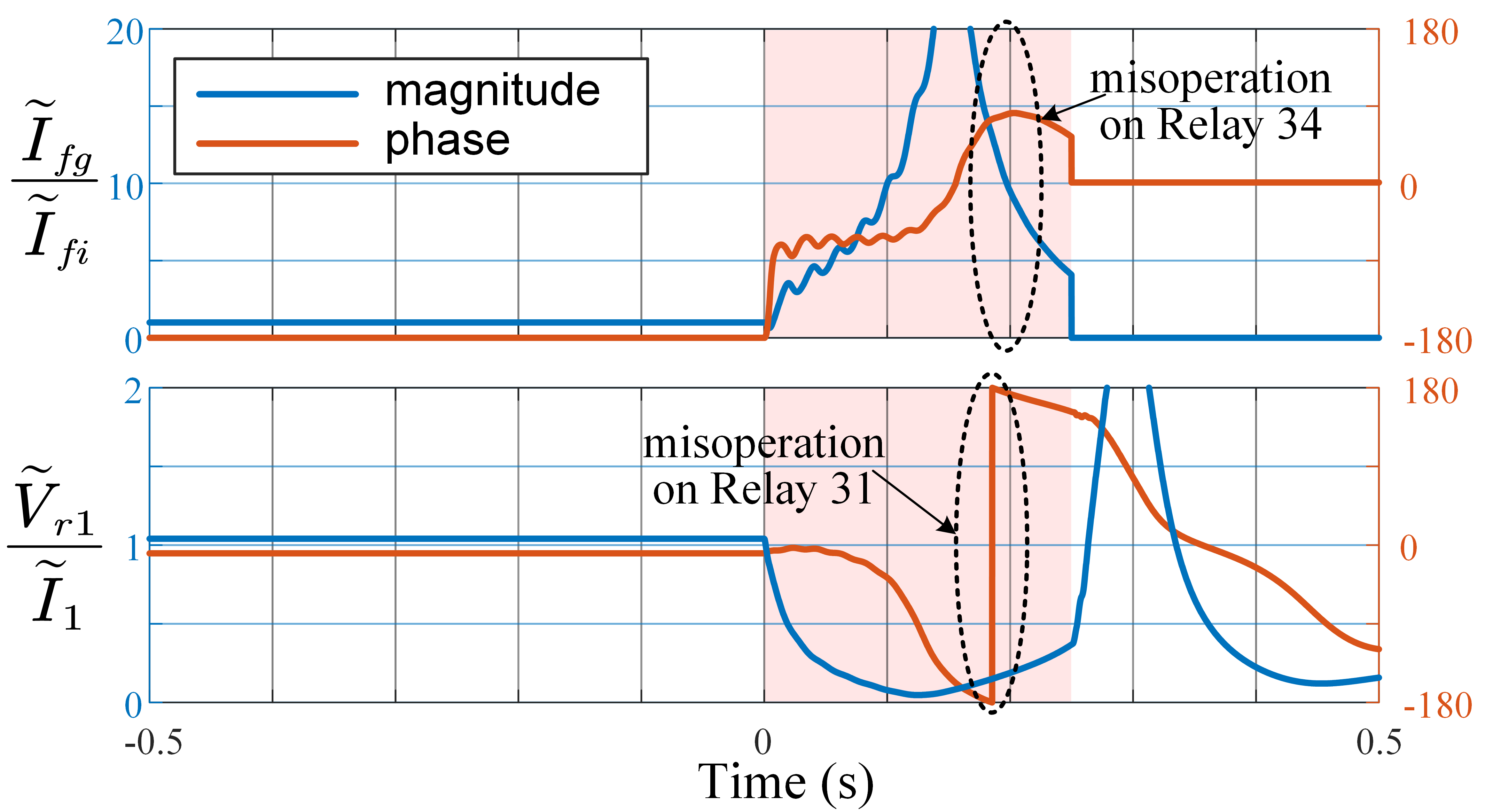}}
\caption{PHIL experiment results of two paralleled GFL inverters under $X_g=0.35$~pu. (continued)}
\vspace{-3mm}
\end{figure} 

\subsection{Large-signal instability mode versus grid inductance}
In the case of $X_g = 0.35$~pu, the calculated metrics are: SR of radius 0.90 rad and a CCT estimated via the SR of $t_\text{SR}=0.212$~s. The fault clearance time is set at two values: $0.212$~s to match the estimated CCT and $0.25$~s to exceed it.

In the case of $X_g = 0.4$~pu the SR is decreased to a radius of 0.53 rad and CCT reduced to $t_\text{SR}=0.131$~s. The fault clearance times are set to $0.131$~s to match and $0.212$~s to exceed.

\begin{figure}[t]
\centering
{\includegraphics[width=0.4\textwidth]{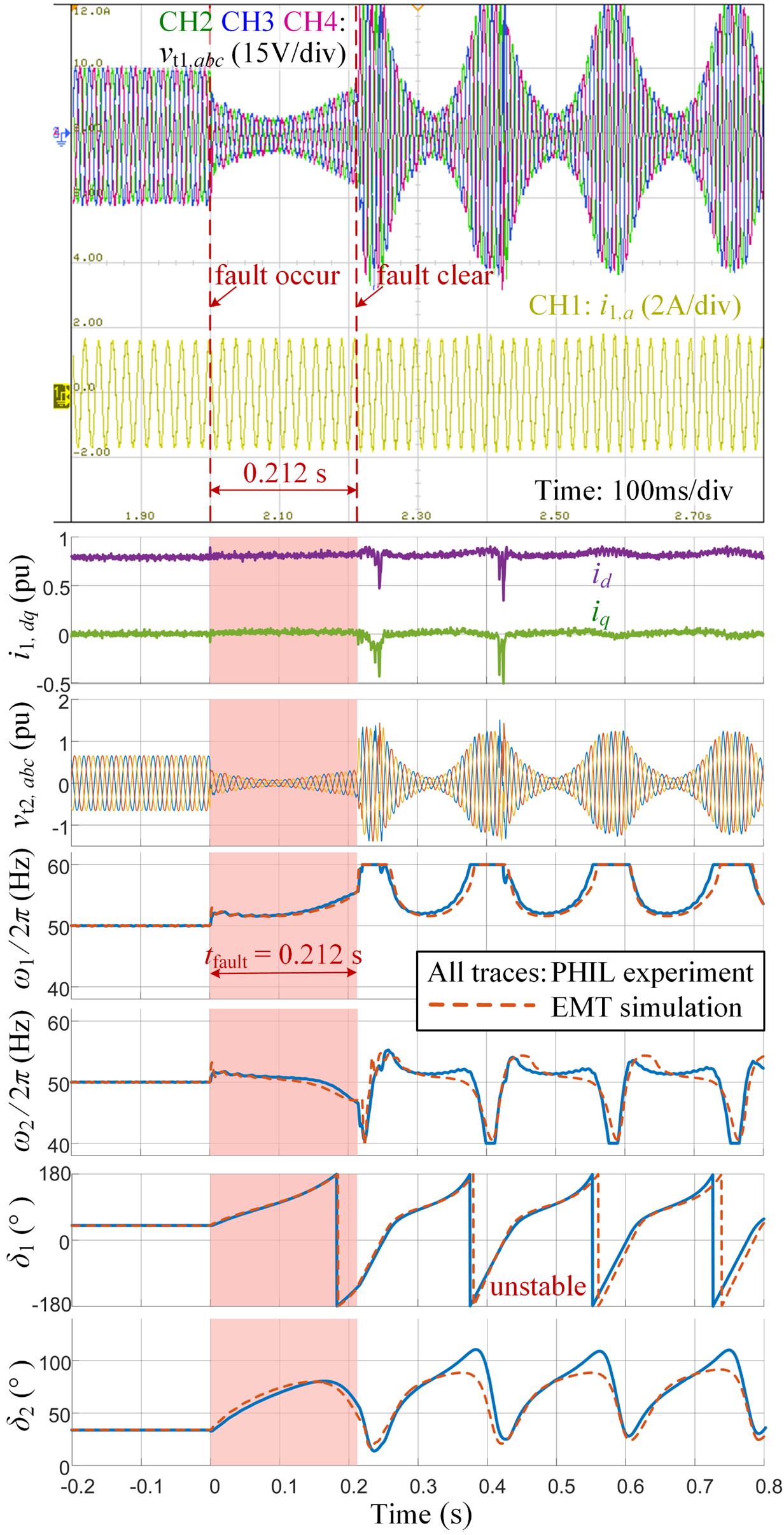}}
\vspace{-2mm}
\caption{PHIL experiment results of two paralleled GFL inverters under $X_g=0.4$~pu, $t_{\text{fault}}=0.212$~s.}
\vspace{-4mm}
\label{Sim_1_3}
\end{figure} 

Using the manifold method the ROA can be drawn in terms of $\delta_1$ and $\delta_2$ phase-plane as shown by the gray shading in \figref{Sim_1_1}. The ROA is shown for the two choices of $X_g$ and two choices of fault clearance time. The pre-fault stable equilibrium is marked with a solid blue dot. During the fault, as the two PLLs are perturbed, the system follows the phase-plane trajectory of the red line according to the $\delta_1$ and $\delta_2$ found in the EMT simulation. Under this is a black dashed line, and closely resembling it, which is the trajectory found by numerical integration of the simplified (theoretical) model. 

The SR is indicated by the red dashed circle and it can be seen that the location at fault clearance (yellow triangle) is on the SR circle if the fault clearance time, is set to $t_\text{fault}=t_\text{SR}$, c.f., sub-figures (a) and (c). When $t_\text{fault}>t_\text{SR}$ the yellow triangle is outside the SR circle, c.f., sub-figures (b) and (d).

Post-fault, the system trajectory (blue solid line) moves to a new\footnote{New because the line impedance is doubled.} SEP marked by an hollow blue dot for $t_\text{fault}=t_\text{SR}$, and diverges for $t_\text{fault}>t_\text{SR}$. The effectiveness of the SR in predicting behavior is confirmed. The divergent trajectories may converge to a different SEP elsewhere on the phase plane but this is considered unstable in practical terms in the same way as pole-slipping of a synchronous machine.PHIL experiment results for the $X_g = 0.35$~pu case and two clearing times are shown in \figref{Sim_1_2}~(a) and (b). In \figref{Sim_1_2}~(a)–(b), the angle ($\delta_1$, $\delta_2$) and angular speed ($\omega_1$, $\omega_2$) traces by EMT simulation are overlaid as orange dashed lines for comparison with experimental measurements. The PHIL results closely match the EMT simulations, thereby validating both the experimental platform and the simulation model. The remaining small discrepancy relative to EMT primarily reflects whether inner-loop control, line dynamics, and the effect of small integral gain $k_i$.

The theoretical model on which SR is based ignored the dynamics of the inner current loop and the integral terms in the PLL so is approximate. However, the comparison of EMT results (in red and blue lines) with the reduced-order model (in the dashed black line) shows that there is little impact on the results. 

In the case of $X_g= 0.4$~pu, Fig.~\ref{Sim_1_1}~(c) and (d), a stable periodic orbit (SPO) is present in the post-fault system and has been shown by a green dashed curve in the phase plane. The SPO is an additional instability mode in this case and if the fault trajectory leaves the ROA, as it does for $t_\text{fault}>t_\text{SR}$ in Fig.~\ref{Sim_1_1}~(d), the trajectory converges to this SPO, leading to continuous divergence of $\delta_1$ and sustained oscillations in $\delta_2$. The results of the PHIL experiment are recorded in \figref{Sim_1_3}, which aligns closely with the EMT simulation. This instability mode is different from the instability mode under $X_g = 0.35$~pu.

Although not the primary focus of this paper, a brief analysis is provided to explain the implications of inverter pole-slipping. The ratio of the fault current from the grid side ($\tilde{I}_{fg}$) and the inverter side ($\tilde{I}_{fi}$) for $t_{\text{fault}} = 0.25$~s is shown in \figref{Sim_1_2}~(c) in terms of phasor. At time 0.2~s, the angle of $\tilde{I}_{fg}$ leads $\tilde{I}_{fi}$ by approximately 90 degrees, which can result in the misoperation of Relay 34 [the location of which is illustrated in \figref{philplatform}~(a)], as noted in \cite{fang2019distanceprotection}. Additionally, the impedance measurement by Relay 1 is shown in \figref{Sim_1_2}~(c), with the measured impedance becoming capacitance, signaling a fault in the transmission line between bus 1 and bus 3 in \figref{philplatform}~(a) (or \figref{fig2}). During this pole-slipping process, the IBR will suffer from reverse power injection, which is not permissible.
\begin{figure}[t]
\centering
\subfloat[$X_g=0.35$~pu, $t_\text{fault}=0.212$~s.]{\includegraphics[width=0.38\textwidth]{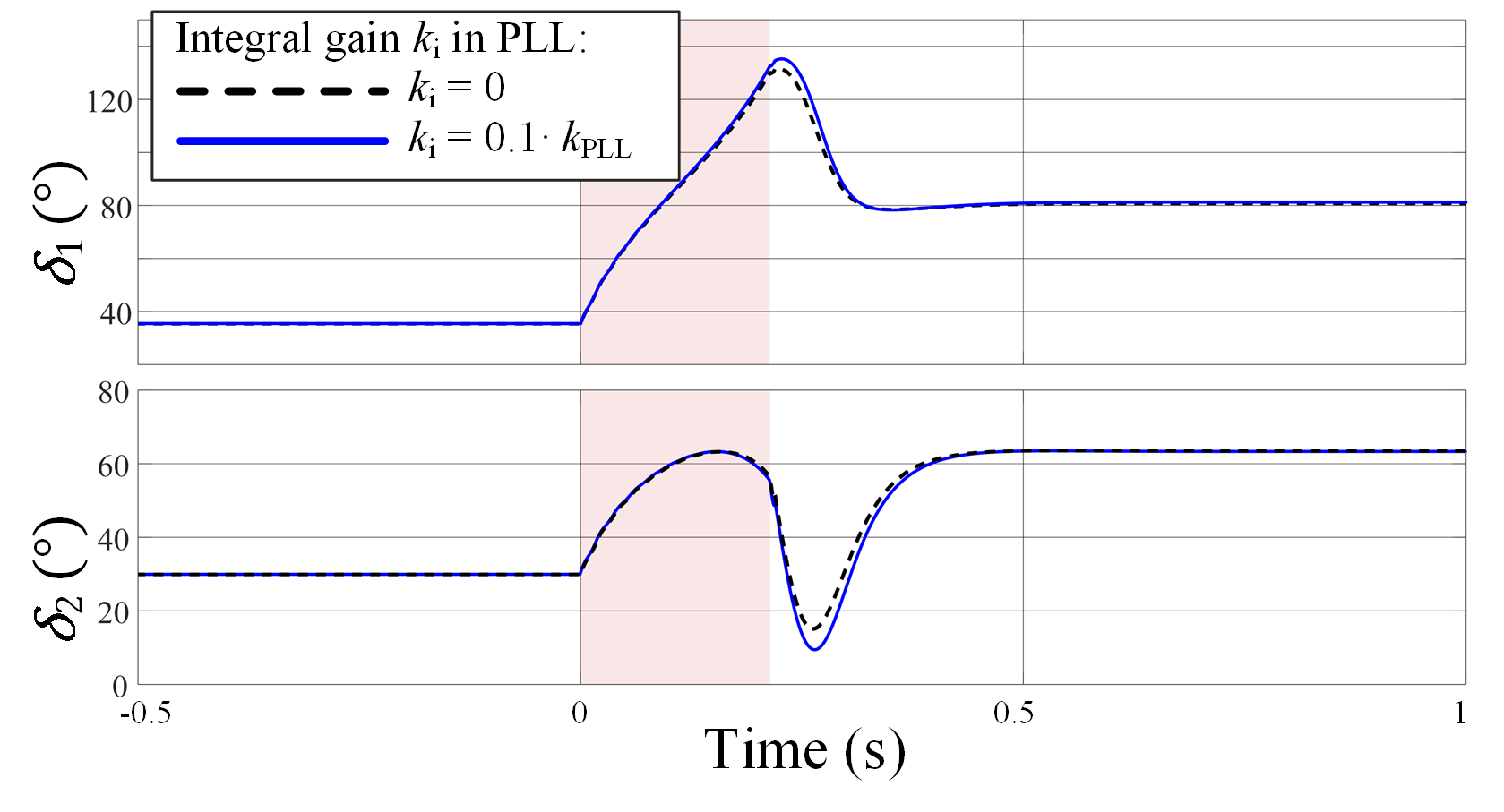}}\\
\subfloat[$X_g=0.4$~pu, $t_\text{fault}=0.131$~s.]{\includegraphics[width=0.38\textwidth]{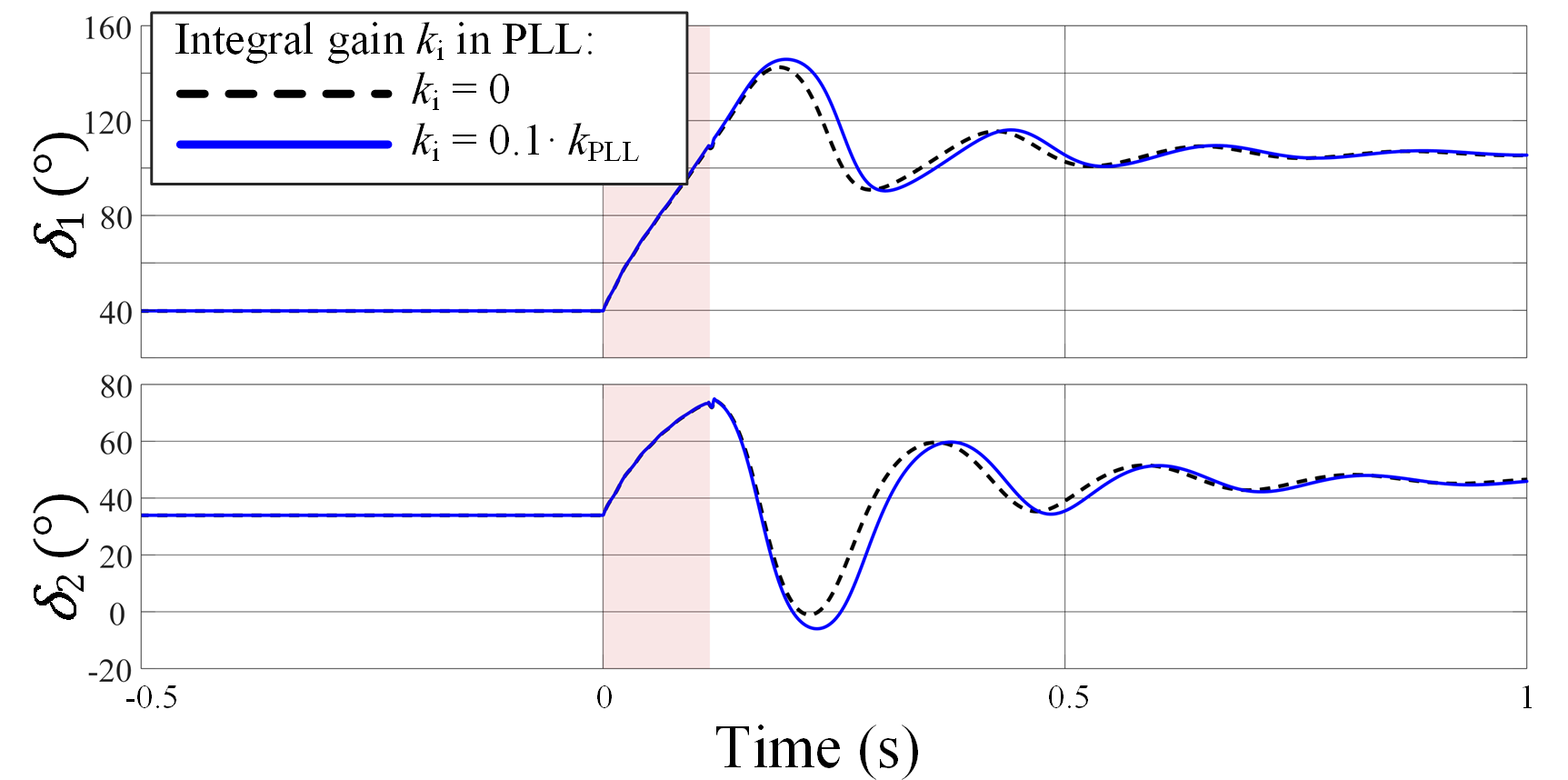}}
\caption{Comparison of EMT simulation results with different $k_i$ in PLL for different $X_g$ and $t_\text{fault}$.}
\vspace{-3mm}
\label{Sim_1_4}
\end{figure} 


\subsection{Validation of the order reduction approach} \label{reduceorder}
Now we validate the accuracy of the order reduction approach introduced in Section II-A. The PLL is configured with different integral gains $k_i$ and the EMT simulation results are presented in \figref{Sim_1_4}. The duration of the faults are shown by the pink shading. For both cases in (a) and (b) with different $X_g$ and $t_\text{fault}$, the EMT simulation shows a good match between the results with and without integral gains ($k_i=0$ and $k_i=0.1k_\text{PLL}$). This indicates that the reduced-order model neglecting the integral control in PLL presents a good approximation, provided that the PLL is designed to be overdamped.

\section{GFL-GFM(GSP) Interaction} \label{PLLGFM}
The effect of voltage support on the large-signal stability of a system that includes a GFL inverter was investigated in this section. Referring to \figref{philplatform}~(a) (or \figref{fig2}), IBR1 is configured as a GFL inverter and IBR2 is set as either a GFM or a GSP inverter. The transmission line parameters are set as follows: $X_1=0.5$~pu, $X_2=0.15$~pu, and the grid impedance $X_g$ is 0.3~pu under pre-fault conditions, increasing to 0.6~pu after the fault cleared. Detailed system parameters are provided in \tabref{TableA2} and \tabref{PHILPF} in Appendix \ref{parameters}.

\begin{figure}[t]
\centering
\subfloat[IBR2 is realized by GFM.]{\includegraphics[width=0.24\textwidth]{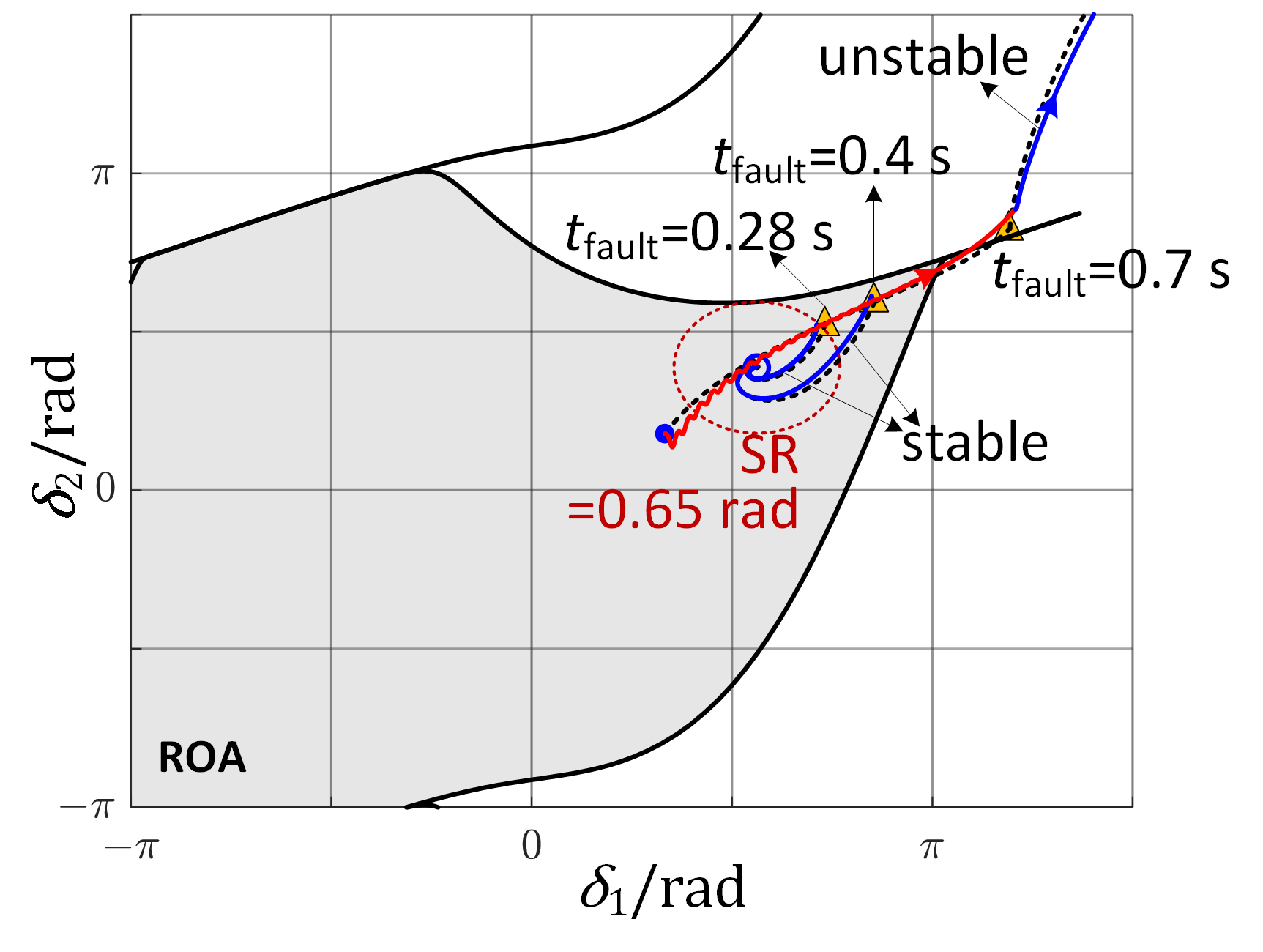}}\\
\subfloat[IBR2 is realized by GSP inverter under $k_v=1$.]{\includegraphics[width=0.234\textwidth]{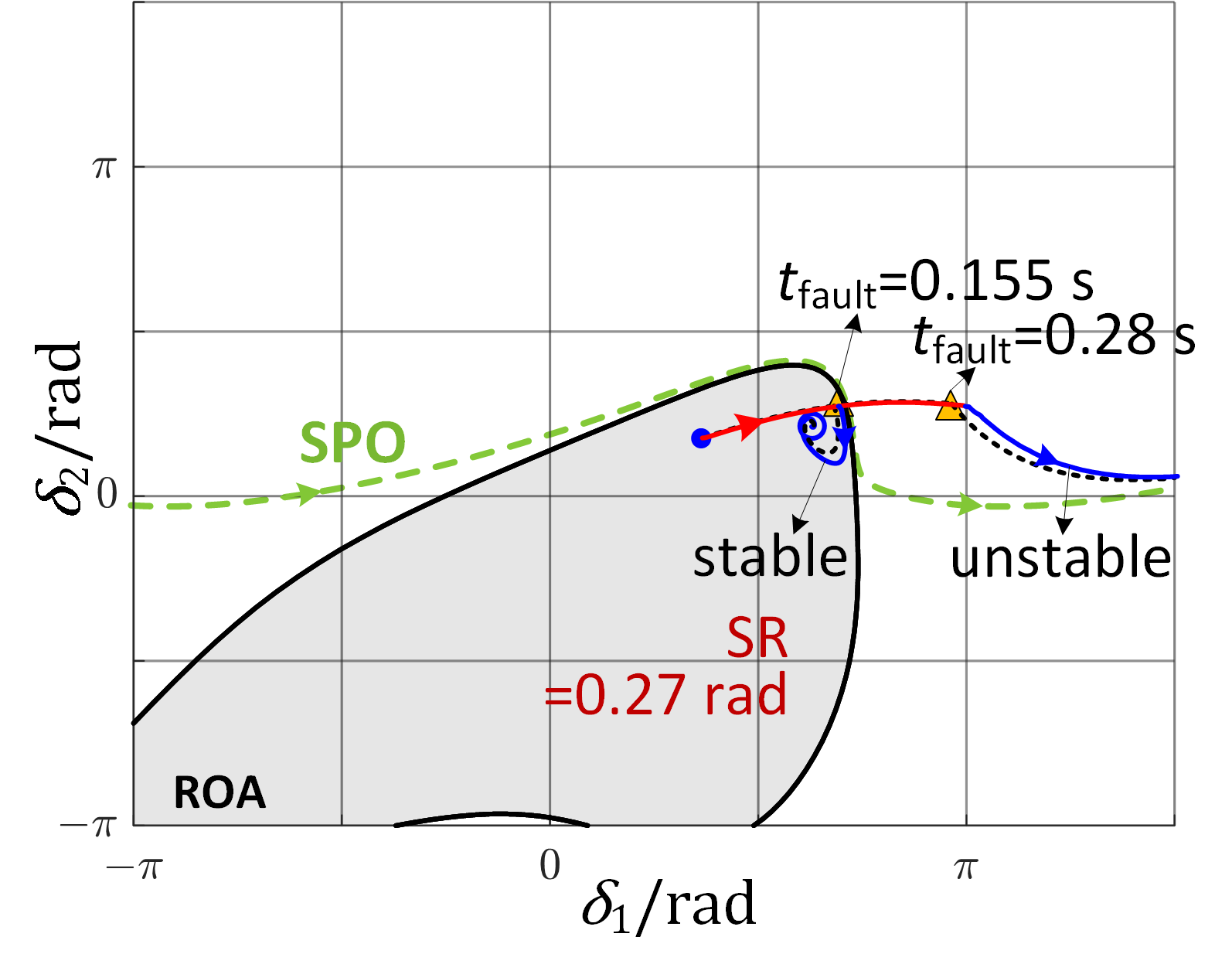}}\quad
\subfloat[IBR2 is realized by GSP inverter under $k_v=4$.]{\includegraphics[width=0.233\textwidth]{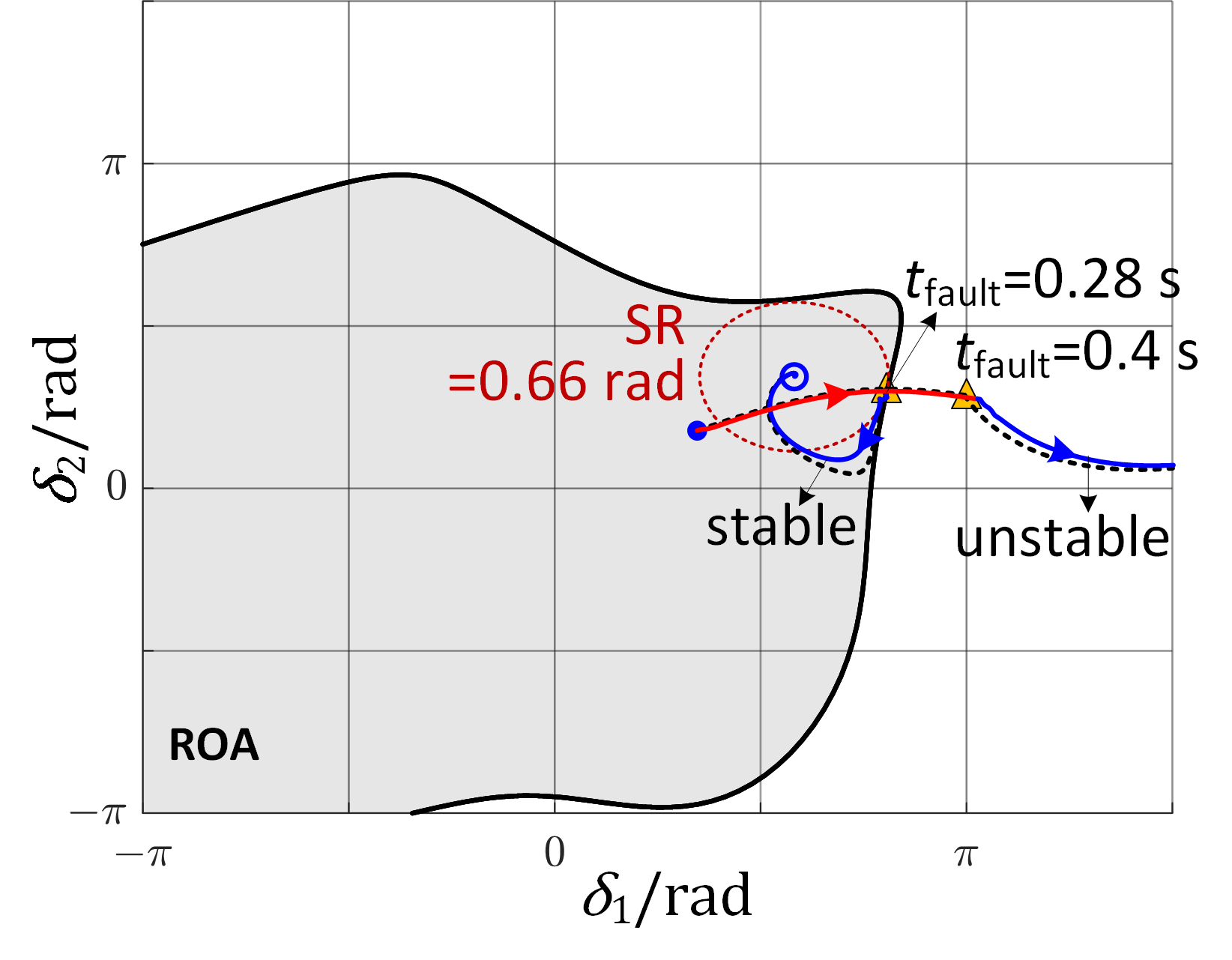}}
\caption{The phase portrait of the system before, during and after the fault for various $t_{\text{fault}}$ as marked on the figures. Color scheme for lines and markers follows that of \figref{Sim_1_1}.}
\vspace{-2mm}
\label{Sim_2_1}
\end{figure}

\subsection{Comparison of GFM and GSP inverters} \label{STATCOM_Comparison} 
Equations \eqref{GFLGFM} and \eqref{GFL+GFLVS} in Appendix \ref{Expression} reveal that when voltage control gain $k_v$ in the GSP inverter is large, both the GFM inverter and the GSP inverter exhibits similar dynamics so similar transient stability of a neighboring GFL inverter might be expected. For a fair comparison, when IBR2 is configured as a GSP inverter, its PLL gain is set to $k_{\text{PLL}2}=k_{\text{droop}}/X_{\Sigma2}$, so that the equivalent gain of IBR2 [$k_2$ in general expression \eqref{general}] remains consistent whether it operates as a GFM or a GSP inverter. A three-phase-to-ground fault is initiated at 0.5~s, with the fault located 0.8 of the distance along $X_g$ from the infinite bus.
\begin{figure}[t!]
\centering
\subfloat[$t_{\text{fault}}=0.28$~s.]{\includegraphics[width=0.4\textwidth]{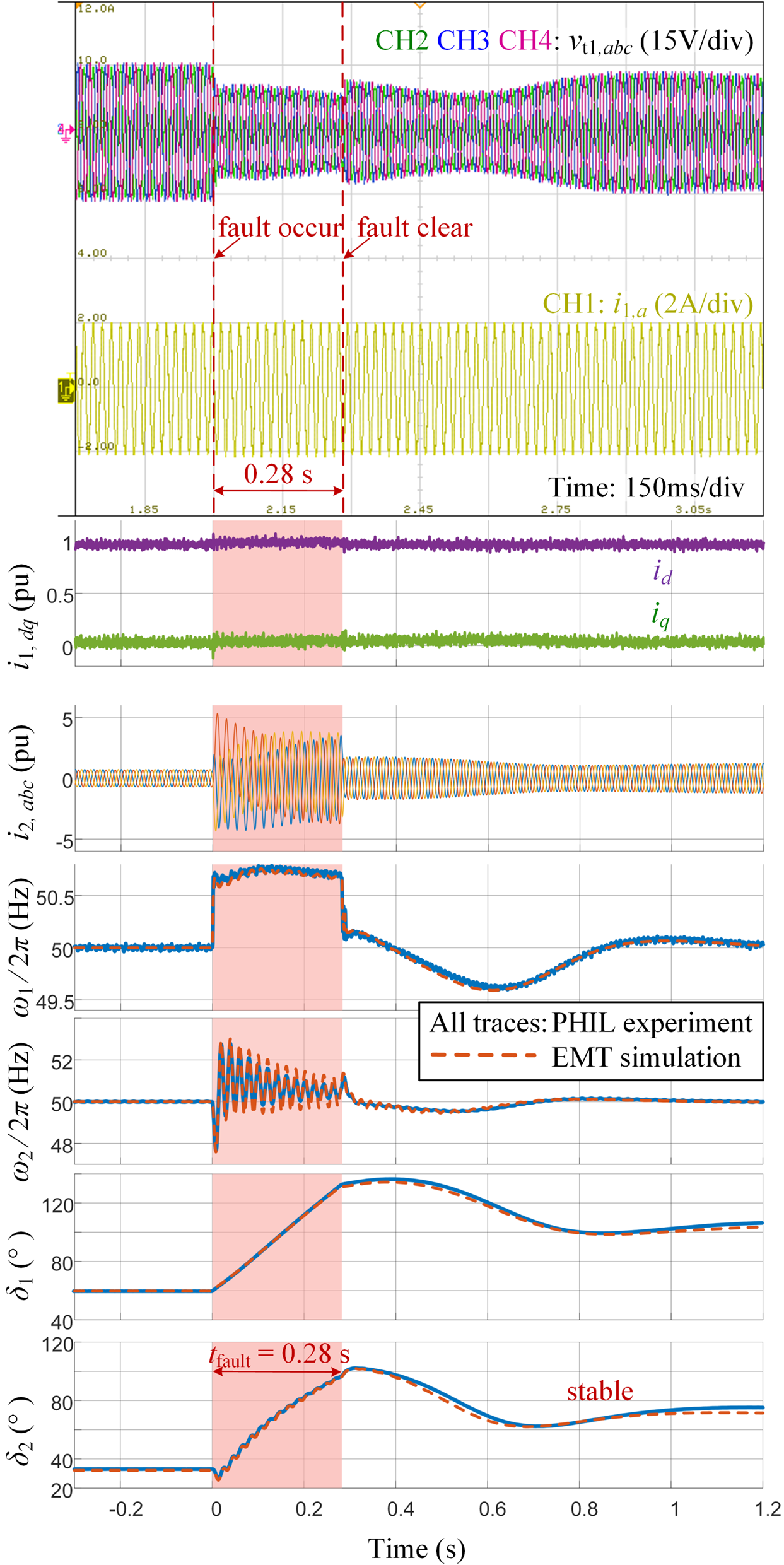}}
\caption{PHIL experiment results for the case where IBR2 is configured as a GFM inverter.}
\vspace{-1mm}
\label{Sim_2_2}
\end{figure}

\begin{figure}[t!]
\centering
\addtocounter{figure}{-1}
\setcounter{subfigure}{1}
\subfloat[$t_{\text{fault}}=0.7$~s.]{\includegraphics[width=0.4\textwidth]{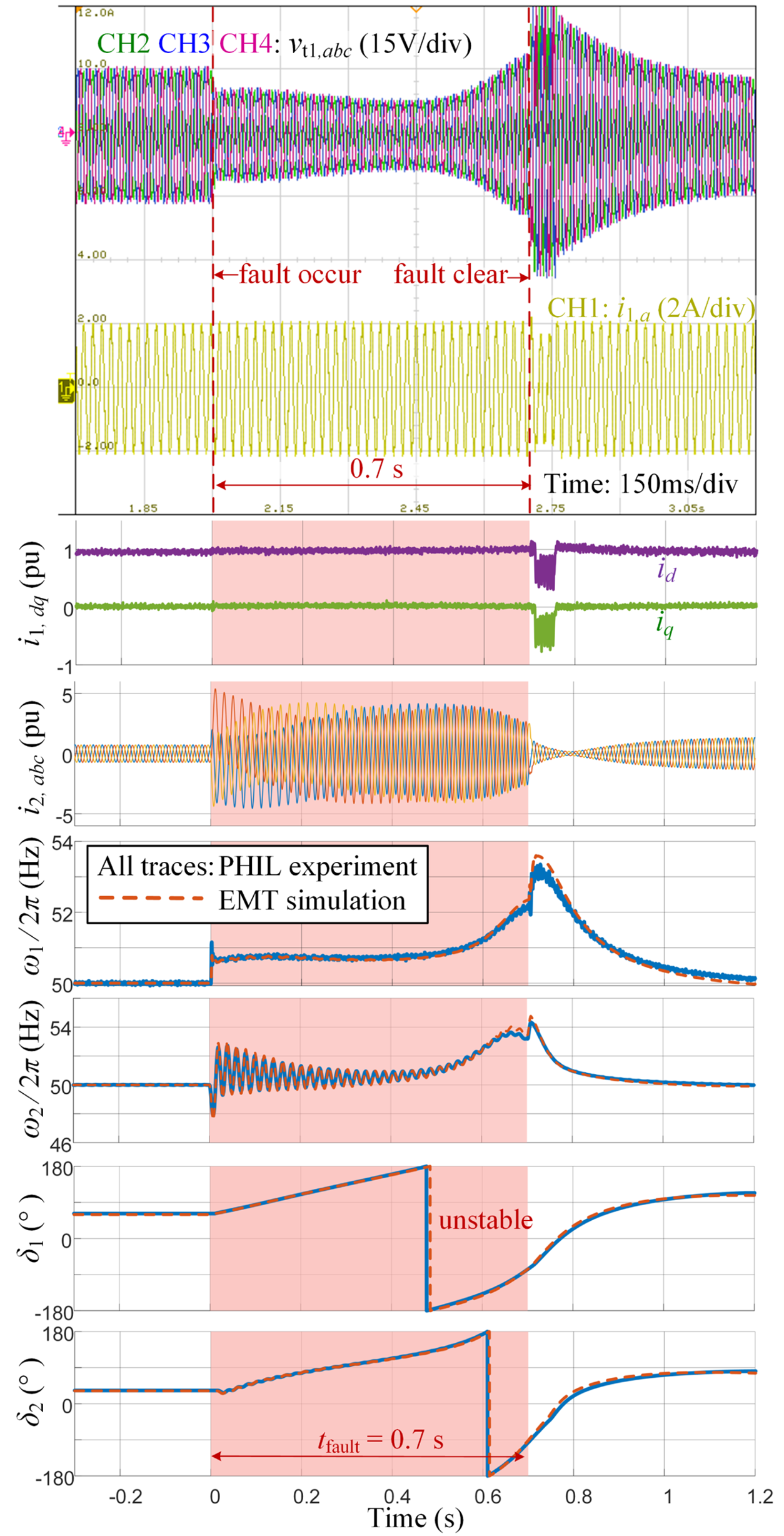}}\\
\subfloat[Measurements of Relay 34 \& 43 under $t_{\text{fault}}=0.7$~s.]{\includegraphics[width=0.42\textwidth]{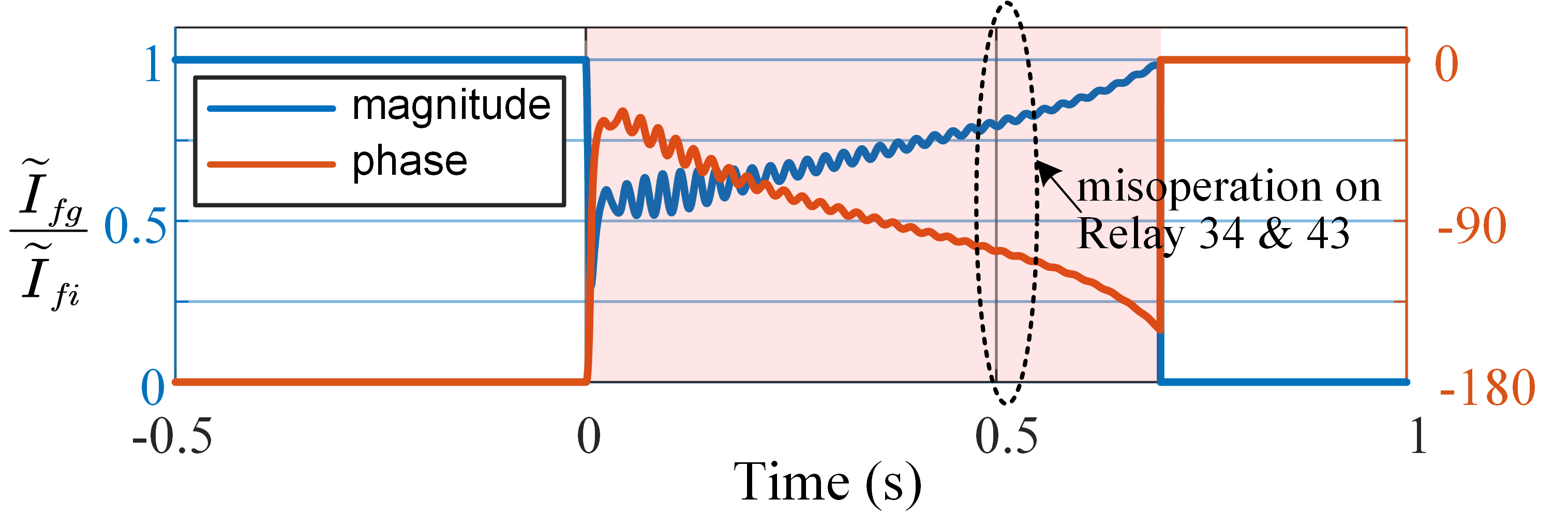}}
\caption{PHIL experiment results for the case where IBR2 is configured as a GFM inverter.(continued)}
\vspace{-1mm}
\end{figure} 

The phase portraits of the system before, during, and after the fault, with IBR2 configured as either a GFM inverter or GSP inverter, are shown in \figref{Sim_2_1}. Again, the ROA is shown as a gray-shaded area, the trajectories of the IBRs obtained from numerical calculations on the simplified analytical model are shown as black dashed lines and the fault-clearing points are marked with yellow triangles. The trajectories obtained from EMT simulations are shown as red lines during the fault and blue lines post-fault. The EMT simulation results closely match their corresponding numerical calculations.

In \figref{Sim_2_1}~(a) shows phase-plane results for IBR2 configured as a GFM inverter. The SR has a radius of 0.65 rad and the CCT estimated by the SR is $t_\text{SR}=0.28$~s. When $t_{\text{fault}}$ is set to this value, the post-fault trajectory of the EMT simulation converges to the post-fault SEP, confirming that $t_\text{SR}$ is a good and slightly conservative estimate of the critical clearing time. The conservativeness is seen from $t_{\text{fault}}=0.4$~s also being stable. The system is unstable at $t_{\text{fault}}=0.7$~s. 

The corresponding PHIL experimental results are shown in \figref{Sim_2_2} with $t_{\text{fault}}=0.28$~s seen to be stable in \figref{Sim_2_2}.~(a) and $t_{\text{fault}}=0.7$~s unstable in \figref{Sim_2_2}.~(b). Furthermore, \figref{Sim_2_2}~(c) shows the phase difference between the currents through Relay 34 ($I_{fi}$) and Relay 43 ($I_{fg}$) exceeds $-90^\circ$ and approaches $-180^\circ$, which would lead to mal-operation of distance protection relay \cite{fang2019distanceprotection} and is why this condition is dangerous.

\begin{figure}[t!]
\centering
\subfloat[Voltage control gain $k_v=1$.]{\includegraphics[width=0.38\textwidth]{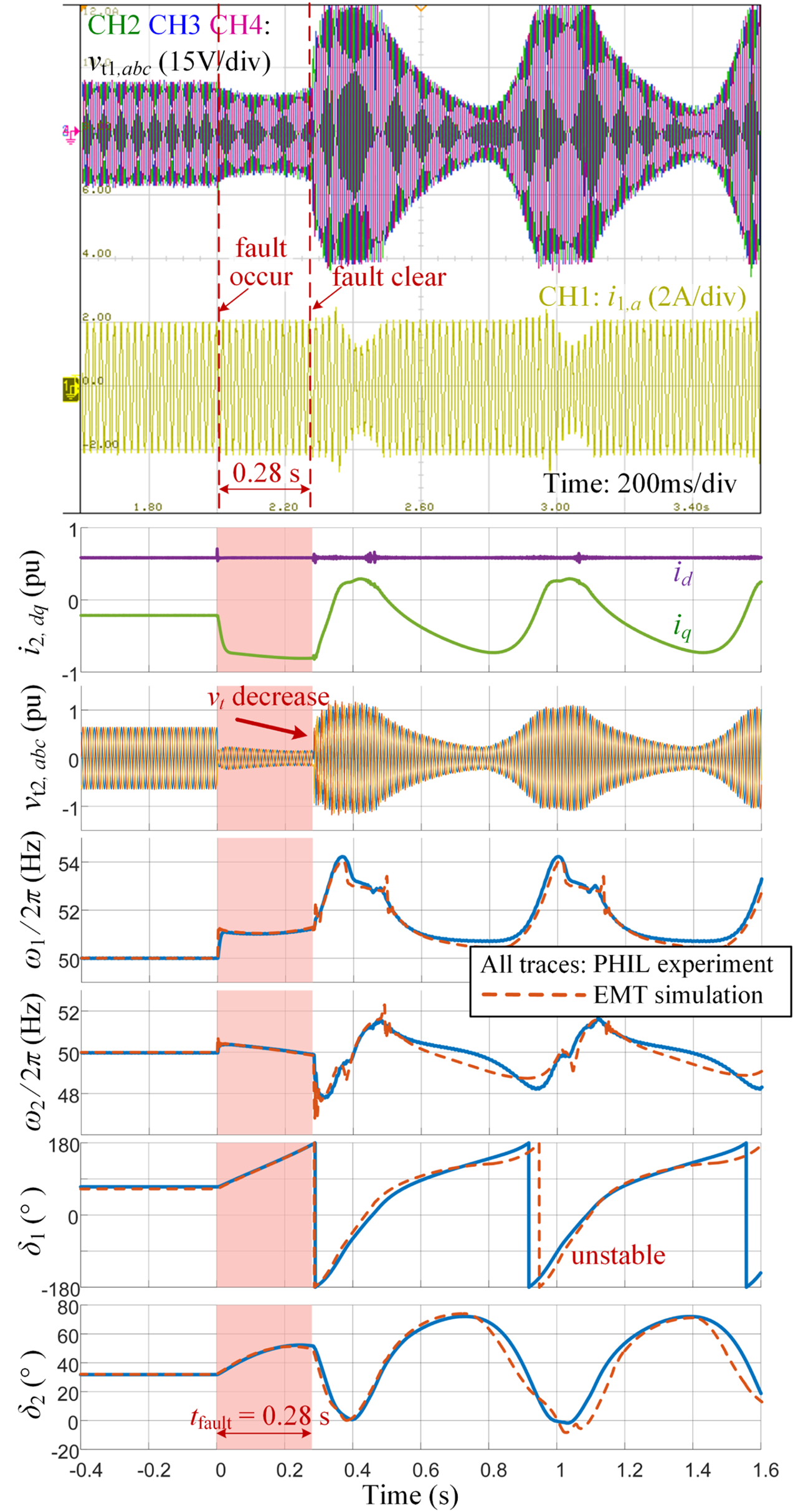}}
\caption{PHIL experiment results under $t_{\text{fault}}=0.28$~s for the case where IBR2 is configured as a GSP with two different values of $k_v$.}
\label{Sim_2_3}
\end{figure} 

\begin{figure}[t!]
\centering
\addtocounter{figure}{-1}
\setcounter{subfigure}{1}
\subfloat[Voltage control gain $k_v=4$.]{\includegraphics[width=0.382\textwidth]{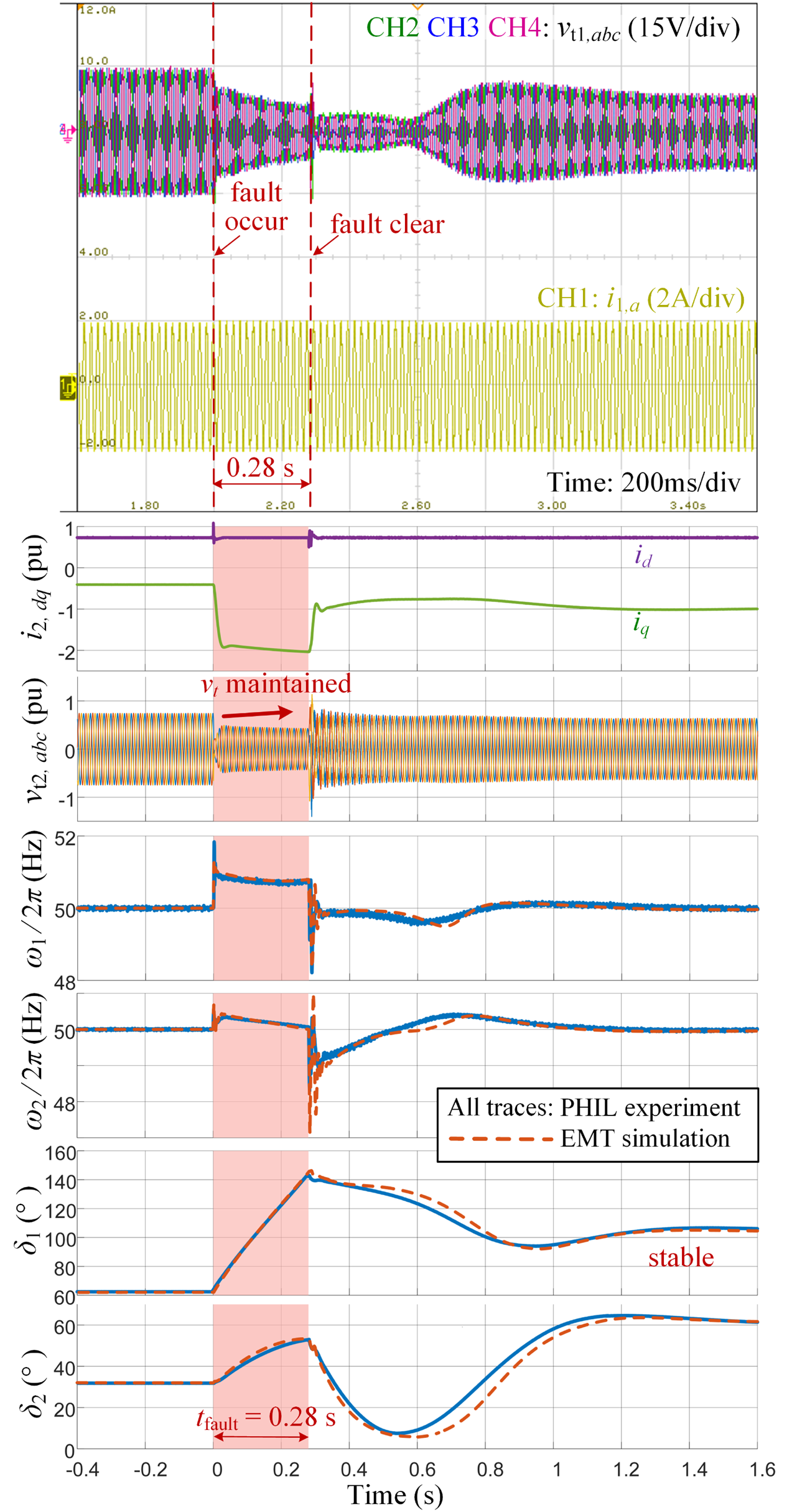}}
\caption{PHIL experiment results under $t_{\text{fault}}=0.28$~s for the case where IBR2 is configured as a GSP with two different values of $k_v$. (continued)}
\vspace{-1mm}
\end{figure} 

\begin{figure}[t]
\centering
\subfloat[$k_{\text{PLL}2}=k_{\text{droop}}=2.5\times2\pi$.]{\includegraphics[width=0.24\textwidth]{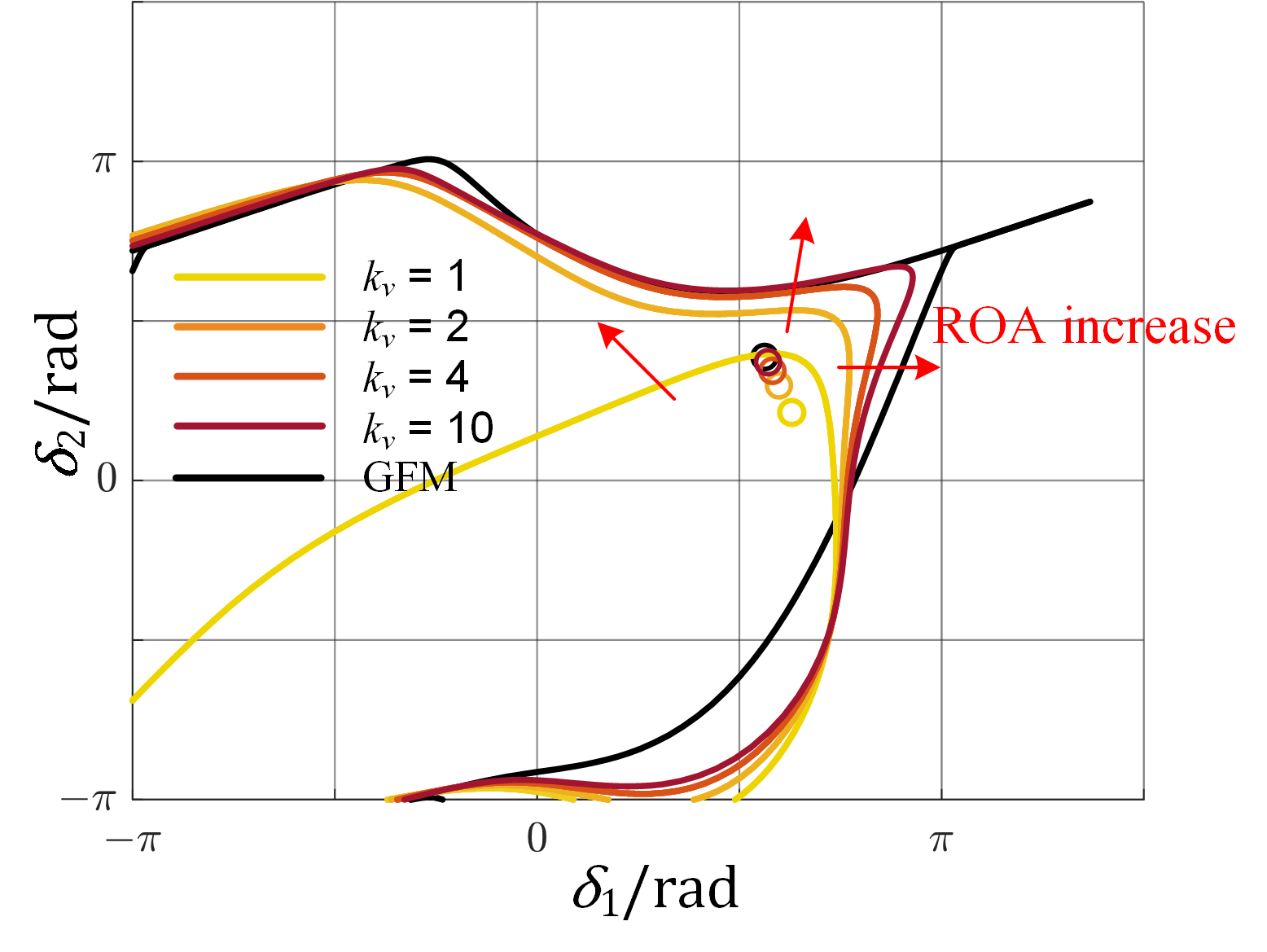}}
\subfloat[$k_{\text{PLL}2}=k_{\text{droop}}=10\times2\pi$.]{\includegraphics[width=0.235\textwidth]{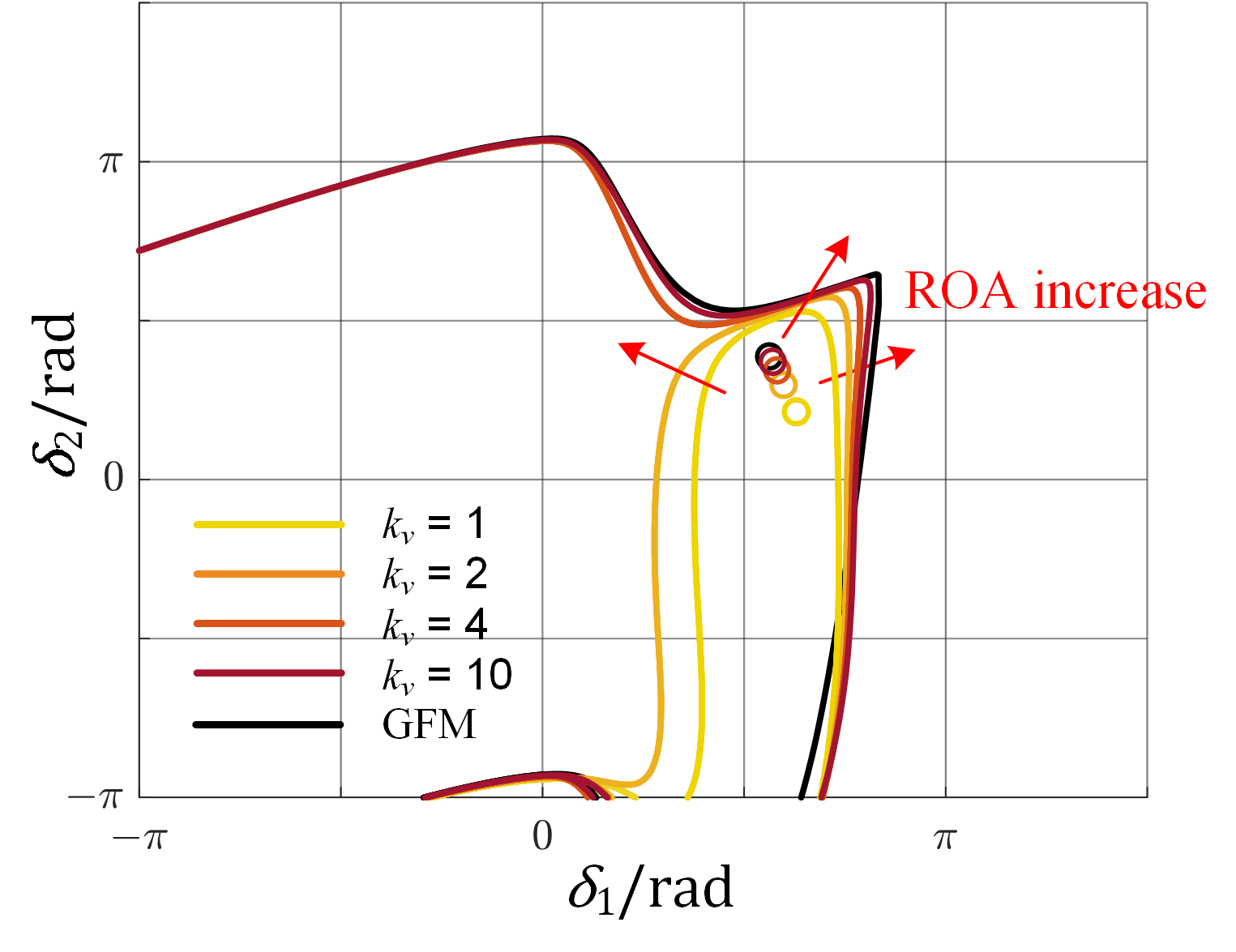}}
\caption{The change of ROA with respect to voltage control gain $k_v$ and PLL bandwidth of GSP inverter.}
\vspace{-2mm}
\label{Sim_2_4}
\end{figure}

Turning now to IBR2 configured as a GSP inverter. \figref{Sim_2_1}~(b) shows results for when the voltage control gain is small, $k_v=1$. Compared to \figref{Sim_2_1}~(a), the SR is much smaller which is 0.27~rad, and $t_\text{SR}$ is 0.155~s. An SPO outside the ROA is plotted in green dashed curves. If $t_{\text{fault}}=0.28$~s, the trajectory converges to this SPO, rendering the system unstable. The corresponding PHIL experiment results, shown in \figref{Sim_2_3}~(a), demonstrate the divergence of $\delta_1$ to positive infinity and oscillation of $\delta_2$. 

If the voltage control gain is increased to $k_v=4$, the ROA is expanded as shown in \figref{Sim_2_1}~(c). The SR at $0.66$~rad becomes comparable to that of the system in which IBR2 operates as a GFM inverter and $t_\text{SR}$ becomes $0.28$~s. A difference is that with the GFM inverter, the ROA is larger along the \( \delta_1 \)-axis. For instance, when $t_{\text{fault}} = 0.4~s$ the system loses stability with the GSP inverter but remains stable when using the GFM inverter despite the near identical SR and $t_\text{SR}$. The PHIL experiment results for $t_{\text{fault}} = 0.28$~s are presented in \figref{Sim_2_3}~(b), showing that a higher $k_v$ provides enhanced voltage support, maintaining the terminal voltage of IBR2 even under severe fault conditions.

\figref{Sim_2_4} illustrates the ROA of the GSP inverters under different voltage control gain $k_v$ and PLL gain $k_{\text{PLL}2}$. As $k_v$ increases, the ROA expands, improving the system’s large-signal stability. Under large $k_v$, the ROA is very near the ROA of the system where IBR2 is a GFM inverter depicted by the black line. The ROA is even closer to that of the GFM case if bandwidth of the PLL2 $k_{\text{PLL}2}$ is high, as seen in \figref{Sim_2_4}~(b). This indicates that a GFL inverter with voltage control can provide very similar transient stability enhancement to a GFM inverter when the PLL dynamics are fast and the voltage control gain $k_v$ is large.

\subsection{Impact of the location of GFM and GSP inverter on transient stability enhancement} \label{STATCOM_location}
The impact on ROA and SR of varying the connection point of the IBR2 acting as a GSP inverter was investigated by testing variations of grid impedance $X_g$ and $X_1$ but with the total transmission line impedance ($X_{\Sigma1} = X_1 + X_g = 1.1 \text{ pu}$) kept constant. The fault type is changed to a remote fault simulated as a voltage sag of 0.1 pu at the infinite bus and the fault duration is set to 0.1~s and 0.25~s. 

\begin{figure}[t]
\centering
\subfloat[GFM near the IBR1, \\$X_1 = 0.2$~pu, $X_g = 0.9$~pu.]{\includegraphics[width=0.233\textwidth]{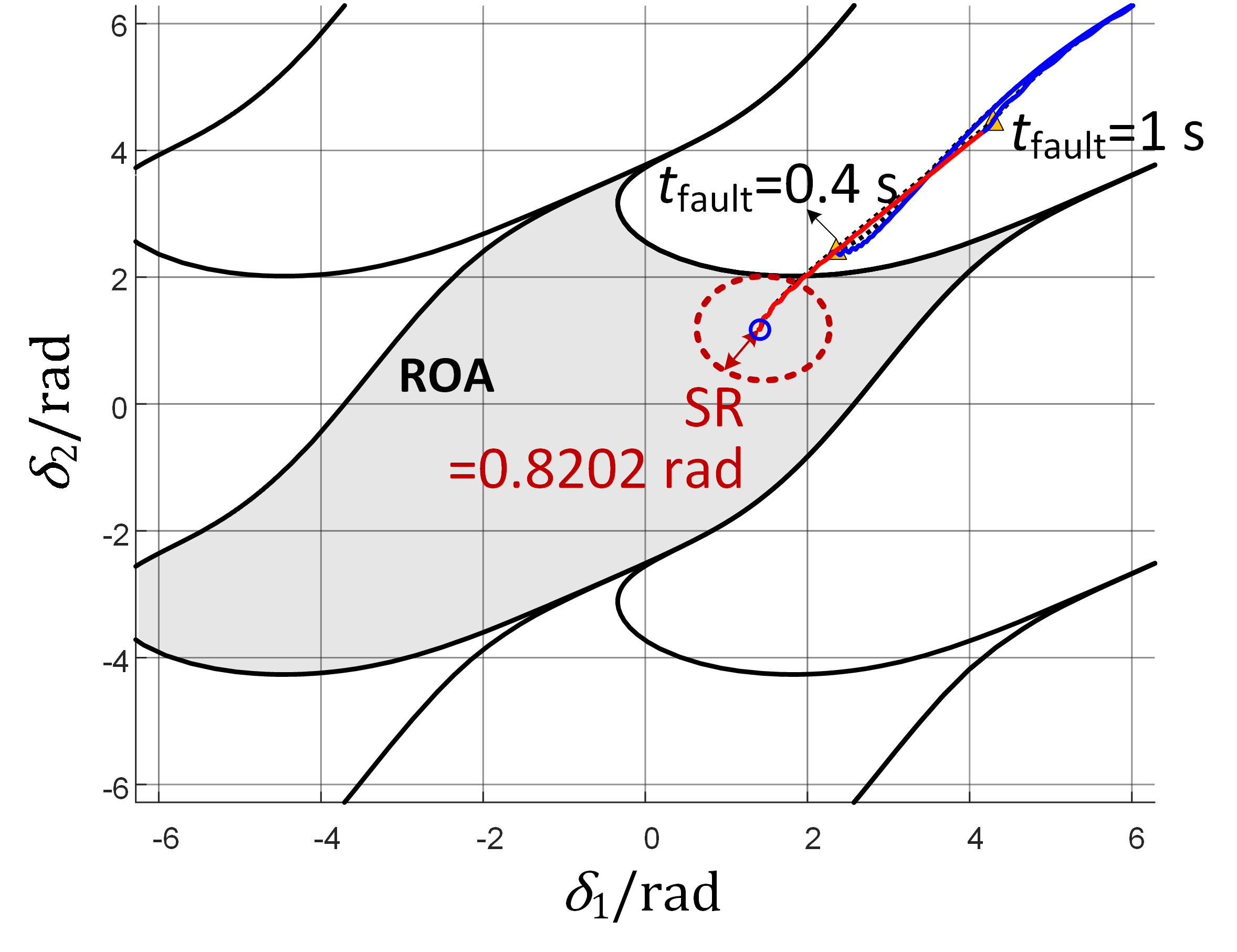}}\quad
\subfloat[GSP near the IBR1, \\$X_1 = 0.2$~pu, $X_g = 0.9$~pu.]{\includegraphics[width=0.233\textwidth]{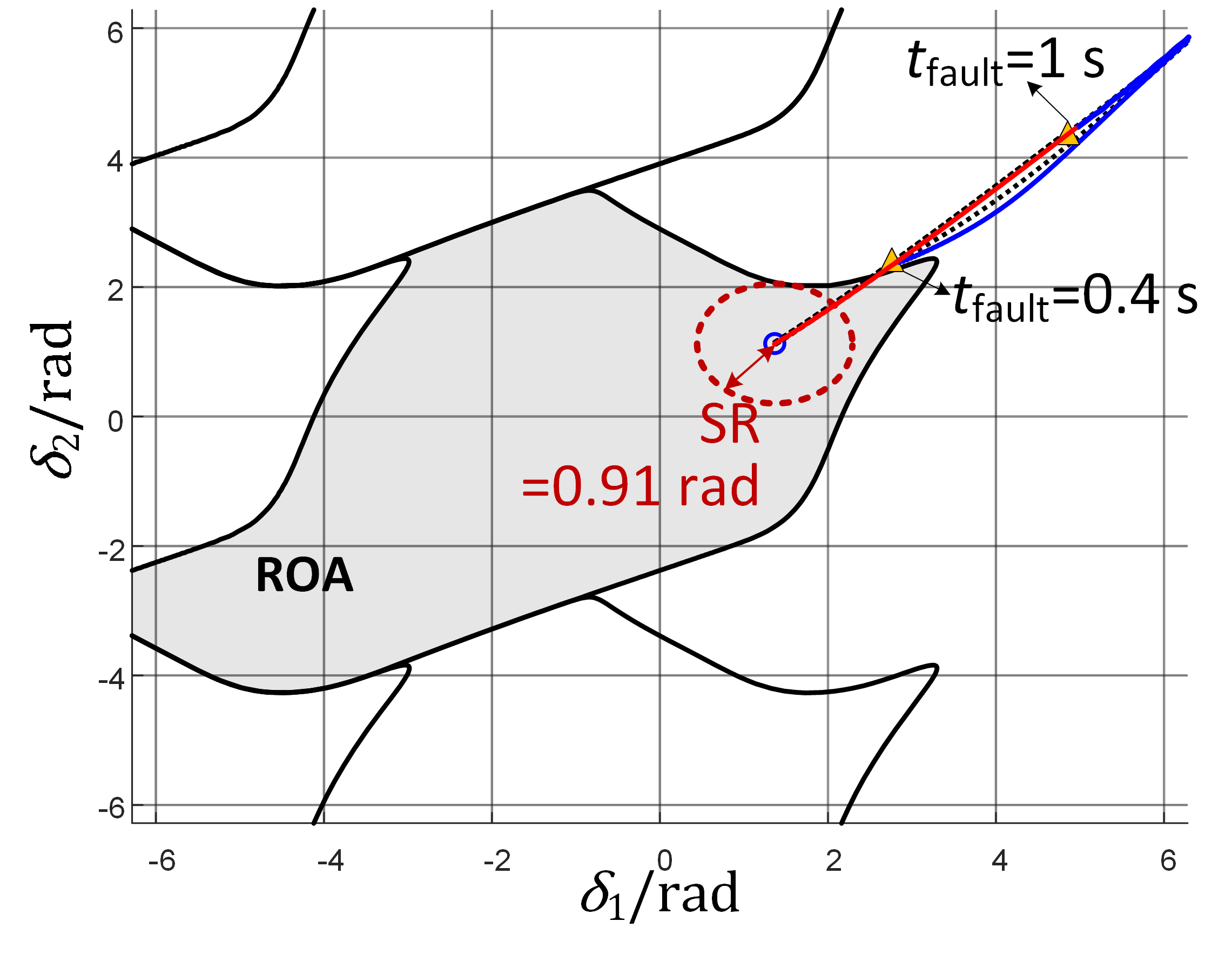}}\\
\subfloat[GFM placed intermediately, \\$X_1 = 0.5$~pu, $X_g = 0.6$~pu.]{\includegraphics[width=0.225\textwidth]{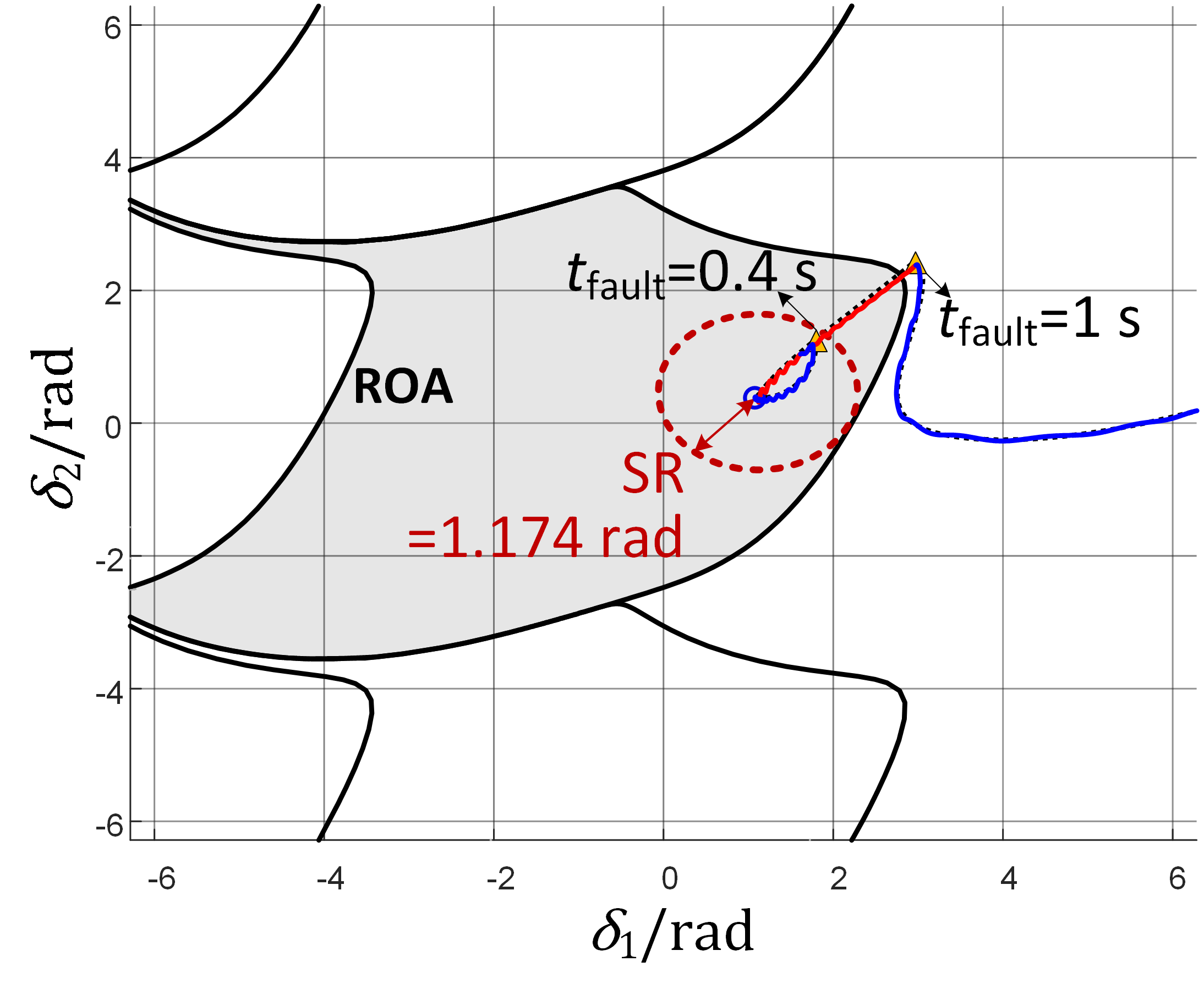}}\quad
\subfloat[GSP placed intermediately, \\$X_1 = 0.5$~pu, $X_g = 0.6$~pu.]{\includegraphics[width=0.238\textwidth]{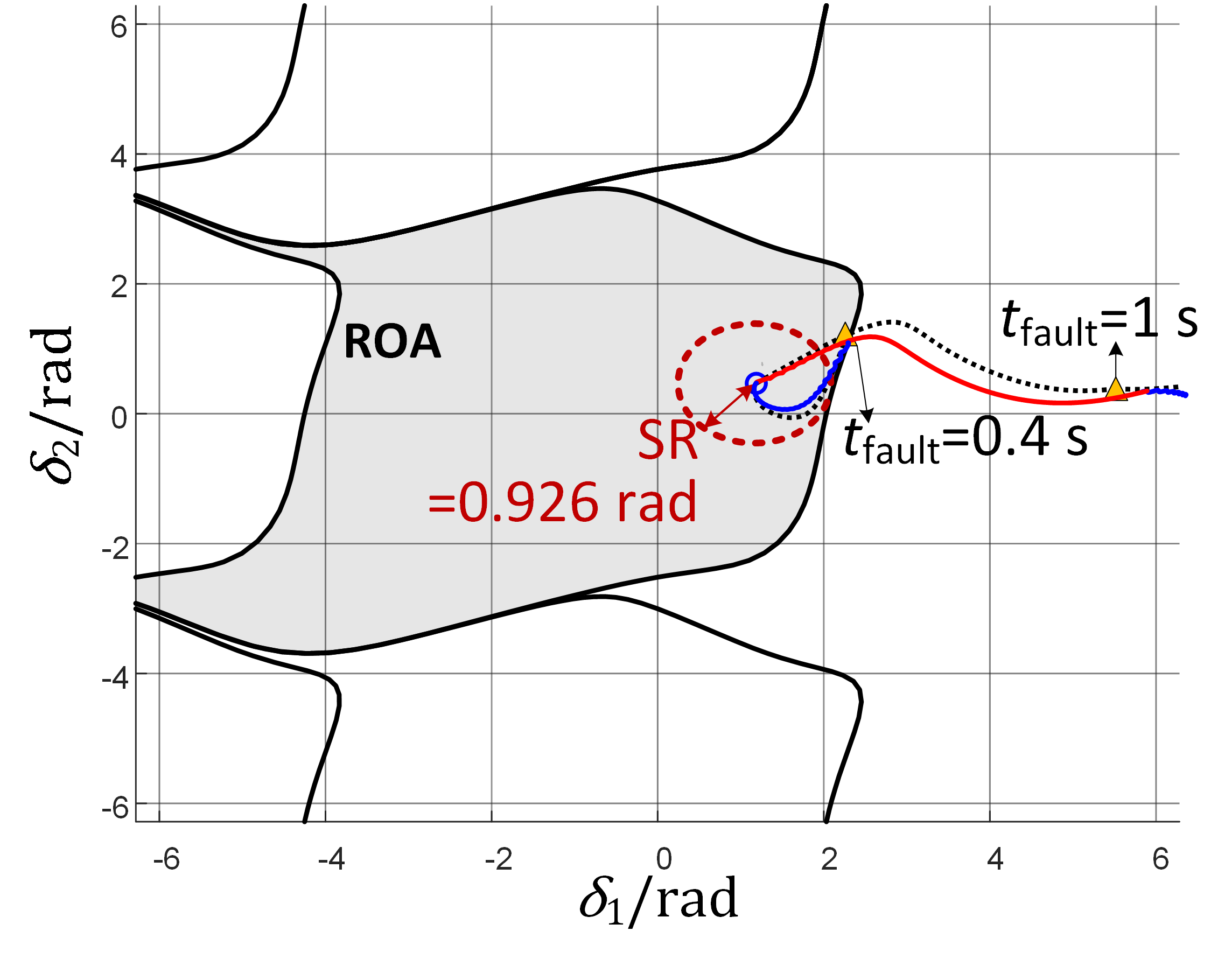}}\\
\subfloat[GFM near the grid, \\$X_1 = 0.9$~pu, $X_g = 0.2$~pu.]{\includegraphics[width=0.23\textwidth]{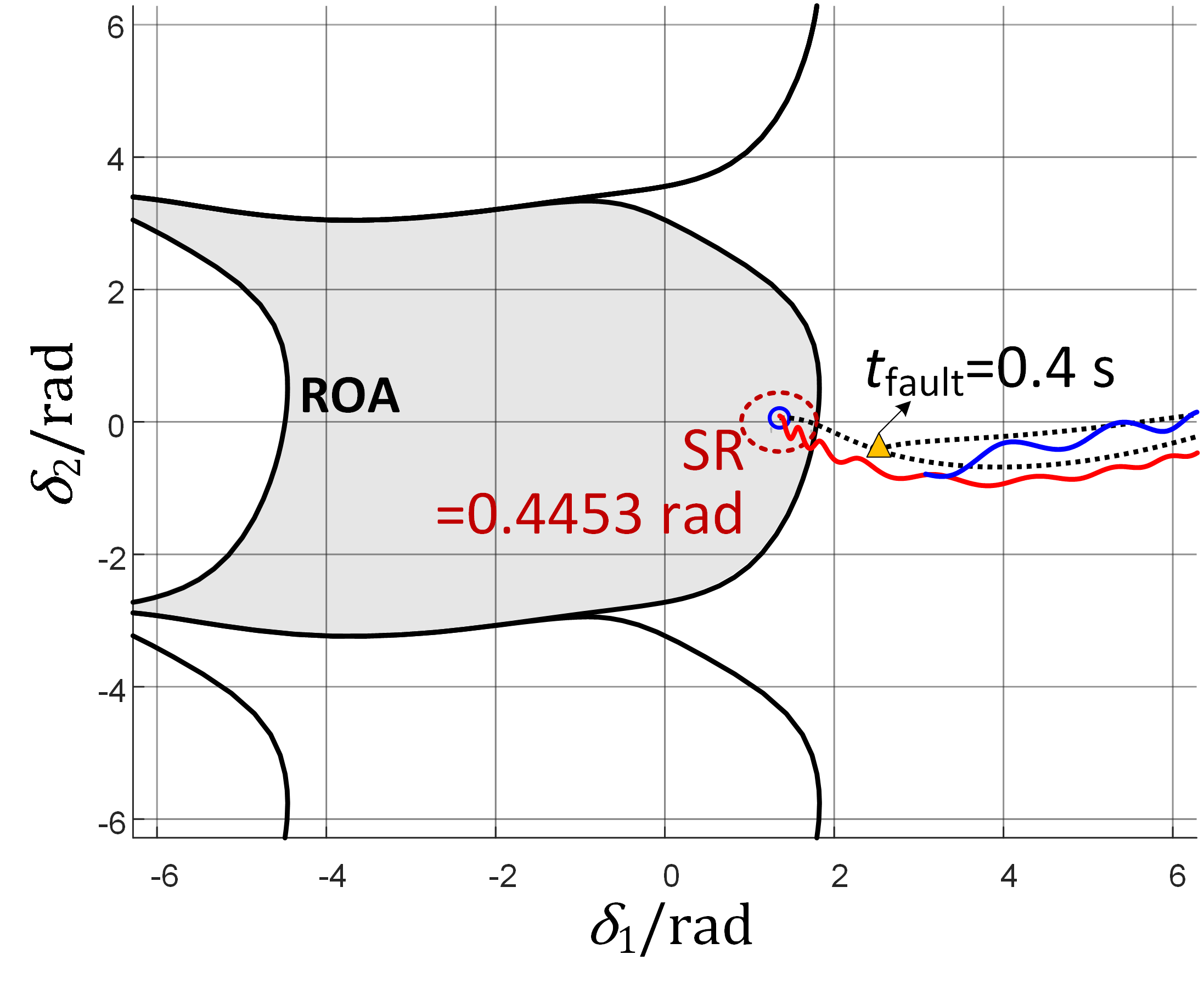}}\quad
\subfloat[GSP near the grid, \\$X_1 = 0.9$~pu, $X_g = 0.2$~pu.]{\includegraphics[width=0.234\textwidth]{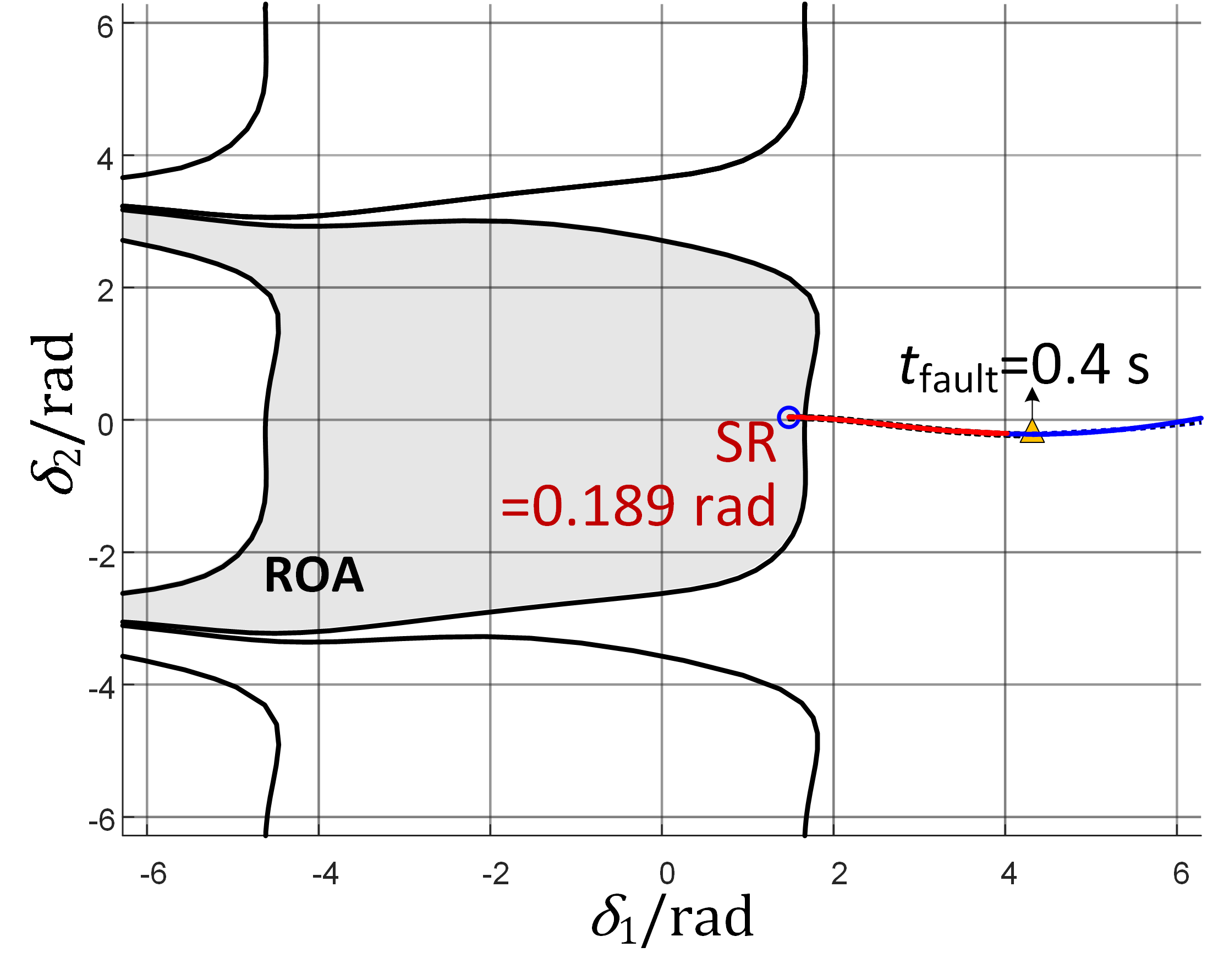}}
\caption{The EMT simulation results, ROA and SR under various locations of IBR2 ($t_{\text{fault}}$~=~0.4~s, 1~s). Color scheme for lines and markers follows that of \figref{Sim_1_1}.}
\label{STATCOMLocation}
\vspace{-0.1cm}
\end{figure} 
The resulting ROA for six cases are illustrated in \figref{STATCOMLocation}. The left column has results for a GFM inverter and the right column for a GSP inverter. The top row is for the IBR2 (GFM  or GSP inverter) near the IBR1 (GFL inverter) and is further away in the middle and bottom rows. Examining \figref{STATCOMLocation} shows that, regardless of whether the
IBR2 is a GFM or GSP inverter, when it is placed in the middle of the transmission line, it is most beneficial to the large-signal stability of the system. When the IBR2 is too near to the grid, its ability to improve voltage regulation is diminished because of proximity to an infinite bus. In the extreme case, when it is connected to the grid directly, there would be no interaction between the IBR2 and IBR1. Conversely, if the IBR2 is positioned too close to IBR1, the impedance between the IBR2 and the grid denoted as $X_g$, becomes large. Since the IBR2 must absorb active power from IBR1 and transfer it to the grid, this large $X_g$  reduces the transmission capacity from IBR2 to the grid and can lead to instability in the IBR2 itself. As illustrated in \figref{STATCOMLocation}~(a) and (b), the SEP is very close to the upper boundary of ROA, which results in small SR and $t_{\text{SR}}$. In particular, when IBR2 is placed at the same location as IBR1, the system even lacks an equilibrium point. Additionally, it is important to note that if there is no IBR2, the system becomes unstable even though the IBR1 is configured as a GSP inverter. This indicates that voltage control in the GSP inverter, while ineffective for improving the stability of the GSP inverter itself, can enhance the stability of other IBRs. The enhancement is maximized if it is placed at the midpoint of the transmission line.

\section{GFM-GSP Interaction} \label{GFLVGFM}

\begin{figure}[b]
\centering
\subfloat[$k_v=0$ (GFL inverter) .]{\includegraphics[width=0.24\textwidth]{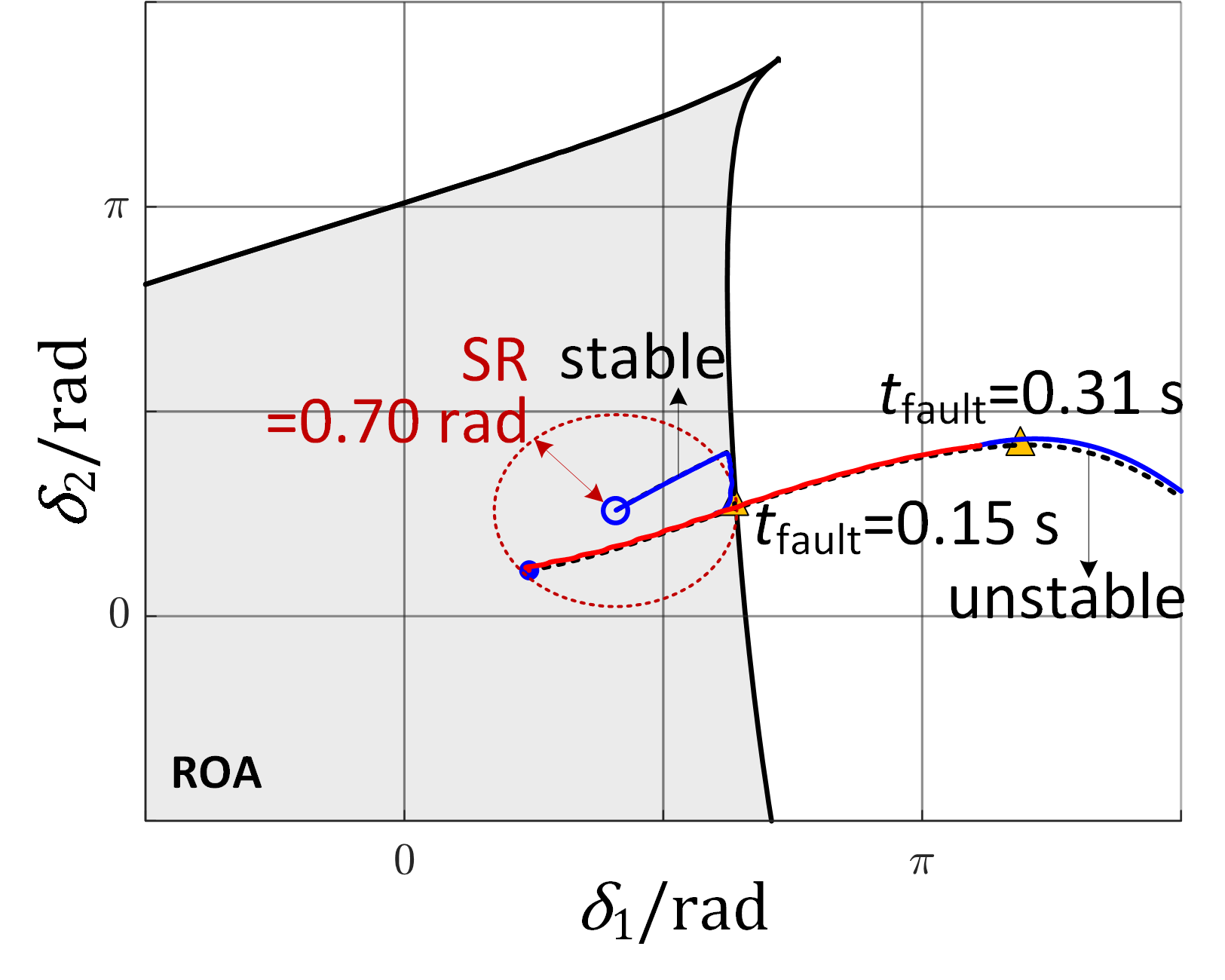}}\\
\subfloat[$k_v=2$ in GSP inverter.]{\includegraphics[width=0.24\textwidth]{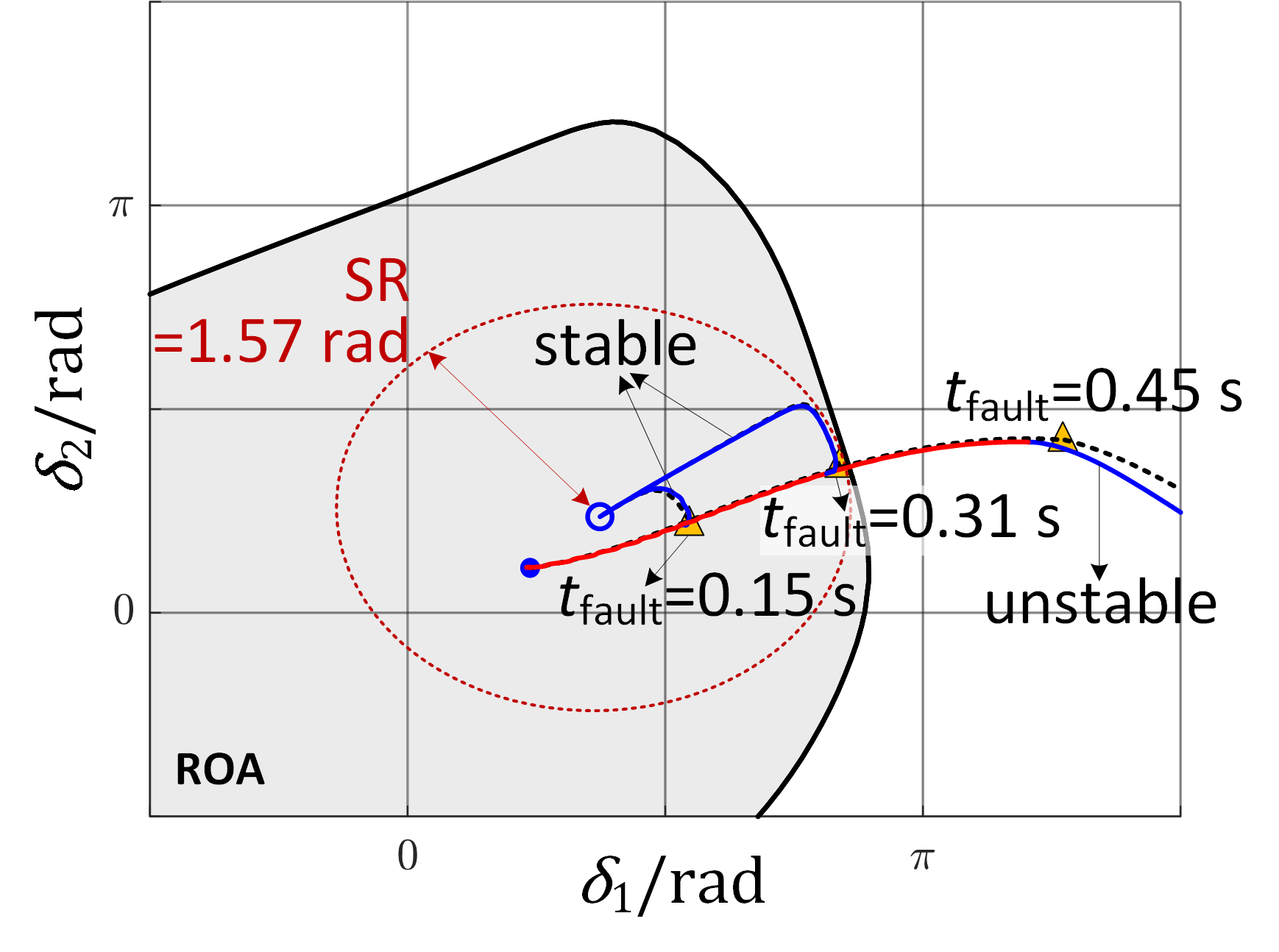}}\quad
\subfloat[$k_v=4$ in GSP inverter.]{\includegraphics[width=0.225\textwidth]{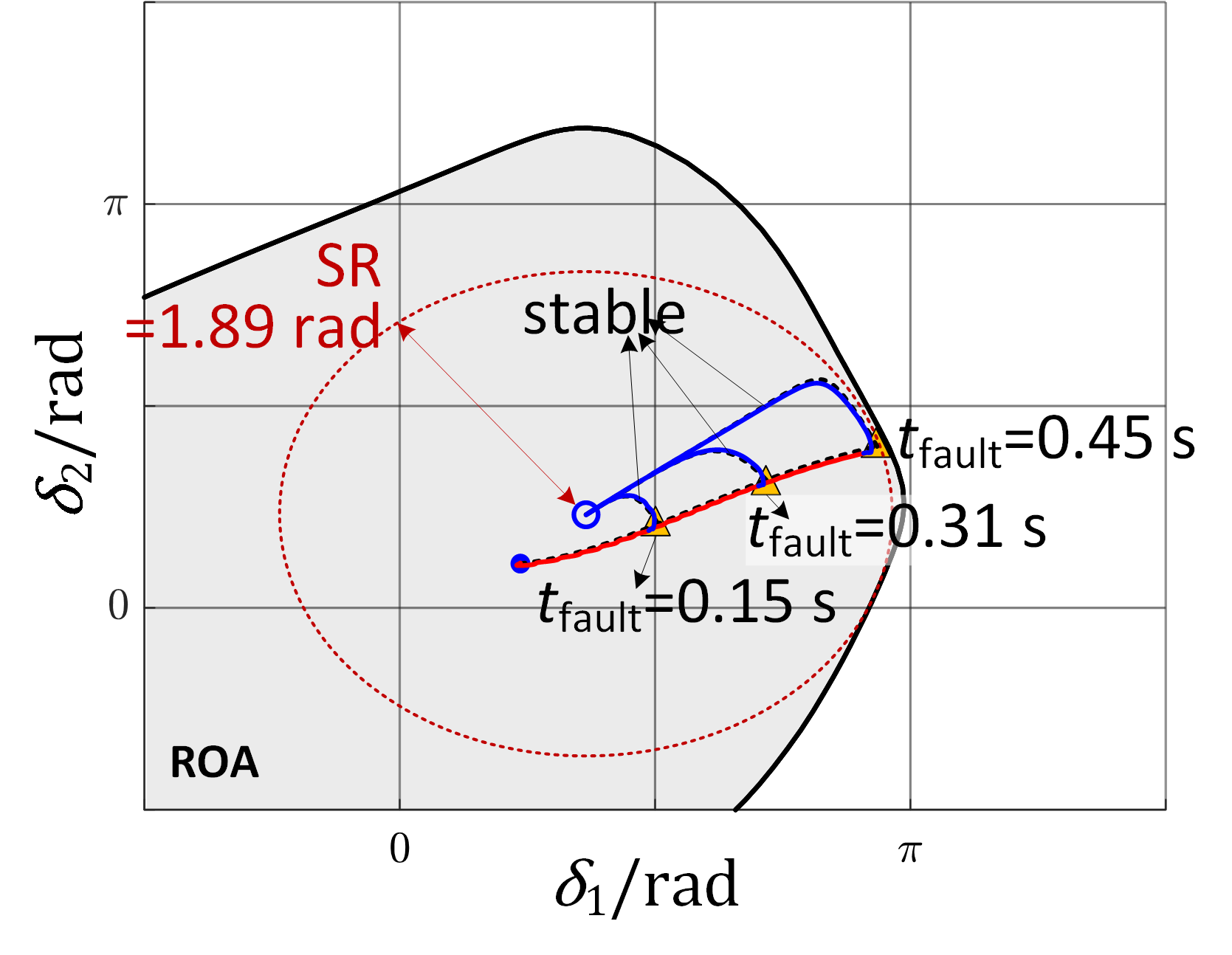}}
\caption{The phase portraits of GFM inverter and GSP inverter with three choices of voltage control gains. Fault duration set to estimated critical clearing time for each case ($t_{\text{fault}}$ = 0.15~s, 0.31~s, and 0.45~s respectively). Color scheme for lines and markers follows that of \figref{Sim_1_1}.}
\label{Sim_3_1}
\vspace{-0.1cm}
\end{figure}

\begin{figure}[t]
\centering
\subfloat[change of ROA]{\includegraphics[width=0.23\textwidth]{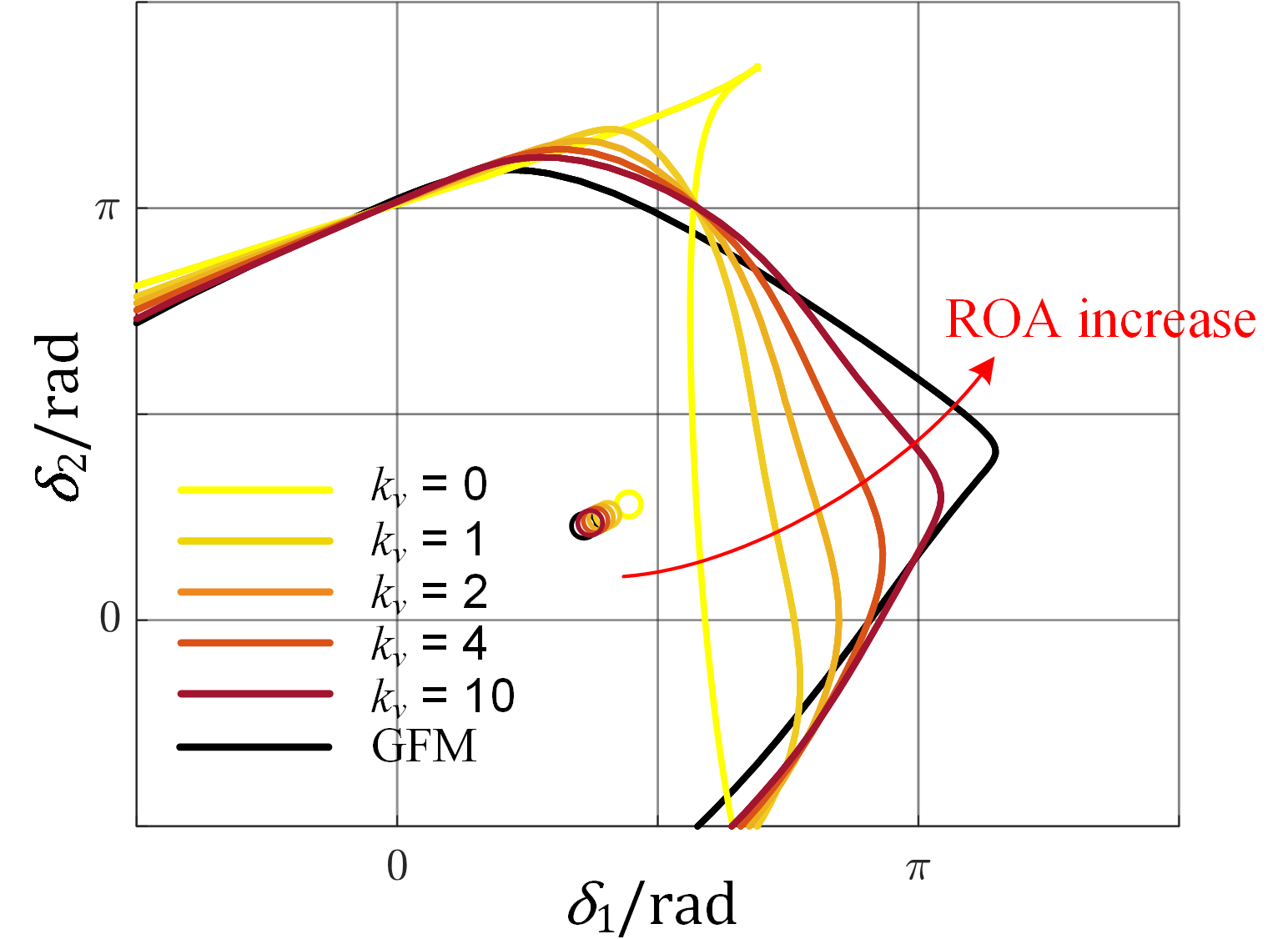}}\quad
\subfloat[change of SR]{\includegraphics[width=0.20\textwidth]{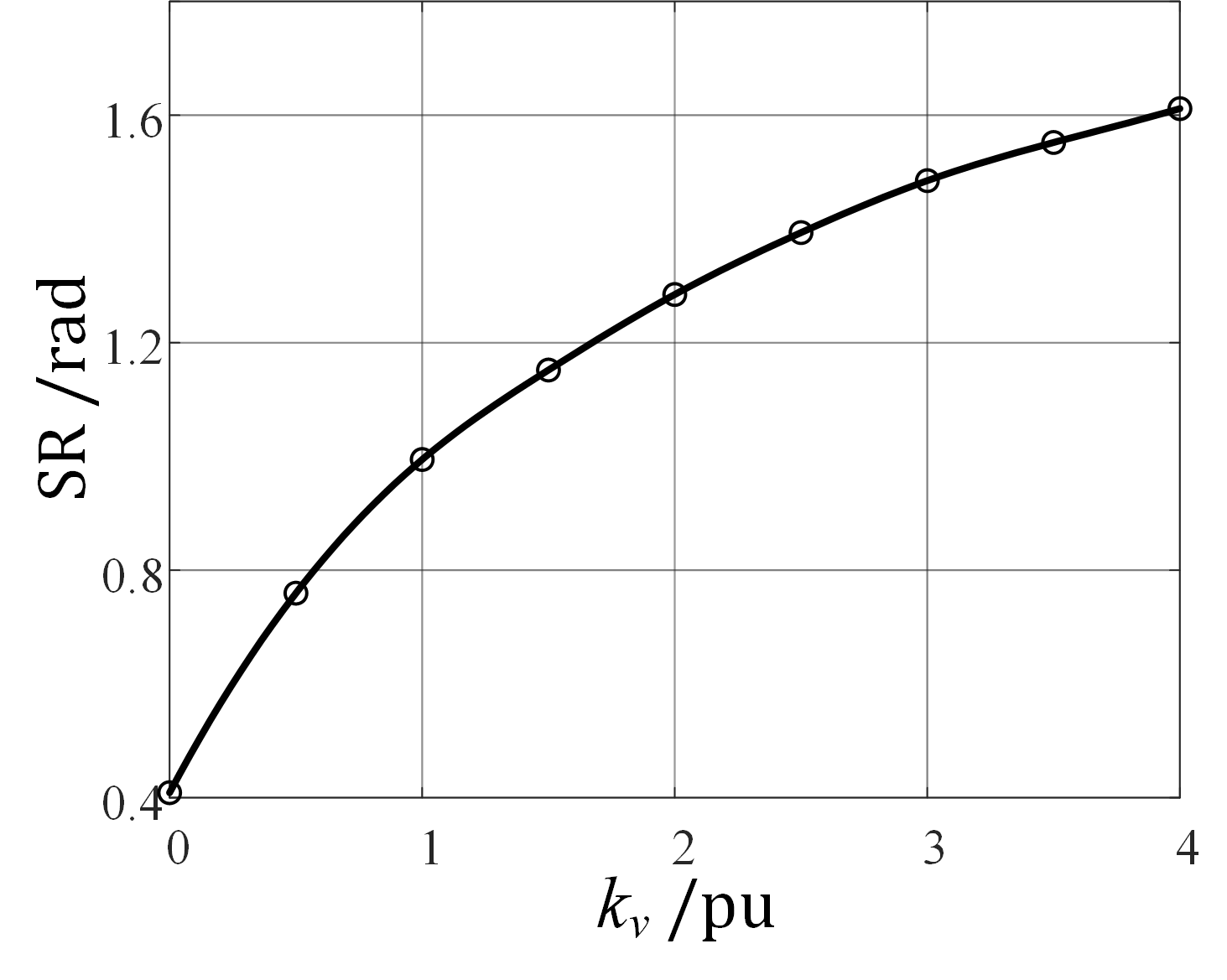}}\\
\subfloat[change of ROA]{\includegraphics[width=0.23\textwidth]{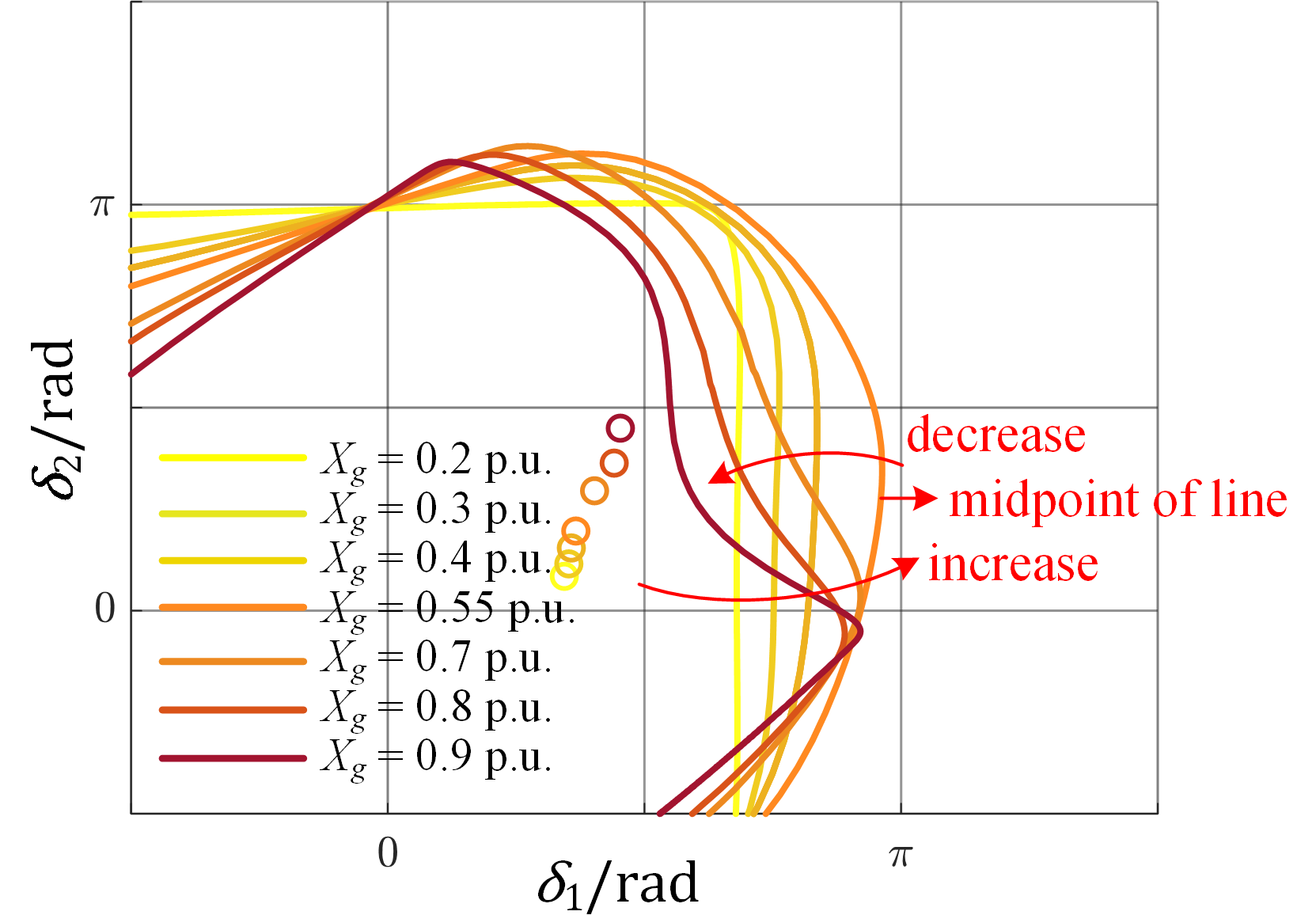}}\quad
\subfloat[change of SR]{\includegraphics[width=0.20\textwidth]{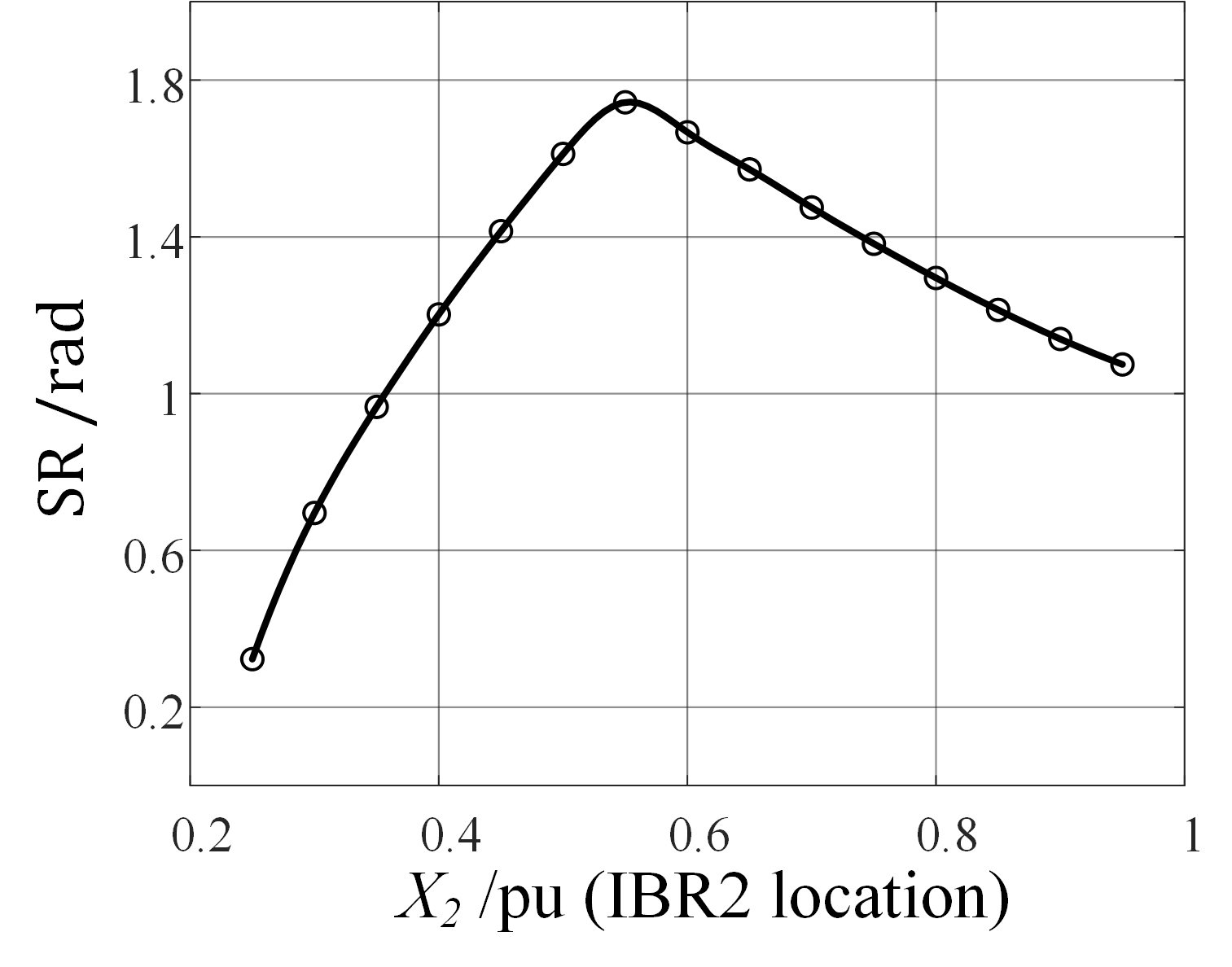}}
\caption{The change of ROA and SR with respect to (a)(b) voltage control gain $k_v$, and (c)(d) location of GSP inverter in the transmission line (The total impedance of transmission line $X_g+X_1=1.1$~pu is kept constant).}
\label{Sim_3_2}
\vspace{-0.1cm}
\end{figure}

This section investigates the impact of voltage support provided by a GSP inverter on the large-signal stability of a GFM inverter. 

Referring to the arrangement in \figref{fig2}, IBR1 is configured as a GFM inverter, feeding power to the grid via the transmission line $X_1 + X_g = 1.1$~pu. The output power of IBR1 is set to $P_{\text{ref}} = 0.8$~pu. IBR2 is configured as a GSP inverter and is connected to the transmission line through quite a small impedance $X_2 = 0.1$~pu. Detailed system parameters are provided in \tabref{TableA3} at Appendix \ref{parameters}. A three-phase-to-ground short-circuit fault is applied 80\% of the distance along the transmission line $X_g$ from the infinite bus.

The effect of voltage control of GSP inverter was tested by comparing performance with three choices of voltage control gains $k_v$ of 0, 2 and 4 and the resulting phase portraits are shown in \figref{Sim_3_1}~(a),~(b) and~(c). The CCTs estimated by SR of the reduced-order model for the three cases are $t_\text{SR} = 0.15$~s, $0.31$~s, and $0.45$~s, and the fault clearance time is set to these values, that is, $t_\text{fault}=t_\text{SR}$. The red solid lines are trajectories while the fault is applied and the blue solid lines are the post-fault trajectories, both from EMT simulation. The trajectories are also found from the integration of the reduced-order model and are shown as black dashed lines. It is seen that increasing $k_v$ leads to a corresponding increase in SR, thereby enhancing the system's large-signal stability. Without voltage control (pure GFL inverter), \figref{Sim_3_1}~(a), the SR and $t_{\text{SR}}$ is much smaller and making the system prone to instability, and adding voltage support in the transmission line of GFM inverters lessens the risk of large-signal instability. The results also further confirm SR can provide conservative and accurate estimations of large-signal stability.

\figref{Sim_3_2} shows how the ROA and SR change with system parameters.  The change of ROA with increasing $k_v$ is shown in subplot (a) (other system parameters remain constant) and subplot (b) is the corresponding change in SR with the system becoming more stable for large $k_v$ and approaching the situation where IBR2 was a GFM inverter. Although the large $k_v$ in the GSP inverter enhances large-signal stability, there is a tension with small-signal stability \cite{8743441} which will require a reduction in voltage control bandwidth which makes high $k_v$ impractical and thus, GFM inverters remain superior to GSP inverters in this context. Subplots (c) and (d) show the effect of the location of the GSP inverter on the ROA and SR. The value of $X_g$ is varied while keeping the total impedance for IBR1, $X_1 + X_g$,  constant at 1.1~pu. The results show that when IBR2 is placed too close to either IBR1 or the infinite bus, the ROA decreases. The ROA and SR are maximized when $X_1 \approx X_g$, consistent with the conclusions drawn in Section \ref{STATCOM_location}.

\section{Conclusions}
This paper recognizes the non-existence of a global energy function for systems containing GFL inverters and discusses the difficulties in large-signal stability analysis that follow from that non-existence. A manifold method was applied to reduced-order models of IBR to give an accurate estimation of the region of attraction (ROA), which defines the limit of large-signal stable operation. The method remains much more computationally efficient compared to direct time-domain integration of system dynamics. The accuracy of the method has been established in comparison to simulation. To evaluate the large-signal stability of multiple inverter systems, an assessment metric termed the stability radius (SR) has been proposed which is a circle centered at a stable equilibrium point (SEP) that exactly touches the ROA. The effectiveness of SR in giving a conservative estimate of the critical clearing time of faults for various combinations of GFL, GFM, and GSP inverters and variations of connection points has been confirmed by comparison with both PHIL experiments and EMT simulations. 

The large-signal stability of pairs of inverters, GFL-GFL, GFL-GFM, and GFL-GSP were assessed with the proposed analysis framework, combining the manifold method with model-order reduction together with the introduction of the SR metric. By accurately depicting the manifold between these inverters, a new large-signal instability mode was identified when the trajectories converged to a stable periodic orbit (SPO). Compared to a GFL inverter, a GSP inverter can provide voltage support, which improves the large-signal stability of the other IBR, whether GFM or GFL, particularly when that inverter is connected to the grid via weak transmission lines. The improvement is maximized if the GSP inverter is placed at the mid-point of the transmission line. When one of the two IBRs is configured as a GFM inverter, compared to a GSP inverter, improved voltage control is observed. Although with high voltage control gain $k_v$, the GSP case can approach the GFM case, but high droop gain is known to degrade the small-signal stability. Although our analysis focused on two-inverter systems, the proposed analysis framework could be extended to systems with more than two inverters.


%% file: figures/Table1.tex
\begin{table*}[htbp]
\centering
\setlength{\tabcolsep}{5.8pt}
\caption{Coefficients of the Expression of Two-Inverter System}\label{Table1}
\renewcommand{\arraystretch}{1.9} 
\begin{tabular}{l|l|cccccc|cccccc}
\hline \hline
IBR1   & IBR2 & $k_1$  & $A_1$  & $B_1$ & $C_1$ & $D_1$ & $\varepsilon$ & $k_2$ & $A_2$ & $B_2$ & $C_2$ & $D_2$ \\ \hline
GFM    & GFM  & $k_\text{droop1}$    &  $\frac{V_1V_2}{X_{\Delta 12}}$  &  $\frac{V_1U_g}{X_{\Delta g1}}$    &   $P_{\text{ref}1}$ &  0 & 0 & $k_\text{droop2}$ &  $\frac{V_1V_2}{X_{\Delta 12}}$   &  $\frac{V_2U_g}{X_{\Delta g2}}$   &  $P_{\text{ref2}}$  &  0 \\ \hline
GFM  & GFL 
& $k_\text{droop}$  &  0  &    $\frac{U_gV_1}{X_{\Sigma1}}$    &   $P_\text{ref}$ &   $\frac{X_gI_{d2}}{X_{\Sigma1}}V_1$ & 0 & $k_\text{PLL}$   &  $\frac{X_g}{X_{\Sigma1}}V_{1}$  &  $\frac{X_1}{X_{\Sigma1}}U_g$  &  $X_{2+1//g}I_{d2}$   &  0 \\ \hline
GFM  & GSP
& $k_\text{droop}$ &  $\frac{V_\text{ref}V_{1}}{X_{\Delta 12}}$  &  $\frac{V_{1}U_g}{X_{\Delta g1}}$ &  $ P_\text{ref}$  &  0  &  $\varepsilon_{mp}$ & $k_\text{PLL}$ &  $\frac{X_g}{X_{\Sigma1}}V_{1}$  &  $\frac{X_2}{X_{\Sigma2}}U_g$  &  $X_{2+1//g}I_{d2}$  &  0 \\ \hline
GFL  & GFL & $k_{\text{PLL}1}$ &  0  & $U_g$      &   $X_{\Sigma1}I_{d1}$  & $X_gI_{d2}$ & 0  &  $k_{\text{PLL}2}$   &  0  & $U_g$   &  $X_{\Sigma1}I_{d1}$  & $X_gI_{d1}$ \\ \hline
GFL  &  GSP
& $k_{\text{PLL}1}$   &  $\frac{X_g}{X_{\Sigma2}}V_\text{ref}$  &   $\frac{X_2}{X_{\Sigma2}}U_g$     &  $X_{1+2//g}I_{d1}$  &  0 & $\varepsilon_{lp}$ &  $k_{\text{PLL}2}$  &  0  &  $U_g$  &  $X_{\Sigma2}I_{d2}$  &  $X_gI_{d1}$ \\ \hline \hline
\end{tabular}
\end{table*}

%% file: appendix.tex
\appendices
\setcounter{table}{0}
\renewcommand{\thetable}{A\arabic{table}}
\section{Derivation of Model for Two-Inverter System} \label{Expression}
\subsection{The expression of Both IBRs are GFM inverters} \label{GFM-GFM}
\begingroup
\setlength{\abovedisplayskip}{6pt}
\setlength{\abovedisplayshortskip}{0pt}
\small
\begin{equation}
\begin{cases}
\dot{\delta}_1=k_{\text{droop}1}[P_{\text{ref1}}-\frac{V_1V_2\sin(\delta _1-\delta _2)}{X_{\varDelta 12}} -\frac{V_1U_g}{X_{\varDelta g1}}\sin \delta _1 ]\\
\dot{\delta}_2=k_{\text{droop}2}[P_{\text{ref2}}-\frac{V_1V_2\sin\left(\delta _2-\delta _1\right)}{X_{\varDelta 12}} -\frac{V_2U_g}{X_{\varDelta g2}}\sin \delta _2]\\
\end{cases}
\label{2GFM}
\end{equation}
\normalsize
\endgroup
where $\delta_1$, $\delta_2$, and $P_{\text{ref1}}$, $P_{\text{ref2}}$ represent the power angles, and the power references of the two GFM inverters, respectively. $V_1$ and $V_2$ are the voltages of two GFM inverters.
\vspace{-1mm}
\subsection{The expression of Both IBRs are GFL inverters} \label{GFL-GFL}
\vspace{-1mm}
\begingroup
\setlength{\abovedisplayskip}{6pt}
\setlength{\abovedisplayshortskip}{0pt}
\small
\begin{equation}
        \left\{ \begin{aligned}
        	\dot{\delta}_1=&k_{\text{PLL}1}\left[ X_{\Sigma1}I_{d1} + X_gI_{d2}\cos \left( \delta _2-\delta _1 \right)  -U_g\sin \delta _1  \right]\\
        	\dot{\delta}_2=&k_{\text{PLL}1}\left[ X_{\Sigma2}I_{d2} + X_gI_{d1}\cos \left( \delta _1-\delta _2 \right) -U_g\sin \delta _2 \right]\\
        \end{aligned} \right.
        \label{2GFL}
\end{equation}
\normalsize
\endgroup
where $\delta_1$, $\delta_2$, and $I_{d1}$, $I_{d2}$ represent the angles of PLL, and the current references on $d$-axis of the two GFL inverters, respectively.

\subsection{The expression of IBR1 is GFL inverter and IBR2 is GFM inverter} 
\vspace{-3mm}
\small
\begin{equation}
\left\{ \begin{aligned}
	\dot{\delta}_1=&k_{\text{PLL}}\left[ X_{1+2//g}I_{d1}-\frac{X_gV_{2}}{X_{\Sigma2}}\sin \left( \delta _1-\delta _2 \right) \right.\\
	&\left. -\frac{X_2U_g}{X_{\Sigma2}}\sin \delta _1 \right]\\
	\dot{\delta}_2=&k_{\text{droop}}\left[ P_{ref}+\frac{X_gI_{d1}V_2}{X_{\Sigma2}}\cos \left( \delta _1-\delta _2 \right) \right.\\
	&\left. -\frac{U_gV_2}{X_{\Sigma2}}\sin \delta _2\right]\\
\end{aligned} \right. 
\label{GFLGFM}
\end{equation}
\normalsize
\vspace{-3mm}
\subsection{The expression of IBR1 is GFL inverter and IBR2 is GSP inverter} 
\small
\begin{equation}
    \left\{ \begin{aligned}
    	\dot{\delta}_1=&k_{\text{PLL}1}\left[X_{1+2//g}I_{d1}-\frac{X_gV_{ref}}{X_{\Sigma2}}\sin \left( \delta_1-\delta_2 \right) \right.\\
         &\left.-\frac{X_2U_g}{X_{\Sigma2}}\sin \delta_1 +\varepsilon_{lp} (k_v,k_{\text{PLL}2},(\delta _1-\delta _2 ))\right]\\
        \dot{\delta}_2=&k_{\text{PLL}2}\left[ X_{\Sigma2}I_{d2} +X_gI_{d1}\cos \left( \delta _1-\delta _2 \right)  -U_g\sin \delta _2 \right]
    \end{aligned} \right. 
\label{GFL+GFLVS}
\end{equation}
\normalsize
where
\small
\begin{equation}
    \begin{aligned}
        \varepsilon_{lp} = &  \epsilon_{v1} g_1\left( \delta _1-\delta _2 \right) +\epsilon_{\text{PLL}} h_1\left( \delta _1-\delta _2 \right)\\
        g_1\left( \delta _1-\delta _2 \right)=&\frac{X_g}{X_{\Sigma2}}V_{ref}\sin \left( \delta _1-\delta _2 \right)\\
        &+\frac{X_g}{X_{\Sigma2}}U_g\cos \delta _2\sin \left( \delta _2-\delta _1 \right)\\ &+\frac{{X_g}^2}{X_{\Sigma2}}I_{d1}\sin \left( \delta _2-\delta _1 \right) ^2 \\
        h_1\left( \delta _1-\delta _2 \right) =&\frac{X_g}{X_{\Sigma2}}\cos \left( \delta _1-\delta _2 \right) \\
    \end{aligned}
    \label{GFL+GFLVS_spA}
\end{equation}
\normalsize
Eq. \eqref{GFL+GFLVS_spA} gives the residual term $\varepsilon_{lp}$, where $\epsilon_{v1}$ is voltage control error defined as $\epsilon_{v1} \triangleq \frac{1}{k_v X_{\Sigma2} + 1}$. $\epsilon_{\text{PLL}}$ is PLL error $\epsilon_{\text{PLL}} \triangleq \dot{\delta}_2 / k_{\text{PLL}2} = v_{q2}$, representing dynamics of the difference between the angle of PLL in IBR2 and its steady state. $g_1$ and $h_1$ are error functions. If $k_{\text{PLL}2}$ is sufficiently large, implying that the PLL dynamics of the GSP inverter are faster than those of IBR1, $\epsilon_{\text{PLL}}$ can be considered negligible. Moreover, if $k_v$ approaches infinity, $\epsilon_{v1}$ approaches zero, indicating that the GSP inverter behaves equivalently to a GFM inverter under ideal voltage regulation. Under this case ($\epsilon_{v1} \approx 0$, $\epsilon_{\text{PLL}1} \approx 0$), the sets of Equations \eqref{GFLGFM}  and \eqref{GFL+GFLVS} are identical, and they share the same form as shown in Eq. \eqref{general} in this paper.

\subsection{The expression of IBR1 is GFM inverter and IBR2 is GSP inverter} 
\small
\begin{equation}
\left\{ \begin{aligned}
    \dot{\delta}_1=&k_{\text{droop}}\left[ P_{ref}-\frac{V_{ref}V_{1}\sin \left( \delta _1-\delta _2 \right)}{X_{\varDelta 12}}\right.\\
    &\left.-\frac{V_{1}U_g\sin \delta _1}{X_{\varDelta g1}}+ \varepsilon_{mp} (k_v, k_{\text{PLL}}, (\delta _1-\delta _2))\right]\\
	\dot{\delta}_2=&k_{\text{PLL}}\left[ X_{2+1//g}I_{d2}-\frac{X_gV_{1}}{X_{\Sigma1}}\sin \left( \delta _2-\delta _1 \right) \right.\\
    &\left. -\frac{X_2U_g}{X_{\Sigma1}}\sin \delta _2 \right]\\
\end{aligned} \right. 
\label{GFM+GFLVS}
\end{equation}
\normalsize
where
\small
\begin{equation}
\scalebox{0.9}{$
    \begin{aligned}
    \varepsilon_{mp} = & \epsilon_{v2} g_2\left( \delta _1-\delta _2 \right) + \epsilon_{\text{PLL}} h_2\left( \delta _1-\delta _2 \right) \\
            g_2\left( \delta _1-\delta _2 \right) =&-\frac{X_g}{X_{\varDelta 12}X_{\Sigma1}}{V_1}^2\cos \left( \delta _1-\delta _2 \right) \sin \left( \delta _1-\delta _2 \right)\\
        &-\frac{X_1}{X_{\varDelta 12}X_{\Sigma1}}V_gV_1\cos \delta _2\sin \left( \delta _1-\delta _2 \right) \\
        &+\frac{V_1V_{ref}}{X_{\varDelta 12}}\sin \left( \delta _1-\delta _2 \right) \\
        h_2\left( \delta _1-\delta _2 \right) =&\frac{V_{1}\cos \left( \delta _1-\delta _2 \right)}{X_{\varDelta 12}}\\
    \end{aligned}
    \label{GFM+GFLVS_spA}$}
\end{equation}
\normalsize
Similarly, Eq. \eqref{GFM+GFLVS_spA} gives the residual term $\varepsilon_{mp}$, where $\epsilon_{v2}$ is voltage control error defined as $\epsilon_{v2} \triangleq {{1}/{\left( k_vX_{2+1//g}+1 \right)}}$. $\epsilon_{\text{PLL}}$ is PLL error, defined as above. $g_2$ and $h_2$ are the corresponding error functions. Likewise, if $k_v$ approaches infinity, $\epsilon_{v2}$ approaches zero, indicating that the GSP inverter behaves equivalently to a GFM inverter under ideal voltage regulation.

\section{Detailed Parameters in \\ PHIL Experiment and EMT Simulation} \label{parameters}
All EMT simulations and code for the calculation of ROA in this paper can be found in \cite{AppendixB}.
\vspace{-6pt}
\begin{table}[h]
\centering
\caption{PHIL Platform Parameters}\label{PHILPF}
\vspace{-6pt}
\renewcommand{\arraystretch}{1.05} 
\begin{tabular}{ll}
\hline\hline
\multicolumn{1}{l|}{Parameters}                                       & Value 
\\ \hline
\multicolumn{1}{l|}{Base capacity $S_b$}                 & 60~V$\cdot$A     \\
\multicolumn{1}{l|}{Base voltage $V_b$}                  & 22.8~V         \\
\multicolumn{1}{l|}{Base current $I_b$}                  & 2.6~A      \\
\multicolumn{1}{l|}{Base impedance $Z_b$}                & 8.64~$\mathrm{\Omega}$     \\
\multicolumn{1}{l|}{Base frequency $\omega_s/2\pi$}      & 50~Hz       \\ 
\multicolumn{1}{l|}{Grid voltage $U_g$}      & 1~pu       \\ 
\multicolumn{1}{l|}{LC filter of IBR1 $L_f$}                     & 1.7~mH (0.0618~pu)       \\ 
\multicolumn{1}{l|}{LC filter of IBR1 $C_f$}                     & 10~$\mu$F (0.027~pu)      \\ 
\multicolumn{1}{l|}{\makecell[lt]{Line impedance $Z_1$}}    &  \makecell[lt]{5.5~mH $+$ 55~m$\mathrm{\Omega}$ \\($0.2\mathrm{j}+0.006$ pu) \\or  13.8~mH $+$ 100~m$\mathrm{\Omega}$ \\($0.5\mathrm{j}+0.01$ pu)}               \\ 
\multicolumn{1}{l|}{\makecell[lt]{DC-link voltage of two 3-ph converters $U_{dc}$}}    & 54~V (2.4 pu)                \\ 

\multicolumn{1}{l|}{\makecell[lt]{Switching frequency of IBR1 \\(3-ph converter \#1) }}   & 10~kHz                 \\ 
\multicolumn{1}{l|}{\makecell[lt]{Switching frequency of power amplifier\\ (3-ph converter \#2)}}      & 50~kHz                  \\ 
\multicolumn{1}{l|}{Timestep of real-time simulation}                    & 20~$\mathrm{\mu}$s                \\ 
\multicolumn{1}{l|}{IBR1 control frequency}                             & 10~kHz                 \\ \hline\hline
\end{tabular}
\end{table}
\vspace{-6pt}
\begin{table}[h]
\centering
\begin{threeparttable}
\caption{Detailed Parameters in GFL-GFL System}
\vspace{-6pt}
\label{TableA1}
\renewcommand{\arraystretch}{1.05}
\begin{tabular}{l|l}
\hline\hline
Parameters                           & Value (pu)   \\ \hline
LC Filter of IBR2 $L_fj+r_f$                  & $0.05\mathrm{j}+0.001$ \hspace{6pt}   \\
LC Filter of IBR2 $C_f$                       & $0.02$         \\
Inner current control loop bandwidth  \hspace{6pt}        & 1 kHz          \\
PLL controller of IBR1 $k_{\text{PLL}1}$      & $10\times2\pi$ \\
PLL controller of IBR1 $k_{i}$     & $2\pi$         \\
PLL controller of IBR2 $k_{\text{PLL}2}$      & $10\times2\pi$ \\
PLL controller of IBR2 $k_{i}$     & $2\pi$         \\
Frequency limit of PLL $\omega_{\text{limit}}$       & $\pm0.2$       \\
Current reference of IBR1 $I_{d1}$            & 0.8            \\
Current reference of IBR2 $I_{d2}$            & 0.4            \\
Line impedance $Z_1$                          & $0.2j+0.006$   \\
Line impedance $Z_2$                          & $0.2j+0.002$   \\
Line impedance $Z_g$                          & $0.35\mathrm{j}$ or $0.4\mathrm{j}$ \\
Fault resistance $R_f$                        & 0.02           \\ \hline\hline
\end{tabular}
\begin{tablenotes}
\item The EMT simulation and PHIL experiment share the same per-unit (pu) parameters.
\end{tablenotes}
\end{threeparttable}
\end{table}
\vspace{-6pt}
\begin{table}[h]
\centering
\begin{threeparttable}
\captionsetup{width=\textwidth, justification=centering}
\caption{Detailed Parameters in GFL - GFM(GSP) System}\label{TableA2}
\vspace{-6pt}
\renewcommand{\arraystretch}{1.05} 
\begin{tabular}{ll}
\hline\hline
\multicolumn{1}{l|}{Parameters}                                                & Value (pu)                 \\ \hline
\multicolumn{1}{l|}{Line mpedance $Z_1$}                                           & $0.5\mathrm{j}+0.01$                   \\
\multicolumn{1}{l|}{Impedance (include virtual one) $Z_2$}                     & $0.1\mathrm{j}+0.002$                  \\
\multicolumn{1}{l|}{Line impedance $2Z_g$}                                          & $0.6\mathrm{j}+0.03$                    \\
\multicolumn{1}{l|}{Fault resistance $R_f$}                                    & 0.001                        \\
\multicolumn{1}{l|}{Fault position (from the infinite bus)}                    & 0.8                          \\ \hline
IBR1 - GFL                                                                     &                              \\ \hline
\multicolumn{1}{l|}{Inner current control loop bandwidth}                      & 1 kHz                        \\
\multicolumn{1}{l|}{PLL controller $k_{\text{PLL}1}$}                          & $2.5\times2\pi$              \\
\multicolumn{1}{l|}{PLL controller $k_i$}                         & $0.2\times2\pi$              \\
\multicolumn{1}{l|}{Frequency limit of PLL $\omega_{\text{limit}}$}            & $\pm0.2$                     \\
\multicolumn{1}{l|}{Current reference $I_{d1}$ }                               & 1                                         \\  \hline
IBR2 - GFM                                                                     &                              \\ \hline
\multicolumn{1}{l|}{$p-\omega$ droop gain $k_{\text{droop}}$}                  & $2.5\times2\pi$              \\
\multicolumn{1}{l|}{Power reference $P_{\text{ref}}$}                                 & \makecell[lt]{0.6 in Section \ref{STATCOM_Comparison}, \\0 in Section \ref{STATCOM_location}}\\
\multicolumn{1}{l|}{AC Voltage $V_{2}$}                                        & 1                            \\
\multicolumn{1}{l|}{Inner voltage control loop bandwidth}                      & 200 Hz                       \\
\multicolumn{1}{l|}{Inner current control loop bandwidth}                      & 1 kHz                        \\
\multicolumn{1}{l|}{LC filter of IBR2 $L_fj+r_f$}                           & $0.2\mathrm{j}+0.02$                   \\
\multicolumn{1}{l|}{LC filter of IBR2 $C_f$}                                & 0.1                          \\ 
\multicolumn{1}{l|}{Current limit $I_{\text{limit}}$}                                 & 5                            \\ 
\multicolumn{1}{l|}{$p-\omega$ droop time constant $\tau_p$}                   &  $1/(25\times2\pi)$                       \\ \hline
IBR2 - GSP                                                            &                              \\ \hline
\multicolumn{1}{l|}{Voltage droop $k_v$}                                       & \makecell[lt]{1 or 4 in Section \ref{STATCOM_Comparison},\\2 in Section \ref{STATCOM_location}} \\
\multicolumn{1}{l|}{Voltage reference $V_{\text{ref}}$}                               & 1                            \\
\multicolumn{1}{l|}{Current reference $I_{d2}$}                                & \makecell[lt]{0.6 in Section \ref{STATCOM_Comparison}, \\0 in Section \ref{STATCOM_location}} \\
\multicolumn{1}{l|}{Voltage droop filter time scale $\tau_v$}                  & $1/(50\times2\pi)$               \\
\multicolumn{1}{l|}{PLL controller $k_{\text{PLL}2}$}                          & $k_{\text{droop}}\cdot\frac{1}{X_{\Sigma2}}$ \\ \hline\hline
\end{tabular}

\begin{tablenotes}
\footnotesize
\item The EMT simulation and PHIL experiment share the same pu parameters.
\end{tablenotes}
\end{threeparttable} 
\end{table}
\vspace{-6pt}
\begin{table}[h]
\centering
\caption{Detailed Parameters in GFM - GSP System }\label{TableA3}
\vspace{-6pt}
\renewcommand{\arraystretch}{1.05} 
\begin{tabular}{ll}
\hline\hline
\multicolumn{1}{l|}{Parameters}                                                & Value (pu)                 \\ \hline
\multicolumn{1}{l|}{Line impedance $Z_1$}                                           & $0.5\mathrm{j}+0.01$                   \\
\multicolumn{1}{l|}{Line impedance $Z_2$}                                           & $0.15\mathrm{j}+0.002$                  \\
\multicolumn{1}{l|}{Line impedance $2Z_g$}                                          & $0.65\mathrm{j}+0.03 $                   \\
\multicolumn{1}{l|}{Fault resistance $R_f$}                                    & 0.001                        \\
\multicolumn{1}{l|}{Fault position (from the infinite bus)}                    & 0.8                          \\ \hline
IBR1 - GFM                                                                     &                              \\ \hline
\multicolumn{1}{l|}{$p-\omega$ droop gain $k_{\text{droop}}$}                           & $2.5\times2\pi$              \\
\multicolumn{1}{l|}{Power reference $P_{\text{ref}}$}                                 & 0.8                          \\
\multicolumn{1}{l|}{AC Voltage $V_{1}$}                                        & 1                            \\
\multicolumn{1}{l|}{Inner voltage control loop bandwidth}                      & 200 Hz                       \\
\multicolumn{1}{l|}{Inner current control loop bandwidth}                      & 1 kHz                        \\ 
\multicolumn{1}{l|}{Current limit $I_{\text{limit}}$}                                 & 5                            \\ 
\multicolumn{1}{l|}{$p-\omega$ droop time constant $\tau_p$}                   &  $1/(25\times2\pi)$ \\ \hline
IBR2 - GSP                                                            &                              \\ \hline
\multicolumn{1}{l|}{voltage control gain $k_v$}                                           & 0, 2, or 4                   \\
\multicolumn{1}{l|}{Voltage reference $V_{\text{ref}}$}                               & 1                            \\
\multicolumn{1}{l|}{Current reference $I_{d2}$}                                & 0.2                          \\
\multicolumn{1}{l|}{Voltage control filter time scale $\tau_v$}                                    & $1/(50\times2\pi)$               \\
\multicolumn{1}{l|}{PLL controller $k_{\text{PLL}2}$}                                 & $6\times2\pi$                  \\ \hline\hline
\end{tabular}
\end{table}

%% file: References.bib
@article{Gu_proceeding,
  author={Gu, Yunjie and Green, Timothy C.},
  journal={Proceedings of the IEEE}, 
  title={{Power System Stability With a High Penetration of Inverter-Based Resources}}, 
  year={2023},
  volume={111},
  number={7},
  pages={832-853},
  doi={10.1109/JPROC.2022.3179826}
}

@article{hu2019large,
  author={Hu, Qi and Fu, Lijun and Ma, Fan and Ji, Feng},
  journal={IEEE Transactions on Power Systems}, 
  title={{Large Signal Synchronizing Instability of PLL-Based VSC Connected to Weak AC Grid}}, 
  year={2019},
  volume={34},
  number={4},
  pages={3220-3229},
  doi={10.1109/TPWRS.2019.2892224}
}

@ARTICLE{Xiongfei_overview,
  author={Wang, Xiongfei and Taul, Mads Graungaard and Wu, Heng and Liao, Yicheng and Blaabjerg, Frede and Harnefors, Lennart},
  journal={IEEE Open Journal of Industry Applications}, 
  title={{Grid-Synchronization Stability of Converter-Based Resources—An Overview}}, 
  year={2020},
  volume={1},
  number={},
  pages={115-134},
  doi={10.1109/OJIA.2020.3020392}
}

@ARTICLE{Genghua_pll,
  author={He, Xiuqiang and Geng, Hua},
  journal={IEEE Transactions on Industry Applications}, 
  title={{PLL Synchronization Stability of Grid-Connected Multiconverter Systems}}, 
  year={2022},
  volume={58},
  number={1},
  pages={830-842},
  doi={10.1109/TIA.2021.3121262}
}

@book{chiang2011direct,
  title={{Direct Methods for Stability Analysis of Electric Power Systems: Theoretical Foundation, BCU Methodologies, and Applications}},
  author={Chiang, Hisao-Dong},
  year={2011},
  publisher={John Wiley \& Sons}
}

@inproceedings{pota1989new,
  author={Pota, H.R. and Moylan, P.J.},
  booktitle={Proceedings of the 28th IEEE Conference on Decision and Control,}, 
  title={{A New Lyapunov Function for Interconnected Power Systems}}, 
  year={1989},
  volume={},
  number={},
  pages={2181-2185 vol.3},
  doi={10.1109/CDC.1989.70554}
}

@article{chiang1989study,
  author={{H.-D. Chiang}},
  journal={IEEE Transactions on Circuits and Systems}, 
  title={{Study of the Existence of Energy Functions for Power Systems with Losses}}, 
  year={1989},
  volume={36},
  number={11},
  pages={1423-1429},
  doi={10.1109/31.41298}
}

@article{bretas2003lyapunov,
  title={{Lyapunov Function for Power Systems with Transfer Conductances: Extension of the Invariance Principle}},
  author={Bretas, Newton G and Alberto, Luis FC},
  journal={IEEE Transactions on Power Systems},
  volume={18},
  number={2},
  pages={769--777},
  year={2003},
  publisher={IEEE}
}

@article{xue1989extended,
  title={{Extended Equal Area Criterion Justifications, Generalizations, Applications}},
  author={Xue, Yusheng and Van Custem, Thierry and Ribbens-Pavella, Mania},
  journal={IEEE Transactions on Power Systems},
  volume={4},
  number={1},
  pages={44--52},
  year={1989},
  publisher={IEEE}
}

@article{fu2022cascading,
  title={{Cascading Synchronization Instability in Multi-VSC Grid-Connected System}},
  author={Fu, Xikun and Huang, Meng and Pan, Shangzhi and Zha, Xiaoming},
  journal={IEEE Transactions on Power Electronics},
  volume={37},
  number={7},
  pages={7572--7576},
  year={2022},
  publisher={IEEE}
}

@article{yi2022transient,
  title={{Transient Synchronization Stability Analysis and Enhancement of Paralleled Converters Considering Different Current Injection Strategies}},
  author={Yi, Xiangtong and Peng, Yelun and Zhou, Quan and Huang, Wen and Xu, Lijia and Shen, Z John and Shuai, Zhikang},
  journal={IEEE Transactions on Sustainable Energy},
  volume={13},
  number={4},
  pages={1957--1968},
  year={2022},
  publisher={IEEE}
}

@article{shen2020transient,
  title={{Transient Stability and Current Injection Design of Paralleled Current-Controlled VSCs and Virtual Synchronous Generators}},
  author={Shen, Chao and Shuai, Zhikang and Shen, Yang and Peng, Yelun and Liu, Xuan and Li, Zuyi and Shen, Z John},
  journal={IEEE Transactions on Smart Grid},
  volume={12},
  number={2},
  pages={1118--1134},
  year={2020},
  publisher={IEEE}
}

@article{me2023transient,
  title={{Transient Stability of Paralleled Virtual Synchronous Generator and Grid-Following Inverter}},
  author={Me, Si Phu and Ravanji, Mohammad Hasan and Mansour, Milad Zarif and Zabihi, Sasan and Bahrani, Behrooz},
  journal={IEEE Transactions on Smart Grid},
  year={2023},
  publisher={IEEE}
}

@article{xue2023transient,
  title={{Transient Stability Analysis and Enhancement Control Strategies for Interconnected AC Systems with VSC-Based Generations}},
  author={Xue, Yicheng and Zhang, Zheren and Zhang, Nan and Hua, Wen and Wang, Guoteng and Xu, Zheng},
  journal={International Journal of Electrical Power \& Energy Systems},
  volume={149},
  pages={109017},
  year={2023},
  publisher={Elsevier}
}

@inproceedings{wang2023fixed,
  title={{Fixed Angle Difference Control for Multi-Converter Grid-Connected System to Improve Transient Stability During Grid Faults}},
  author={Wang, Yukun and Zhu, Donghai and Yang, Yihang and Zou, Xudong and Hu, Jiabing and Kang, Yong},
  booktitle={2023 IEEE 6th International Electrical and Energy Conference (CIEEC)},
  pages={4351--4356},
  year={2023},
  organization={IEEE}
}

@inproceedings{zhao2018transient,
  title={{Transient Stability Analysis of Grid-Connected VSIs via PLL Interaction}},
  author={Zhao, Jiantao and Huang, Meng and Zha, Xiaoming},
  booktitle={2018 IEEE International Power Electronics and Application Conference and Exposition (PEAC)},
  pages={1--6},
  year={2018},
  organization={IEEE}
}

@article{wu2019design,
  title={{Design-Oriented Transient Stability Analysis of PLL-Synchronized Voltage-Source Converters}},
  author={Wu, Heng and Wang, Xiongfei},
  journal={IEEE Transactions on Power Electronics},
  volume={35},
  number={4},
  pages={3573--3589},
  year={2019},
  publisher={IEEE}
}

@article{li2022whole,
  title={{Whole-System First-Swing Stability of Inverter-Based Inertia-Free Power Systems}},
  author={Li, Yitong and Gu, Yunjie},
  journal={arXiv preprint arXiv:2207.03292},
  year={2022}
}

@inproceedings{zhao2020reactive,
  title={{Reactive Power Compensation Control of PV Systems for Improved Power Transfer Capability in Weak Grid}},
  author={Zhao, Chen and Sun, Dan and Nian, Heng and Fan, Yue},
  booktitle={2020 12th IEEE PES Asia-Pacific Power and Energy Engineering Conference (APPEEC)},
  pages={1--5},
  year={2020},
  organization={IEEE}
}

@article{netz2006grid,
  title={{Grid Code—High and Extra High Voltage. E. ON Netz GmbH}},
  author={Netz, EON},
  journal={Tech. Rep},
  year={2006}
}

@article{Xinshuo2022,
   author = {Wang, Xinshuo and Wu, Heng and Wang, Xiongfei and Dall, Laurids and Kwon, Jun Bum},
   title = {{Transient Stability Analysis of Grid-Following VSCs Considering Voltage-Dependent Current Injection during Fault Ride-Through}},
   journal = {IEEE Transactions on Energy Conversion},
   pages = {1-13},
   ISSN = {0885-8969
1558-0059},
   DOI = {10.1109/tec.2022.3204358},
   year = {2022},
   type = {Journal Article}
}

@book{chiang2015stability,
  title={{Stability Regions of Nonlinear Dynamical Systems: Theory, Estimation, and Applications}},
  author={Chiang, Hsiao-Dong and Alberto, Lu{\'\i}s FC},
  year={2015},
  publisher={Cambridge University Press}
}

@article{fang2019distanceprotection,
   author = {Fang, Yu and Jia, Ke and Yang, Zhe and Li, Yanbin and Bi, Tianshu},
   title = {{Impact of Inverter-Interfaced Renewable Energy Generators on Distance Protection and an Improved Scheme}},
   journal = {IEEE Transactions on Industrial Electronics},
   volume = {66},
   number = {9},
   pages = {7078-7088},
   ISSN = {0278-0046
1557-9948},
   DOI = {10.1109/tie.2018.2873521},
   year = {2019},
   type = {Journal Article}
}

@ARTICLE{Zaborszky1988,
  author={Zaborszky, J. and Huang, G. and Zheng, B. and Leung, T.-C.},
  journal={{IEEE Transactions on Automatic Control}}, 
  title={{On the Phase Portrait of a Class of Large Nonlinear Dynamic Systems Such 
  as the Power System}}, 
  year={1988},
  volume={33},
  number={1},
  pages={4-15},
  doi={10.1109/9.356}
}

@ARTICLE{XiuqiangMultiGFM,
  author={He, Xiuqiang and Pan, Sisi and Geng, Hua},
  journal={{IEEE Transactions on Energy Conversion}}, 
  title={{Transient Stability of Hybrid Power Systems Dominated by Different Types of Grid-Forming Devices}}, 
  year={2022},
  volume={37},
  number={2},
  pages={868-879},
  doi={10.1109/TEC.2021.3113399}
}

@INPROCEEDINGS{Quisepe_distance,
  author={Quispe H, Juan Carlos and Villarroel-Gutiérrez, Héctor and Orduña, Eduardo},
  booktitle={2020 IEEE PES Transmission \& Distribution Conference and Exhibition - Latin America (T \& D LA)}, 
  title={{Analyzing Short-Circuit Current Behavior Caused by Inverter-Interfaced Renewable Energy Sources. Effects on Distance Protection}}, 
  year={2020},
  volume={},
  number={},
  pages={1-6},
  doi={10.1109/TDLA47668.2020.9326226}}

@inproceedings{wang2023transient,
  title={{Transient Overvoltage Analysis of Grid-Following VSCs during Fault Recovery}},
  author={Wang, Xinshuo and Wu, Heng and Wang, Xiongfei},
  booktitle={2023 11th International Conference on Power Electronics and ECCE Asia (ICPE 2023-ECCE Asia)},
  pages={2404--2409},
  year={2023},
  organization={IEEE}
}

@ARTICLE{8743441,
  author={Huang, Linbin and Xin, Huanhai and Li, Zhiyi and Ju, Ping and Yuan, Hui and Lan, Zhou and Wang, Zhen},
  journal={IEEE Transactions on Smart Grid}, 
  title={{Grid-Synchronization Stability Analysis and Loop Shaping for PLL-Based Power Converters With Different Reactive Power Control}}, 
  year={2020},
  volume={11},
  number={1},
  pages={501-516},
  doi={10.1109/TSG.2019.2924295}}

@ARTICLE{9309174,
  author={Fu, Xikun and Sun, Jianjun and Huang, Meng and Tian, Zhen and Yan, Han and Iu, Herbert Ho-Ching and Hu, Pan and Zha, Xiaoming},
  journal={IEEE Transactions on Power Electronics}, 
  title={{Large-Signal Stability of Grid-Forming and Grid-Following Controls in Voltage Source Converter: A Comparative Study}}, 
  year={2021},
  volume={36},
  number={7},
  pages={7832-7840},
  doi={10.1109/TPEL.2020.3047480}}

@ARTICLE{8632731,
  author={Taul, Mads Graungaard and Wang, Xiongfei and Davari, Pooya and Blaabjerg, Frede},
  journal={IEEE Transactions on Power Electronics}, 
  title={{An Overview of Assessment Methods for Synchronization Stability of Grid-Connected Converters Under Severe Symmetrical Grid Faults}}, 
  year={2019},
  volume={34},
  number={10},
  pages={9655-9670},
  doi={10.1109/TPEL.2019.2892142}}

@ARTICLE{Gu_reduction,
  author={Gu, Yunjie and Bottrell, Nathaniel and Green, Timothy C.},
  journal={IEEE Transactions on Power Electronics}, 
  title={{Reduced-Order Models for Representing Converters in Power System Studies}}, 
  year={2018},
  volume={33},
  number={4},
  pages={3644-3654},
  doi={10.1109/TPEL.2017.2711267}}

@misc{AppendixB,
  title        = {},
  howpublished = {[Online]. Available: \url{https://github.com/Future-Power-Networks/Publications/blob/main/LargeSignal}},
  year         = {},
  note         = {}
}

@ARTICLE{GSP,
  author={Khajehoddin, Sayed Ali and Karimi-Ghartemani, Masoud and Ebrahimi, Mohammad},
  journal={IEEE Transactions on Industrial Electronics}, 
  title={Grid-Supporting Inverters With Improved Dynamics}, 
  year={2019},
  volume={66},
  number={5},
  pages={3655-3667},
  doi={10.1109/TIE.2018.2850002}}

@online{InsRealm,
  author       = {{InsRealm Technology Co., Ltd.}},
  title        = {Official website},
  year         = {2025},
  url          = {https://www.insrealm.co.uk/},
  note         = {Accessed: Sep. 17, 2025}
}
